\RequirePackage{lineno}
\documentclass[aps,onecolumn,tightenlines,superscriptaddress,floatfix,nofootinbib,twoside,longbibliography]{revtex4-1} 

\usepackage{etoolbox}
\patchcmd{\section}
  {\centering}
  {\raggedright}
  {}
  {}
\patchcmd{\subsection}
  {\centering}
  {\raggedright}
  {}
  {}

\patchcmd{\subsubsection}
  {\centering}
  {\raggedright}
  {}
  {}

\pdfoutput=1

\usepackage{enumitem}
\usepackage{aas_macros}
\def\jcap{{JCAP}}

\usepackage{hyperref}
\usepackage[usenames,dvipsnames]{xcolor}
\hypersetup{
    colorlinks = true,
    citecolor = {MidnightBlue},
    linkcolor = {BrickRed},
    urlcolor = {BrickRed}
}

\usepackage{latexsym}
\usepackage{graphicx}
\usepackage{url}
\usepackage{mathrsfs,amssymb,amsmath,amsfonts,amstext,graphics, multirow}
\usepackage[export]{adjustbox}
\usepackage{epstopdf}
\usepackage{appendix}
\usepackage{color,slashed}
\usepackage{tikz}
\usepackage{cancel}
\usepackage{mciteplus}
\usepackage{verbatim}
\usepackage[normalem]{ulem} 
\usetikzlibrary{arrows,shapes}
\usetikzlibrary{trees}
\usetikzlibrary{matrix,arrows} 
\usepackage{xspace}
\usepackage{wrapfig}
\usepackage{rotating}
\usepackage{dcolumn}
\usepackage{bm}
\usepackage{longtable}
\usepackage{changepage} 
\usepackage[binary-units=true]{siunitx} 

\newcommand{\omegahi}{\Omega_{\rm HI}}
\newcommand{\neff}{N_{\rm eff}}
\newcommand{\fnl}{f_{\rm NL}}

\newcommand{\stagetwo}{Stage~{\sc ii}}
\newcommand{\stageone}{Stage~{\sc i}}

\usepackage{colortbl}

\usepackage{mdframed}
\usepackage{times}
\usepackage{pbox}
\usepackage{wrapfig}
\usepackage{tcolorbox}

\newenvironment{WrapText}[1][r]
  {\wrapfigure{#1}{0.5\textwidth}\tcolorbox}
  {\endtcolorbox\endwrapfigure}

\def\beq{\begin{equation}}
\def\eeq{\end{equation}}

\def\bea{\begin{eqnarray}}
\def\eea{\end{eqnarray}}

\newcommand{\vk}{\mathbf{k}}
\newcommand{\kpar}{k_{\parallel}}
\newcommand{\kperp}{k_{\perp}}

\def\startframe{\begin{mdframed}[topline=false,rightline=false,bottomline=false,linewidth=0.8mm,linecolor=lightgray]
\hspace*{-\parindent}}
\def\endframe{\end{mdframed}}

\usepackage[utf8]{inputenc}											
\usepackage{relsize}

\newcolumntype{C}[1]{>{\centering\let\newline\\\arraybackslash\hspace{0pt}}m{#1}}

\usepackage{etoolbox}

\newtoggle{DEBUG}
\togglefalse{DEBUG}

\iftoggle{DEBUG}{
\linenumbers
\def\as#1{{\bf \textcolor{blue}{[{AS:#1}]}}}
\def\mw#1{{\bf \textcolor{blue}{[{MW:#1}]}}}
\def\dm#1{{\bf \textcolor{green}{[{DM:#1}]}}}
\def\sjf#1{{\bf \textcolor{purple}{[{SF: #1}]}}}
\def\fv#1{{\bf \textcolor{orange}{[{FV: #1}]}}}
\def\acl#1{{\bf \textcolor{red}{[{ACL:#1}]}}}
\def\kmb#1{{\bf \textcolor{olive}{[{KMB:#1}]}}}
\def\ao#1{{\bf \textcolor{magenta}{[{AO:#1}]}}}
\def\ec#1{{\bf \textcolor{magenta}{[{EC:#1}]}}}
\def\eja#1{{\bf \textcolor{CadetBlue}{[{EJA:#1}]}}}
\def\cs#1{{\bf \textcolor{cyan}{[{CS:#1}]}}}
\def\pvm#1{\bf \textcolor{yellow}{[{PM:#1}]}}
\def\bw#1{{\bf \textcolor{blue}{[{BW:#1}]}}}
}{
\def\as#1{}
\def\mw#1{}
\def\dm#1{}
\def\sjf#1{}
\def\fv#1{}
\def\acl#1{}
\def\kmb#1{}
\def\ao#1{}
\def\ec#1{}
\def\eja#1{}
\def\cs#1{}
\def\pvm#1{}
\def\bw#1{}
}

\makeatletter
\def\l@subsubsection#1#2{}
\renewcommand{\p@subsection}{}
\renewcommand{\p@subsubsection}{}
\makeatother

\AtBeginDocument{%
	\newwrite\bibnotes
	\def\bibnotesext{Notes.bib}
	\immediate\openout\bibnotes=\jobname\bibnotesext
	\immediate\write\bibnotes{@CONTROL{REVTEX41Control}}
	\immediate\write\bibnotes{@CONTROL{%
			apsrev41Control,author="08",editor="1",pages="1",title="0",year="1"}}
	\if@filesw
	\immediate\write\@auxout{\string\citation{apsrev41Control}}%
	\fi
}%

\newcommand{\ParisSud}{LAL, Universit\'{e} Paris-Sud, 91898~Orsay~Cedex, France \& CNRS/IN2P3, 91405~Orsay, France}

\newcommand{\AmsterdamAstro}{Anton Pannekoek Institute for Astronomy, University of Amsterdam, 1098~XH Amsterdam, The Netherlands}

\newcommand{\BNL}{Brookhaven National Laboratory, Upton, NY~11973, USA}

\newcommand{\Caltech}{California Institute of Technology, Pasadena, CA~91125, USA}

\newcommand{\FNAL}{Fermi National Accelerator Laboratory, Batavia, IL~60510, USA}

\newcommand{\IAS}{Institute for Advanced Study, Princeton, NJ~08540, USA}

\newcommand{\UNIPD}{Dipartimento di Fisica e Astronomia ``G. Galilei'', Universit\`a degli Studi di Padova, 35131~Padova, Italy}
\newcommand{\JPL}{Jet Propulsion Laboratory, California Institute of Technology, Pasadena, CA~91109, USA}

\newcommand{\UCBP}{Department of Physics, University of California Berkeley, Berkeley, CA~94720, USA}
\newcommand{\UCB}{Department of Astronomy, University of California Berkeley, Berkeley, CA~94720, USA}

\newcommand{\McGill}{McGill University, Montreal, QC~H3A~2T8, Canada}
\newcommand{\MIT}{Massachusetts Institute of Technology, Cambridge, MA~02139, USA}

\newcommand{\ORNL}{Oak Ridge National Laboratory, Oak Ridge, TN~37831, USA}

\newcommand{\PI}{Perimeter Institute, Waterloo, ON~N2L~2Y5, Canada}

\newcommand{\SLAC}{SLAC National Accelerator Laboratory, Menlo Park, CA~94025, USA}

\newcommand{\UBC}{University of British Columbia, Vancouver, BC~V6T~1Z1, Canada}

\newcommand{\UCSD}{University of California San Diego, La Jolla, CA~92093, USA}

\newcommand{\CITA}{Canadian Institute for Theoretical Astrophysics, University of Toronto, Toronto, ON~M5S~3H8, Canada}
\newcommand{\SISSA}{International School for Advanced Studies~(SISSA), 34136~Trieste, Italy}
\newcommand{\RAL}{Radio Astronomy Laboratory, University of California Berkeley, Berkeley, CA~94720, USA}
\newcommand{\kavli}{Kavli Institute for Cosmology, Cambridge CB3~0HA, UK}
\newcommand{\ioa}{Institute of Astronomy, University of Cambridge, Cambridge CB3~0HA, UK}
\newcommand{\damtp}{DAMTP, University of Cambridge, Cambridge CB3~0WA, UK}
\newcommand{\RUG}{Kapteyn Astronomical Institute, University of Groningen, 9700~AV Groningen, The~Netherlands}
\newcommand{\VSI}{Van Swinderen Institute for Particle Physics and Gravity, University of Groningen, 9747~AG~Groningen, The~Netherlands}
\newcommand{\CCA}{Center for Computational Astrophysics, Flatiron Institute, New York, NY~10010, USA}
\newcommand{\Yale}{Department of Physics, Yale University, New Haven, CT~06520, USA}
\newcommand{\WVU}{CSEE, West Virginia University, Morgantown, WV~26505, USA}
\newcommand{\WVUGWAC}{Center for Gravitational Waves and Cosmology, West Virginia University, Morgantown, WV~26505, USA}
\newcommand{\UWMadison}{Department of Physics, University of Wisconsin-Madison, Madison, WI~53706, USA}
\newcommand{\StonyBrook}{Stony Brook University, Stony Brook, NY~11794, USA}
\newcommand{\UWaterloo}{Department of Physics and Astronomy, University of Waterloo, Waterloo, ON~N2L~3G1, Canada}
\newcommand{\INFN}{National Institute for Nuclear Physics (INFN), 34127~Trieste, Italy}
\newcommand{\YorkU}{York University, Toronto, ON~M3J~1P3, Canada}

\begin{document}

\pagestyle{plain}
\title{\texorpdfstring{\Large Cosmic Visions Dark Energy: \\Inflation and Early
Dark Energy with a \stagetwo\ \\ Hydrogen Intensity Mapping Experiment}{Cosmic 
Visions Dark Energy: Inflation and Early Dark Energy with a \stagetwo\ Hydrogen 
Intensity Mapping Experiment}}


\collaboration{\textbf{Cosmic Visions 21$\,$cm Collaboration}}

\author{R\'{e}za Ansari}
\affiliation{\ParisSud}

\author{Evan J. Arena}
\affiliation{\BNL}
\affiliation{\StonyBrook}

\author{Kevin Bandura}
\affiliation{\WVU}
\affiliation{\WVUGWAC}

\author{Philip Bull}
\affiliation{\UCB}
\affiliation{\RAL}

\author{Emanuele Castorina}
\affiliation{\UCBP}

\author{Tzu-Ching Chang}
\affiliation{\JPL}
\affiliation{\Caltech}

\author{Shi-Fan Chen}
\affiliation{\UCBP}

\author{Liam Connor}
\affiliation{\AmsterdamAstro}

\author{Simon Foreman}
\affiliation{\CITA}

\author{Josef Frisch}
\affiliation{\SLAC}

\author{Daniel Green}
\affiliation{\UCSD}

\author{Matthew C. Johnson}
\affiliation{\YorkU}
\affiliation{\PI}

\author{Dionysios Karagiannis}
\affiliation{\UNIPD}

\author{Adrian Liu}
\affiliation{\UCB}
\affiliation{\RAL}
\affiliation{\McGill}

\author{Kiyoshi W. Masui}
\affiliation{\MIT}

\author{P. Daniel Meerburg}
\affiliation{\kavli}
\affiliation{\ioa}
\affiliation{\damtp}
\affiliation{\RUG}
\affiliation{\VSI}

\author{Moritz M{\"u}nchmeyer}
\affiliation{\PI}

\author{Laura B. Newburgh}
\affiliation{\Yale}

\author{Andrej Obuljen}
\affiliation{\SISSA}
\affiliation{\INFN}
\affiliation{\UWaterloo}

\author{Paul O'Connor}
\affiliation{\BNL}

\author{Hamsa Padmanabhan}
\affiliation{\CITA}

\author{J. Richard Shaw}
\affiliation{\UBC}

\author{Chris Sheehy}
\affiliation{\BNL}

\author{An\v{z}e Slosar}
\thanks{Corresponding author. E-mail: \href{mailto:anze@bnl.gov}{anze@bnl.gov}}
\affiliation{\BNL}

\author{Kendrick Smith}
\affiliation{\PI}

\author{Paul Stankus}
\affiliation{\ORNL}

\author{Albert Stebbins}
\affiliation{\FNAL}

\author{Peter Timbie}
\affiliation{\UWMadison}

\author{Francisco Villaescusa-Navarro}
\affiliation{\CCA}

\author{Benjamin Wallisch}
\affiliation{\UCSD}
\affiliation{\IAS}

\author{Martin White}
\affiliation{\UCB}

\date{\today}

\maketitle
\cleardoublepage
\tableofcontents

\clearpage

\section*{Version history}
\noindent

This document is updated as the \stagetwo\ concept evolves and new results come to the fore. Release versions are described in the table below.

\vspace*{1cm}

\begin{tabular}{c|C{5cm}|p{10cm}}
  Version & Release Date, arXiv entry & Comments\\
\hline
  v2.0    & Jul 2019, \href{https://arxiv.org/pdf/1810.09572v3}{arXiv:1810.09571v3} &
                                                                                            Updated to match the PUMA APC submission to the Astronomy and Astrophysics Decadal Survey \cite{PUMAAPC}.  Stage 2 concept is 32000 dish interferometric array of hexagonally close-packed (at 50\% fill factor) 6m dishes operating at $z=0.3-6$. Modeling of system performance across ultra-wide band is made more realistic. Science section has been updated with new science opportunities.\\

  v1.0    & Oct 2018, \href{https://arxiv.org/pdf/1810.09572v1}{arXiv:1810.09571v1} &
                                                                                            Completed Version submitted to DOE: Stage 2 concept is 256$\times$256 interferometric array of 6m dishes operating at $z=2-6$\\
  
  \hline
\end{tabular}

\cleardoublepage\null\newpage



\section*{Preamble}
\noindent
The Department of Energy (DOE) of the United States government has
tasked several Cosmic Visions committees to work with relevant
communities to make strategic plans for the future experiments in the
Cosmic Frontier of the High Energy Physics effort within the DOE
Office of Science. The Cosmic Visions Dark Energy committee was the
most open-ended, with a broad effort to study periods of accelerated
expansion in the Universe, both early and late, using surveys. It has
conducted two community workshops and produced two white
papers~\citep{2016arXiv160407626D,2016arXiv160407821D,2018arXiv180207216D}.

In \cite{2016arXiv160407626D} and \cite{2016arXiv160407821D},
intensity mapping of large scale structure was discussed as one
possible new observational avenue for the DOE's dark energy
program. In the intervening years, an informal working group has been
working towards charting a science case and the research and
development (R\&D) path towards a successful experimental program. The
working group has engaged in regular teleconferences and organized one
community meeting.\footnote{Tremendous Radio Arrays,
  \url{https://www.bnl.gov/tra2018/}} This white paper summarizes
the work of this group to date.

In the July 2019 (version 2) of this document, we have upgraded the
\stagetwo\ concept to match the Packed Ultra-wideband Mapping Array
(PUMA\footnote{\url{https://www.puma.bnl.gov}}) telescope project proposal
submitted to APC call of Decadal Survey on Astronomy and
Astrophysics \cite{PUMAAPC}. Most plots have been upgraded to reflect the PUMA
configuration. Nevertheless, for the purpose of this document, we
still refer to the concept as the \stagetwo, since PUMA is just one
possible incarnation of this concept and others with somewhat
different trade-offs are also plausible future experiments.

\section*{Executive Summary}

In the next decade, two flagship DOE dark energy projects will be
nearing completion: (i)~DESI, a highly multiplexed optical
spectrograph capable of measuring spectra of 5000 objects
simultaneously on the 4m Mayall telescope; and (ii)~LSST, a 3~Gpixel
camera on a new 8m-class telescope in Chile, enabling an extreme
wide-field imaging survey to 27th magnitude in six filters.  DESI will
perform a redshift survey of 20-30 million galaxies and quasars to
$z\sim 3$ to measure the expansion history of the Universe using
baryon acoustic oscillations and the growth rate of structure using
redshift-space distortions \cite{DESI2016}.  Prominent among LSST's
science goals are the study of dark energy/dark matter through
gravitational lensing, galaxy and galaxy cluster correlations, and
supernovae \cite{LSST_Science_Book}.

This white paper proposes a revolutionary post-DESI, post-LSST dark
energy program based on intensity mapping of the redshifted $21\,$cm
emission line from neutral hydrogen out to redshift $z \sim 6$ at
radio frequencies. Proposed intensity mapping survey has the unique capability to
quadruple the volume of the Universe surveyed by the optical programs
(see Fig.~\ref{fig:volModes}), providing a percent-level measurement
of the expansion history to $z \sim 6$ and thereby opening a window
for new physics beyond the concordance $\Lambda$CDM model, as well as
significantly improving precision on standard cosmological parameters.
In addition, characterization of dark energy and new physics will be
powerfully enhanced by multiple cross-correlations with optical
surveys and cosmic microwave background measurements.

The rich dataset produced by such intensity mapping instrument will be
simultaneously useful in exploring the time-domain physics of fast
radio transients and pulsars, potentially in live ``multi-messenger''
coincidence with other observatories.

\smallskip
The core Dark Energy/Inflation science advances enabled by this
program are the following\footnote{See Section \ref{sec:sciencecase} for quantitative forecasts.}:

\begin{itemize}

\item Measure the expansion history of the universe using a single instrument spanning
  redshfits $z=0.3-6$. This will complement existing measurements at
  low redshift while opening up new windows at high redshifts.

\item Measure the growth rate of structure formation in the Universe
  over the same wide range of redshift as expansion history and thus
  constrain modifications of gravity over a wide range of scales and
  times in cosmic history.
  
\item Observe, or constrain, the presence of inflationary relics
in the primordial power spectrum, improving existing constraints by an
order of magnitude.

\item Observe, or constrain, primordial non-Gaussianity with unprecedented
precision, improving constraints on several key numbers by an order of magnitude.
\end{itemize}

\noindent
Detailed mapping of the enormous, and still largely unexplored, volume
of space observable in the mid-redshift ($z\sim 2-6$) range will thus
provide unprecedented information on fundamental questions of vacuum
energy and early-universe physics. Radio surveys are unique in their
sensitivity and efficiency over this redshift range. The
lower-redshift component ($z\sim 0.3-2$) will offer ample
cross-correlation opportunities with existing tracers, including
optical number density, weak lensing, gravitational waves, supernovae,
and other tracers of structure.  The full spectrum of scientific
possibilities enabled by these cross-correlations is impossible to
foresee at this stage.

\smallskip

The field of $21\,$cm intensity mapping is currently in its infancy.
Intensity mapping experiments now underway, or proposed, fall into two
main classes: those targeting the so-called ``Epoch of Reionization" (EoR)
at redshift~$z\sim7-20$, and those attempting to observe in the
low-redshift range where dark energy begins to dominate the expansion
rate around~$z\sim 1$. In addition, there are currently operating and
proposed large-aperture, high--angular-resolution radio telescopes
targeting a range of redshifts with a limited field
of view, appropriate for observations of individual astrophysical
objects. The program proposed here will fill the redshift gap for
intensity mapping experiments, overlap in survey area with precursor
experiments, and take advantage of their progress in addressing the
challenges of beam calibration, receiver stability, and foreground
component separation. Early science results and operational
practicalities from all of these programs will inform the design
decisions for next-generation $21\,$cm surveys.

In this document, we lay out a long-term program in three
overall stages (see Table \ref{tab:roadmap}).  Stage~{\sc i} will consist of targeted R\&D, finalizing
and elaborating the science case, and collaboration building, 
which we foresee as the main activities through the early 2020s.
This time frame will also see first-generation dedicated intensity
mapping experiments release their first datasets. This work will enable \stagetwo, the construction and operation of 
a new, US-led, dedicated radio facility to accomplish the 
science mission centered on $21\,$cm intensity mapping in the $z\sim 0.3-6$ range,
starting in the mid-2020s and running through the early 2030s.
The promises and challenges of this \stagetwo\ experiment are the 
main subject of this paper (see Sections \ref{sec:sciencecase} and \ref{sec:challenges}).  
We designate \stagetwo\ to refer to an aspirational but currently
speculative program of extending $21\,$cm intensity mapping to the
pre-stellar ``Dark Ages'' at $z\gtrsim 30$, which
could begin in the 2030s; see Section \ref{sec:dark-ages}  for discussion and physics promise.

A new approach to achieving these science goals is now possible thanks
to the explosive growth of wireless communications technology enabled
by mass-produced digital RF microelectronics and software-defined
radio techniques. These developments have already resulted in
spectacular results in science such as the first imaging of a black
hole by the Event Horizon Telescope (add citation), which has been
enabled by the ultra-wide band electronics and fast digital processing
very similar in spirit to what is required for a successful \stagetwo\
experiment proposed in this document. It is safe to assume that these electronic
components will continue to decline in price over the years leading to
a construction project. We argue that radio offers the most practical and cost-effective
platform for a highly-scaled next-generation survey instrument.

Expertise within the DOE Office of High Energy Physics (OHEP) network
can be leveraged to address the needs of the radio frequency intensity mapping
program. The principal reasons why this program naturally belongs to
the DOE network are not only that the science goals address topics
that are traditionally in the cosmic frontier of the DOE OHEP, but
also that the difficulty in these measurements calls for an approach
involving a single large collaboration tightly integrating
experimental design, construction, analysis and simulation.  This way
of operating has been a traditional strength of the DOE program.
There are also concrete synergies at the level of existing expertise
within DOE, namely: RF analog and digital techniques for accelerator
control and diagnostics; comprehensive detector calibration
methodology; high-throughput, high-capacity data acquisition; and
large-scale computing for simulations and data analysis. These are
coupled with management-side capabilities, including facility
operations (with partner agencies) and management of large-scale
detector construction projects.

From the standpoint of both physics return and engineering feasibility, we believe that a strong case can be made for including a large scale $21\,$cm intensity mapping experiment in the DOE's Cosmic Frontier program in the late 2020's timeframe.

\clearpage

\section{Introduction}

\subsection{Overview and Scientific Promise}

The 2014 Particle Physics Project Prioritization Panel (P5) report
``Building for Discovery'' contained five goals, of which three are
amenable to study through cosmological probes. These three are: i)
pursue the physics associated with neutrino mass; ii) identify the new
physics of dark matter; and iii) understand cosmic acceleration: dark
energy and inflation. New knowledge in cosmology that will help us
address these topics is acquired by mapping and studying ever
increasing volumes of the Universe with improved precision and
systematics control. No cosmological theory can predict the locations
of individual galaxies or cosmic voids, but such theories can predict
the statistical properties of the observed fields, such as correlation
functions and their evolution with redshift. Studying fluctuations in
the gravitational potential and associated density contrast across
space and time thus forms the bedrock of cosmological analysis. Since
cosmological constraints are inherently statistical, measurements over
increased cosmological volume will lead to tighter bounds. Galaxy
surveys at optical wavelengths have been exploring large scale
structure (LSS) over increasingly large volumes and are a mature and
well tested-technique. However, to keep increasing the maximum
redshift and thus measure ever incerasing volumes of the Universe at
the same rate, we need a different, higher through-put technique.

\begin{WrapText}
 {\bf Overview: \stagetwo\ 21cm intensity mapping survey}
 \begin{itemize}[leftmargin=*]
   \item Large-volume cosmological survey optimized for BAO, structure growth and bispectrum science, covering half the sky at $z=0.3-6$.
   \item Main science goals:
     \begin{itemize}[leftmargin=1em]
     \item  Expansion history and physics of dark energy in
       pre-acceleration era
       
     \item Growth of structure and modified gravity over past 13 billion years
     \item Inflationary features in
       primordial power spectrum
     \item Non-Gaussianity of
       primordial fluctuations
     \end{itemize}
   \item Reference design:
     \begin{itemize}[leftmargin=1em]
     \item 32,000 dishes arranged in hexagonal close-packed array with 50\% fill factor, using FFT
       correlation 
     \item Individual elements are  $6\,$m diameter and can track in the N-S direction
     \item Room-temperature dual-polarization receivers, covering
       $200 - 1100$ MHz.
     \end{itemize}
   \item 5 years on-sky time, targeted at construction start $\sim 2027$
  \end{itemize}
\end{WrapText}

In this report we advocate a novel technique: 3D mapping of cosmic structure using the aggregate emission of many galaxies in the (redshifted) $21\,$cm line of neutral hydrogen as a tracer of the overall matter field. Although currently less mature than optical techniques, we will argue that the coming decade is an ideal time to make large $21\,$cm surveys a reality. Such surveys will allow us to probe to higher redshifts with higher effective source number densities for a smaller investment. They scale better in cost by relying on Moore's law in a way that optical surveys cannot. However, these methods need to be developed and validated, and this document aims to set the roadmap for this research. 

In the field of low-redshift $21\,$cm cosmology, one attempts to
measure the fluctuations in the number density of galaxies across
space \cite{2006astro.ph..6104P}. Galaxies typically emit at many
wavelengths: their optical emission is mostly integrated starlight,
while their emission at low RF frequencies is in synchrotron radiation
and also the $21\,$cm line of neutral hydrogen.  This emission comes
from the (hyperfine) transition of electrons from the triplet to the
singlet spin state; the narrow width of the resulting $21.11$\,cm
line, along with its isolation from other features, allows it to be
readily and unambiguously identified in the galaxy's radio
spectrum. Detection of this line in a galaxy spectrum then allows the
galaxy's redshift to be determined with an error that is negligible
for any standard cosmological analysis.

In the intensity
mapping technique, the intention is not to resolve individual
galaxies. Instead, one designs radio interferometers with angular
resolution limited to scales relevant for studying the large-scale
structure traced by those galaxies. In each 3D resolution element
(voxel), given by the coarse angular pixel and considerably finer
frequency resolution, emission from many galaxies is averaged to boost
the signal-to-noise. Even without resolving individual objects, we
can still trace the fluctuations in their number density across space
and redshift on sufficiently large scales. This allows us to put the
experimental signal-to-noise where it really matters for cosmology: on large spatial
scales, where our theoretical modeling is most robust.

\begin{figure}
  \centering
  \includegraphics[width=0.7\linewidth]{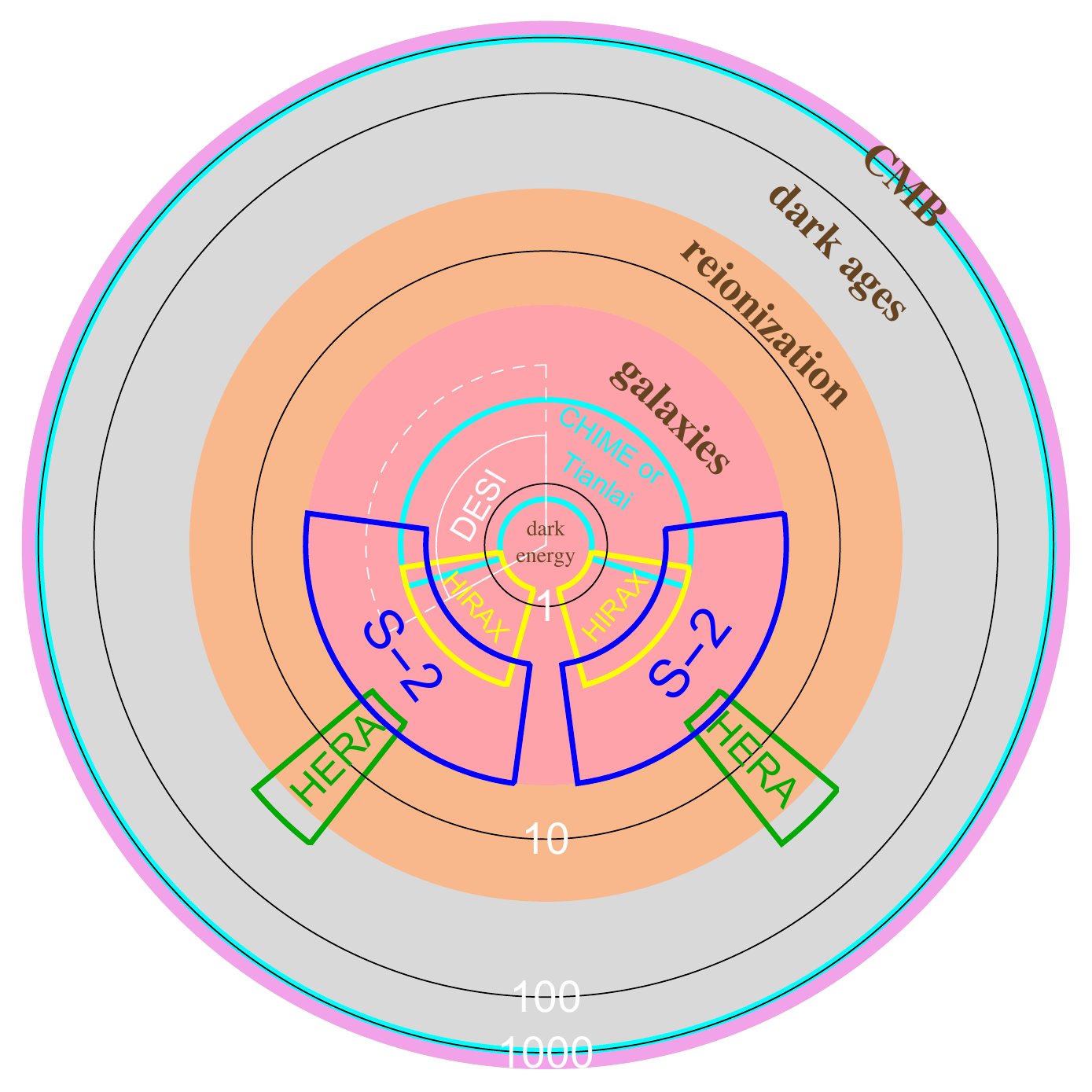}

  \caption{Plotted is a schematic 2D representation of the observable
    universe where the \emph{area} is proportional to the \emph{comoving volume} and
    the distance from center monotonically increases with distance
    from Earth.  Different epochs are color coded: the epoch of
    galaxies ($z<6$) pink; the epoch of reionization ($6<z<20$) orange;
    the dark ages ($20<z<700$) gray; the epoch of the last scattering
    ($700<z<1300$) cyan; and the early universe ($z>1300$) purple.  The
    volumes surveyed by various current experiments with dense redshift
    space sampling are outlined, including the DESI optical
    spectroscopic survey of galaxies (white) and quasars (white
    dashed); HI intensity mapping surveys of the intergalactic
    medium during the epoch of reionization (HERA; green) and lower-redshift galaxies (CHIME/Tianlai; cyan);
    HIRAX (yellow); and the $21\,$cm Stage~{\sc ii}  project proposed here
    (blue). The wedge sizes give rough representations of the covered
    volume.}
  \label{fig:vol}
\end{figure}

All neutral hydrogen in the universe below redshift of $z\sim 150$ is
in principle amenable to $21\,$cm observations. This includes the
large volumes at $z\gtrsim 30$, the so-called ``Dark Ages" before the
first luminous objects were created; at $6\lesssim z \lesssim 30$,
when these first objects formed and reionized the universe; and at
$2\lesssim z \lesssim 6$, after reionization but difficult to fully
map with large optical surveys.  (See Figure~\ref{fig:vol} for a
visual comparison of the volumes accessible to different kinds of
observations and in different epochs of cosmic history.)  In the Dark
Ages and post-reionization era, the $21\,$cm signal is a theoretically
well-understood tracer of cosmic structure, and any science amenable
to study through statistics of cosmic fields can be studied using this
technique. However, the Dark Ages pose a formidable challenge (to say
the least), for several reasons related to the low frequencies at
which the associated observations must take place. Thus, we have
identified the post-reionization era at $2\lesssim z \lesssim 6$ as
the most natural target for a dedicated next-generation $21\,$cm
instrument, although we will briefly discuss the high-redshift promise
in Section \ref{sec:intro-highz}.

In this white paper, we have not attempted to optimize the many design choices that must go into such an instrument. Rather, we have chosen a configuration that, while somewhat ambitious, is expected to comfortably fit within the cost profile of a typical DOE OHEP project, and performed a first round of forecasts for the scientific capabilities of this configuration. This exercise has allowed us to identify a trio of key science results that could be obtained by an instrument broadly in line with our chosen specifications, and also to explore a range of other applications of such an instrument.

\smallskip
The remainder of this introduction is as follows:
\begin{itemize}
\item In Section~\ref{sec:intro-science}, we summarize the three key science results, and a set of ancillary capabilities, associated with our fiducial instrument, which we have dubbed a ``Stage~{\sc ii}'' $21\,$cm experiment.
\item In Section~\ref{sec:observ-univ-with}, we briefly introduce the
  basic mode of operation of radio telescopes in order to set the context.

\item In Section~\ref{sec:intro-posteorexp}, we review the landscape of operational or planned post-reionization $21\,$cm surveys, and place a Stage~{\sc ii} experiment in that context.
\item In Section~\ref{sec:intro-pract-limit}, we introduce and discuss the practical challenges of implementing a Stage~{\sc ii} $21\,$cm experiment.
\item Finally, in Section~\ref{sec:intro-roadmap}, we lay out a provisional roadmap for a three-stage $21\,$cm program, building from ``Stage~{\sc i}" (current experiments) through Stage~{\sc ii} and beyond.
\item In Section~\ref{sec:intro-advantages}, we describe the synergies
  between optical surveys and $21\,$cm experiments and unique
  advantages of each.

\end{itemize}
The main text of the paper is devoted to more detailed discussions of
the various science cases (Section~\ref{sec:sciencecase}), the
challenges and opportunities associated with Stage~{\sc ii}
(Section~\ref{sec:challenges}), and a brief foray into observations
21-cm beyond redshift of $z=6$ (Section~\ref{sec:intro-highz}), with a discussion
of current epoch of reionization experiments (Section~\ref{sec:cosmic-dawn-epoch}) followed by a discussion of the exciting
potential of the Dark Ages a probe of cosmology
(Section~\ref{sec:dark-ages}). We conclude in
Section~\ref{sec:conclusions}.

\subsection{Science capabilities of a large-scale \texorpdfstring{$21\,$cm}{21cm} experiment}
\label{sec:intro-science}

The starting point for a \stagetwo\ concept was the realization that
the same instrument could help achieve three high-impact science
objectives that are deeply connected to some of the biggest problems
in fundamental physics. These are:

\renewcommand{\theparagraph}{\underline{\normalfont{\bf A\arabic{paragraph}}}}
\setcounter{secnumdepth}{4}

\startframe
\paragraph{\bf Characterize the expansion history in the
  pre-acceleration era to the same precision as low-redshift measurements.}
The precision of expansion history measurements in the low-redshift
era using the BAO technique (see Section \ref{sec:meas-expans-hist}
for a technical description) is close to its theoretical limit due to
the finite amount of large-scale information available per redshift.
However, the measurement landscape deteriorates very fast for
$z\gtrsim 2$, and will not be satisfactory in this range for the
foreseeable future. It is imperative to measure the expansion history
to better than percent level all the way to $z=6$, which allows
measurement of the energy density in the \emph{dark energy component}
with the precision of 10\% at those redshifts. In the pre-acceleration
era, this is a very difficult measurement, because the total energy
density and thus expansion history of the Universe is dominated by the
matter density. Consequently, signatures of dark energy are expected
to be small in a minimal $\Lambda$CDM Universe.  There is, however,
strong theoretical motivation to explore this particular era, since
theoretical explanations for the minimal $\Lambda$CDM Universe
generally suffer from extreme fine-tuning issues. Alternative
explanations to $\Lambda$CDM have generic signatures in the
$2 \lesssim z \lesssim 6$ range, and percent-level expansion
measurements within this range will impose stringent constraints on
such theoretical models, which are otherwise unconstrained.  Note that
the \stagetwo\ experiment will characterize the expansion history over
the full redshift range starting at $z\sim 0.3$. However, the
extragalactic sky at $z\lesssim 2$ range will be mostly measured by a
combination of Euclid and DESI, allowing only modest \stagetwo\
improvements in this range. On the other hand, our measurements will
represent an important cross-check of the results from the same volume
using a fundamentally different tracer.
\endframe

\startframe
\paragraph{\bf Characterize the growth rate of structure in the
  pre-acceleration era to the same precision as low-redshift measurements.}
Similar to expansion history, the \stagetwo\ experiment will also
measure the growth of fluctuations across the redshift range from
$0.3<z<6$, with out determinations becoming particularly relevant in
the high redshift regime $z\gtrsim 2$. The method employed is similar
to the redhift-space distortion measurements in galaxy cluster, but
relies on the growth  signatures in the weakly non-linear regime
\cite{CastorinaWhite19,2019arXiv190411923M}. In combination with expansion history over
the same range of redshift, we would be able to potentially detect difference
between the growth of structure and expansion rate, one of the smoking
guns of modified gravity which would give a unique insight into the
dark energy phenomenon.

\endframe

\startframe
\paragraph{\bf Constrain or measure inflationary relics in the shape
  of features present in the primordial power spectrum.}  Sufficiently
sharp features in the primordial power spectrum survive mild
non-linear evolution and biasing, and are predicted in various
inflationary models. The amplitude, frequency and phase of the feature
are all indicative of the mechanism that sourced the initial seeds of
structure. If found, they would present a breakthrough discovery and
unique opportunity in the attempt to understand the physics of the
early Universe. It would be highly informative to constrain or detect the presence of
oscillatory features with frequencies $\omega_\mathrm{lin}>50\,$Mpc and amplitude
smaller than $10^{-3}$ relative to the inflationary power-law spectrum.

\endframe

\startframe
\paragraph{\bf Constrain or measure the equilateral and orthogonal bispectrum of large-scale structure with unprecedented precision.}  
Primordial non-Gaussianities are generically predicted by non-minimal
inflationary models of the early Universe. The size and shape of
primordial non-Gaussianities would be indicative of the number of
fields present as well as the strength of interactions and
self-interactions of the field or multiple fields driving inflation.
The huge amount of clean, large-scale statistics from the volume
accessible to a high-redshift survey presents a unique opportunity to
put unprecedented constraints on non-Gaussianities that are sensitive
to the dynamics during inflation. Specifically, the three-point
correlation function of Fourier modes of the density field (the
so-called bispectrum) is amenable to measurement using high-redshift
LSS surveys, and its amplitude in different configurations
(corresponding to the three points forming squeezed, equilateral, or
folded triangles) is directly connected to different inflationary
models. Moreover, these types of non-Gaussianities (equilateral and
orthogonal) cannot be constrained using bias constraints in the power
spectrum and are therefore not amenable through cosmic variance
cancellation techniques that are forecasted to put stringent
constraints on squeezed non-Gaussianities. In other words, a
high-redshift survey of the universe will most likely present the only
viable opportunity to improve over CMB constraints.  \endframe

All three objectives described above could be achieved with a
next-generation 21$\,$cm experiment, which we designate a \stagetwo\
experiment. Our fiducial configuration consists of a 50\% fill factor
hexagonally close-packed array of 32,000 $6\,$m dishes, operating from
200 to $1100\,$MHz. This configuration is an ambitious but realizable
expansion over the current generation $21\,$cm experiments.
Section~\ref{sec:fiducial} contains a technical arguments motivating
this particular choice of fiducial experimental parameters.  The
precise configuration of the array and other experimental details are
expected to evolve and be further developed depending on key science
targets and experience obtained with predecessors of a Stage~{\sc ii}
$21\,$cm experiment. However, having an explicit experiment allows us
to make concrete forecasts that set the context for further
optimization.

The objectives outlined above directly follow from the ability of
$21\,$cm emission to obtain a pristine picture of large-scale
structure with essentially no tracer shot-noise.
In the following, we list some of the other new capabilities that will be enabled by a Stage~{\sc ii} experiment:\\

\setcounter{paragraph}{0}
\renewcommand{\theparagraph}{\underline{\normalfont{\bf B\arabic{paragraph}}}}

\setlength{\parindent}{0cm}

\paragraph{\bf Add a new tracer at $z<2$}
By the time the \stagetwo\ experiment becomes a reality, the volume of
the universe at redshift $z\lesssim2$ will be mapped by the current
and upcoming experiments using numerous tracers. These include
spectroscopic galaxies from DESI, photometric galaxies from LSST,
velocity measured by Hubble diagram residuals from type Ia supernovae,
gravitational wave siren sources, weak gravitational lensing from both
LSST and CMB, etc. Adding a new tracer will enable numerous new
studies, some of which we discuss in the Section
\ref{sec:cross-corr-stud}, but the full breadth of potential new
science remains will likely be only fully understood as the field
evolves over the coming decade.
\\[-0.8em]

\paragraph{\bf Quadruple the observed volume at an increased fidelity.} 
The volume between $z=2$ and $z=6$ is approximately three times the
volume between $z=0$ and $z=2$, and contains structures whose
clustering statistics are easier to predict than at lower redshifts
(see Figure~\ref{fig:volModes}). $21\,$cm intensity mapping can probe
this volume with a very high effective number of sources, allowing for
straightforward extraction of cosmological information from these
measurements. While we have identified several well-motivated uses of the large
number of linear modes present in this volume as our main scientific
goals, other, yet to be discovered, statistical quantities describing
and constraining fundamental physics are also likely to improve
equally due to generic $\sqrt{N}$ scaling of error on any derived statistical quantity.
\\[-0.8em]


\paragraph{\bf Measure information from scales and redshifts not directly present in the survey.} 
Couplings between different Fourier modes of the cosmic density field will allow us to reconstruct modes that are not directly present in the survey through their effects on the observed small-scale modes. In particular, the tidal effect of large-scale modes on the small-scale power will give access to the large-scale modes (which may otherwise be obscured by foregrounds in certain scenarios). Furthermore, gravitational lensing effects on small scales will provide information about lower-redshift structure. Three-dimensional $21\,$cm observations will provide several source ``screens" for lensing analyses; the signal to noise of a joint analysis of all such screens will exceed that for the next generation CMB lensing reconstruction in cross-correlation.
  \\[-0.8em]

\paragraph{\bf Improve measurements of 
    parameters that encode deviations from the minimal cosmological model,
     including neutrino mass, radiation content of the
    early universe, and curvature.} $21\,$cm observations can, in
  conjunction with other synergistic measurements, aid in constraining
  these parameters. In particular, we should
  achieve an independent detection of the neutrino masses and constrain the
  radiation content to within a factor of a few of the guaranteed
  correction due to electron-positron annihilation.
  \\[-0.8em]

  \paragraph{\bf Potentially directly detect the expansion of the Universe.} The
  Universe expands at the Hubble rate and in principle this expansion
  can be detected by observing sources drift in redshift over the time
  of experiment. The advantage of radio observations is that the
  clocks stable enough to drive the digitization circuits at the
  required time stability are nearly off-the-shelf equipment.
  \\[-0.8em]

  \paragraph{\bf Explore the physics of fast radio bursts (FRBs).}
  This instrument will also likely detect millions of FRBs as we
  discuss in Section~\ref{sec:ancill-scienc-time}. The physics of FRBs
  is currently poorly understood, but in some models they could act as
  standard candles or alternatively their dispersion measure in
  conjunction with kinetic Sunyaev-Zeldovich effect measurements from
  CMB could open another possible window into the expansion and growth
  history of the universe.
  \\[-0.8em]

  \paragraph{\bf Explore modified gravity using pulsars.} The same
  instrument that can be used for cosmology will also be able to
  observe numerous pulsars and study general relativity through
  precision changes in pulsar timings.
  \\[-0.8em]

\setlength{\parindent}{1em}

Using our fiducial Stage~{\sc ii} $21\,$cm configuration, we will perform a detailed exploration of all possible science targets identified above in Section
\ref{sec:sciencecase}.

\subsection{Observing the universe with a radio telescope}
\label{sec:observ-univ-with}

Radio telescopes observe the electromagnetic radiation at radio
frequencies and for 21$\,$cm this means at frequencies below 1.42GHz. A
traditional single-dish radio telescope contains a focusing element,
typically a parabola that focuses the incoming radiation onto a radio
receiving element. Such parabola coherently adds all radio waves
coming from a given direction. Such a telescope can observe a single
pixel in the sky at once and the bigger the parabola, the higher is its
resolution, with the sky response function scaling as roughly
$\lambda/D$, where $\lambda=\lambda_0(1+z)$ is the observing
wavelength (redshifted from rest-frame $\lambda_0$) and $D$ is
the parabola size. Because radio wavelengths are very long (compared
to typical optical wavelengths, for example), the size of the
reflector needs to be very large to achieve a fine resolution.

\begin{figure}[t]
  \centering
  \includegraphics[width=0.75\linewidth]{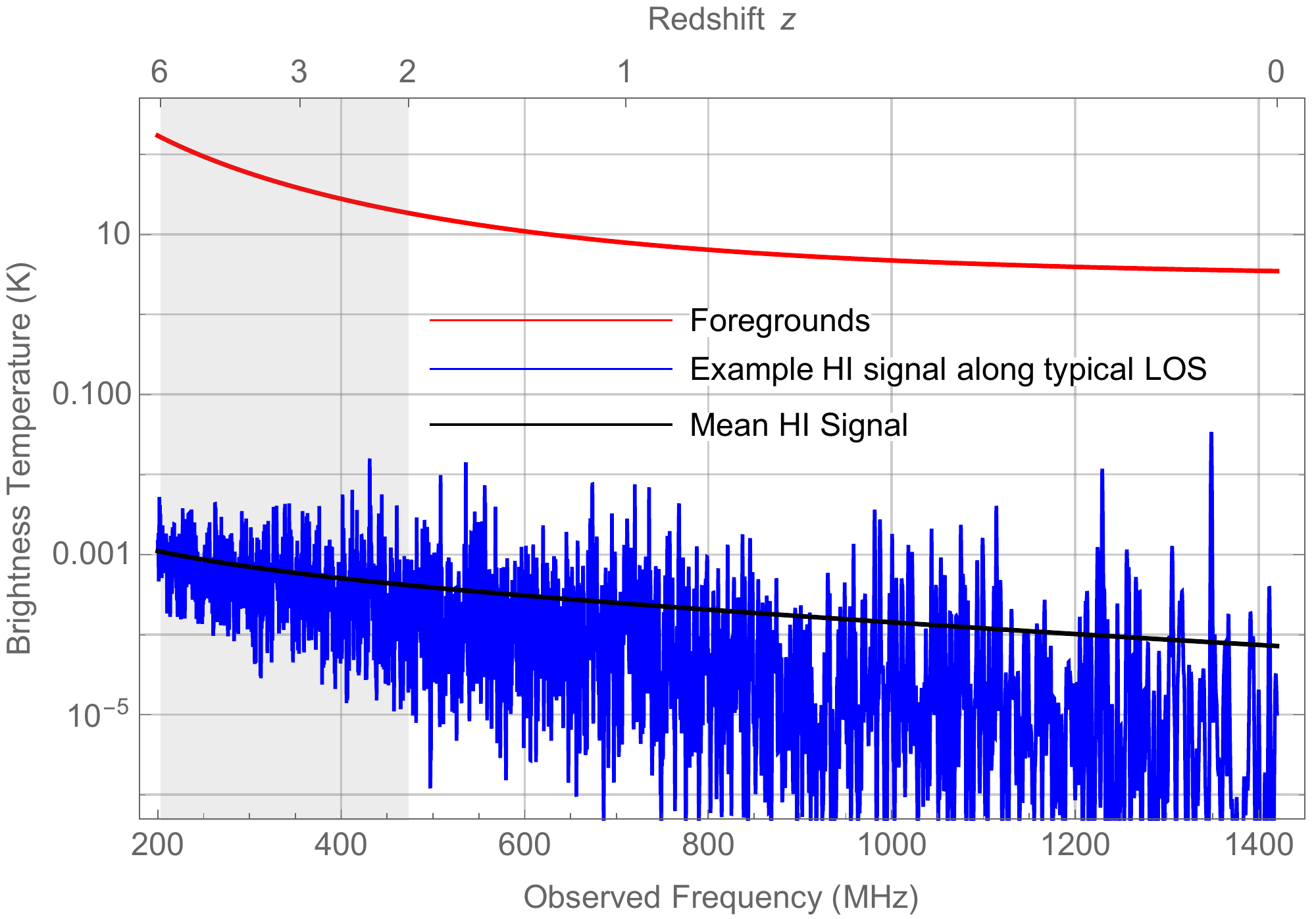}
  \caption{Illustration of 21cm emission spectra, showing observed
  brightness temperature as a function of observed frequency and source 
  redshift. 
  {\em Red:}~Average emission from galactic foreground sources (see 
  Equation~\ref{eq:sky_temperature}) varying between a brightness
  of a few~K to a few hundred~K.
  {\em Black:}~Mean signal from cosmological HI, following
  Eq.~\ref{eq:21cm_brightness}, smaller by about five orders 
  of magnitude; 
  {\em Blue:}~Example realization of the HI signal
  that would be seen with one beam along a typical line of sight.
  At low redshifts the matter signal is dominated by a few peaks,
  indicating the growth of structure; while at earlier times the
  fluctuations around the mean are smoother.  The grey band 
  highlights the redshift range $z$=2--6.
  }
  \label{fig:schem}
\end{figure}

In Figure \ref{fig:schem} we schematically show the signal observed by one such
single-receiver pointed at a typical direction on the sky (and assuming it could
observe signal from 200MHz to 1420Mhz). The signal would be dominated by the
emission from our own galaxy -- shown as the red line. This emission is very strong,
but at the same time very smooth, which gives us a handle at subtracting it. The
blue lines illustrates what the 21-cm signal would look like: at low redshift it would
correspond to individual over-densities traced by small objects, while at high
redshift the structure in the radial directions blurs into a continuous field.

It has long been recognized, that instead of combining the signals by
optically adding them, one can add them electronically. This concept,
known as aperture synthesis (for which the Nobel prize was awarded in
1974 to Martin Ryle) led to a class of instruments called radio
interferometers. In such telescopes, the collecting area of a single
dish is replaced with several individual smaller elements, that do not
need to be, but are are often smaller dishes themselves. Signals form
individual receivers are combined and allow one to synthesize an
effective dish whose total collecting area is the sum of individual
collecting areas and whose resolution matches that of a dish
with the same size as the largest separation between individual
elements. But the most important advantage is that multiple beams can
be synthesized concurrently which can cover all of the primary
field of view of individual elements. This can lead to an exponential
increase in sensitivity compared to traditional single-element dishes.

In order to perform aperture synthesis, the signal from every pair of
elements needs to be correlated and hence the difficulty increases as
the square of the number of individual elements forming an
interferometer. Therefore, traditional interferometers employed at
most a few tens of elements. In the 21$^{\rm st}$ century, however,
digital technology allows the possibility of doing the signal
combination digitally, leading to telescopes made of thousands of
receiving elements. This progression in technology moved the
complexity first from the problems of mechanical engineering in making
large receiver dishes to that of building and replicating analogue
electronics and finally to processing massive amounts of digital
data. As we will see later, part of this white-paper continues this
trend by arguing for digitization as soon as possible after the signal
enters the system.

\subsection{Post-reionization \texorpdfstring{$21\,$cm}{21cm} surveys: the state of the art}
\label{sec:intro-posteorexp}

$21\,$cm cosmology has only been made possible recently through developments in infrastructure (e.g.\ high-throughput computing and
commodification of low noise radio-frequency technology) that allow for correlations at full bandwidth
  at the necessary scale. Tools and techniques have been developing rapidly, and the first steps towards extracting cosmological information from $21\,$cm
 observations have already been demonstrated.

The first detection of the redshifted $21\,$cm emission in the intensity
mapping regime was achieved by Chang~et~al.\ in 2010~\cite{2010Natur.466..463C}. The measured 3D field,
obtained from the Green Bank Telescope (GBT) 800~MHz receiver, spans
the redshift range of $z = 0.53$ to~1.12 and overlaps with 10,000
galaxies in the DEEP2 survey~\cite{2001defi.conf..241D} in spatial and redshift
distributions.  This enabled a cross-correlation measurement on~9$h^{-1}$ Mpc scales at a 4$\sigma$ significance level. 
This detection was the first verification that the
$21\,$cm intensity field at $z \sim 1$  traces the
distribution of optical galaxies, which are themselves known tracers of the
underlying matter distribution. It presents an important proof of concept for the
intensity mapping technique as a viable tool for studies of
large-scale structure.

A continuing observing campaign to expand the GBT $21\,$cm IM survey in
both sensitivity and spatial coverage has yielded two subsequent
publications: an updated cross-power spectrum at $z \sim 0.8$~\cite{2013ApJ...763L..20M} between $21\,$cm and optical galaxies in the WiggleZ survey~\cite{2010MNRAS.401.1429D}, and an upper limit on the $21\,$cm auto-power spectrum~\cite{2013MNRAS.434L..46S}. Combining the cross- and
auto-power spectrum measurements yields a $\sim$3-$\sigma$
measurement on the combination of the cosmic HI abundance~$\Omega_{\rm HI}$ and bias~$b_{\rm HI}$ parameters, $\Omega_{\rm HI} b_{\rm HI} = 0.62^{+0.23}_{-0.15} \times 10^{-3}$~\cite{2013MNRAS.434L..46S}.  Further analysis of 800 hours of GBT observations
taken during 2010-2015 is currently ongoing.

No experiment has detected the $21\,$cm power spectrum in
auto-correlation. While this should be possible with non-dedicated
experiments in terms of statistical significance, the instrumental
challenges are currently too large. However, this situation should
change with the advent of dedicated instruments. 

There are currently five main experiments that are presently being
built or are in the commissioning phase to measure LSS with the
$21\,$cm intensity mapping technique with dedicated instrumentation:
CHIME in Canada, HIRAX in South Africa, Tianlai in China, OWFA in
India, and BINGO, a UK/Brazil experiment. In addition, there are
several smaller efforts dedicated to R\&D, such as BMX at Brookhaven
National Laboratory and PAON at Nan\c{c}ay in France. We list the main
properties of these instruments in Table \ref{tab:current}. These
small-scale experiments will teach us about the viability of the
intensity mapping technique, for example by providing testbeds for
calibration, foreground removal, and RFI mitigation techniques.

Of the listed experiments, CHIME is currently the most advanced, and
has recently upgraded from a prototype to the full instrument. It
consists of 4 cylindrical radio antennas with no moving parts,
observing the entire accessible sky which passes above it as the Earth
rotates. It operates from 400-800~MHz, equivalent to mapping LSS
between redshift $z =0.75$ to 2.5.  We expect the first cosmology
results from CHIME in the next 3 years, which should include
foreground removal or mitigation techniques for intensity mapping
measurements of LSS in $21\,$cm emission. Note that CHIME has already
shown promise related to one of its other science goals, having
recently announced the first detection of a low-frequency fast radio
burst~\cite{2018ATel11901....1B}.

\begin{table}
  \centering
  \begin{tabular}{cccccrc}
    Name & Optimized & Steerable & Type  & Elements & Redshift& First light \\
    \hline
\\     \multicolumn{1}{p{2.5cm}}{\uline{Existing w data:}} \\
    GBT & N & Y & Single Dish & 1\,dual-pol on $100\,$m dish & $\sim$0.8 & 2009  \\
\\    \multicolumn{1}{p{3.3cm}}{\uline{Dedicated experiments:}} \\
    CHIME & Y & N & Cylinder Interferometer & 1024 dual-pol over 4 cyl   & 0.75 -- 2.5 & 2017  \\
    HIRAX & Y & limited & Dish Interferometer & 1024 dual-pol\,$\times$\,$6\,$m dishes
                                                & 0.75 -- 2 & 2020  \\
    TianLai Dish & Y & Y & Dish Interferometer & 16 dual-pol\,$\times$\,$6\,$m dishes& 0 -- 1.5 & 2016 \\
    TianLai Cylinder & Y & N & Cylinder Interferometer & 96 dual-pol
                                                         over 3 cyl   & 0 -- 1.5 & 2016 \\
    OWFA & N & Y & Cylinder Interferometer &  264 single-pol& $\sim$ 3.4$\pm$0.3 & 2019 \\
    BINGO & Y & N & Single Dish  & $\sim$60 dual-pol sharing 
                                   $\sim$50\,m dish & 0.12 -- 0.45 & 2020 \\
\\    \multicolumn{1}{p{2.5cm}}{\uline{Dedicated R\&D:}} \\
    BMX & Y & N & S. Dish + Interferometer & 4 dual-pol\,$\times$4\,m
                                             off-axis dishes & 0 -- 0.3 & 2017 \\
    NCLE & Y & N & Satellite  & 3$\times$5\,m monopole ant. at
                                Earth-Moon $L_2$ & $>17$ &  2018\\
    PAON-4 & Y & limited & Dish Interferometer & 4 dual-pol\,$\times$5\,m dishes & 0 -- 0.14 & 2015 \\                            
\\    \multicolumn{1}{p{2.5cm}}{\uline{Non-dedicated:}} \\
    MeerKAT & N & Y & Single-Dish & 64 dual-pol\,$\times$\,13.5\,m dishes & 0 -- 1.4 & 2016 \\
    SKA-MID & N & Y & Single-Dish & $\sim 200$ dual-pol\,$\times$\,15\,m dishes & 0 -- 3 & 2023 \\
\\    \multicolumn{1}{p{2.5cm}}{\uline{Proposed Here:}} \\
    {\bf Stage {\sc ii}} & Y & limited &  Dish Interferometer & 32,000
                                                                dual-pol\,$\times$\,
                                                               6\,m dishes & 0.3 -- 6 & $<$2030 \\

  \end{tabular}
  \caption{List of current and planned experiments. The ``First
    light'' column refers to first light for $21\,$cm observations for
    non-dedicated experiments. In the ``Optimized'' column, we note
    whether the telescope has been designed with intensity mapping as
    its primary scientific goal. The HIRAX and PAON-4 dishes can only
    be steered by manual human intervention.  For MeerKAT and SKA-MID,
    dishes will likely be used in a single-dish mode, with
    interferometric capability used only for gain calibration.  }
  \label{tab:current}
\end{table}

\begin{figure}
  \centering
  \includegraphics[width=0.7\linewidth, trim = 0 15 0 0]{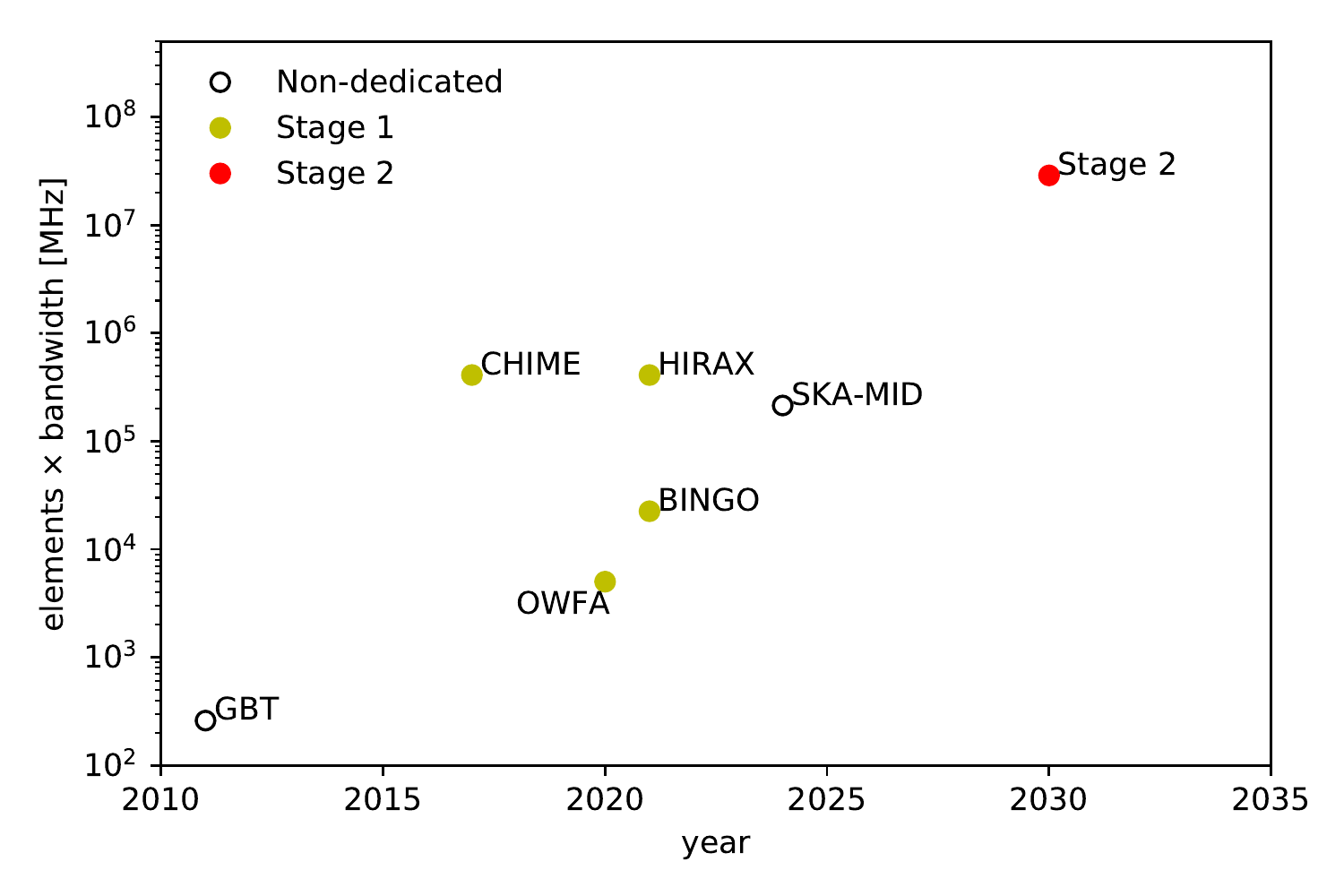}

  \caption{Representation of improvements from current-generation to
    future proposed experiments in a figure of merit analogous to
    optical etendue measure:  number of
    receiving elements $\times$ bandwidth. See text for discussion.}
  \label{fig:stages}
\end{figure}

Another experiment often mentioned in this context is the
SKA\footnote{\url{https://www.skatelescope.org/}} (Square Kilometre 
Array). The SKA1-MID mid-frequency dish array is a formidable instrument, but 
is optimized for a variety of radio astronomy goals other than intensity mapping. 
In many aspects the comparison is similar to new generation of extremely 
sensitive optical telescopes that have mirror-sizes exceeding 30m, but are 
nevertheless not competitive for survey-science optical cosmology due their 
small field of view and focus on diffraction-limited imaging of
individual objects. For intensity mapping, SKA1-MID suffers from a similar
mismatch in scales to which it is sensitive compared to the proposed
\stagetwo\ experiment. While it will typically act as an interferometer with 
several hundred large dish elements, the baseline distribution best matches 
the scales relevant to imaging of individual objects rather than intensity
mapping of large-scale structure. As a workaround, the SKA1-MID array will 
instead be used as a collection of single dishes for intensity mapping,
perhaps using interferometry only as a calibration tool. This will have 
relatively poor angular resolution at $z \gtrsim 1$ however, leaving it 
sensitive mostly to only the radial BAO feature \citep{Villaescusa-Navarro:2016kbz}.
Additionally, individual elements of SKA1-MID are highly capable fully-steerable 
dishes that can operate up to 14\,GHz. Dedicated designs for $21\,$cm intensity 
mapping survey science typically use transiting arrays instead, since one wants 
to maximize the sky coverage rather than point at objects of interest, and reduce 
mechanical costs; $21\,$cm intensity mapping also requires considerably lower maximum 
frequencies and corresponding dish-surface accuracy requirements (i.e.~500\,MHz 
for our \stagetwo\ experiment and never higher than the frequency of $z=0$ neutral 
hydrogen at 1420\,MHz). It is clear that the SKA1-MID instrument has been optimized for different 
science goals and has therefore embarked on a different set of trade-offs to an optimal $21\,$cm experiment. As such, it will not be directly competitive with
dedicated instruments for many of the science cases discussed in this document, 
and thus does not present an obstacle to DOE for entry into this field. The same is true for the SKA1-LOW instrument, which partially overlaps in frequency coverage with our proposed \stagetwo\ experiment (i.e. at the high redshift end of our band), but has a greater focus on Cosmic Dawn and Epoch of Reionization science, and will not be competitive with \stagetwo\ for BAO measurements for example (see \cite{2016ApJ...817...26B} for cosmological forecasts for SKA1-MID and SKA1-LOW surveys). Nevertheless, as the largest and most complex radio astronomical facility to be constructed in advance of \stagetwo, we expect SKA to offer a number of valuable lessons in terms of calibration and data analysis techniques, computing infrastructure, and data management.

In Figure~\ref{fig:stages}, we plot the same information as Table~\ref{tab:current}, but compressed
into in a figure of merit analogous to optical etendue measure:
\begin{equation}
\text{FoM} = \text{number of receiving elements} \times \text{bandwidth}\ .
\label{eq:fom}
\end{equation}
This equation is motivated by the expression for the system
temperature contribution to noise (see Eq.~\ref{eq:pnoise} in Appendix~\ref{app:forecasts-noise}) and it is necessarily a very crude
simplification. Most importantly, it does not take into account the
surface area of reflector material and would naturally drive you
towards a field of dipole antennas at fixed cost. While this might be
the right answer in the absence of systematic effects, the current
consensus is that some directionality of individual elements is desirable.
Moreover, a compact interferometric array with the same
figure of merit will in general perform better than a traditional
radio array with the same figure of merit for the science discussed in
this paper. Finally, observing at different central frequency affects
the result in a non-trivial way:  the sky noise is lower at
higher frequencies, but the volume per unit bandwidth is larger
at lower frequencies and the Universe is more linear at higher redshifts.

\begin{figure}
  \centering
  \begin{tabular}{cc}
    \includegraphics[width=0.45\linewidth]{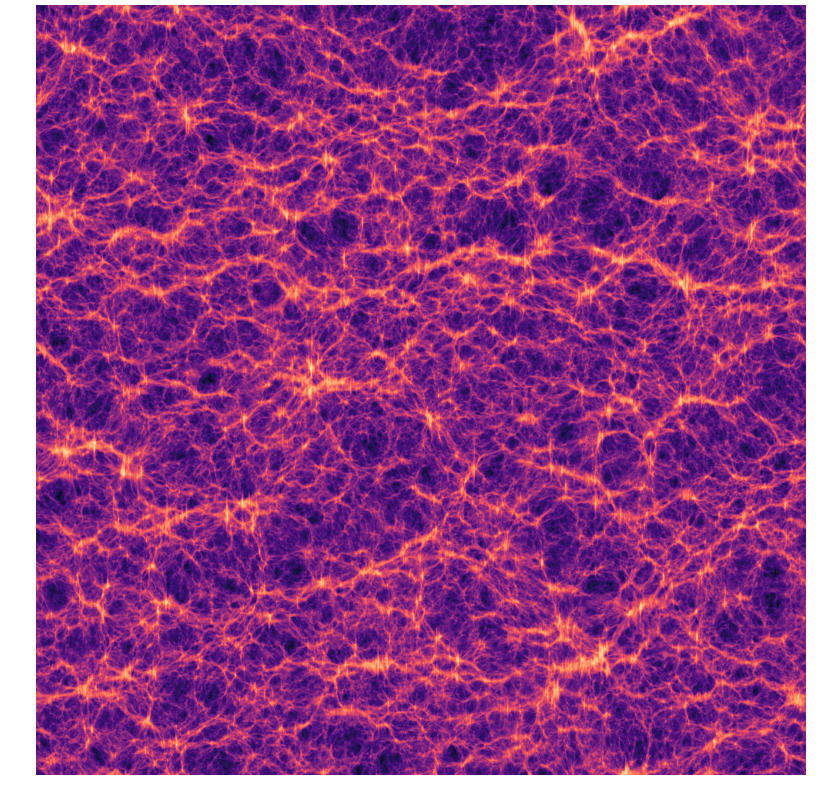} &
    \includegraphics[width=0.45\linewidth]{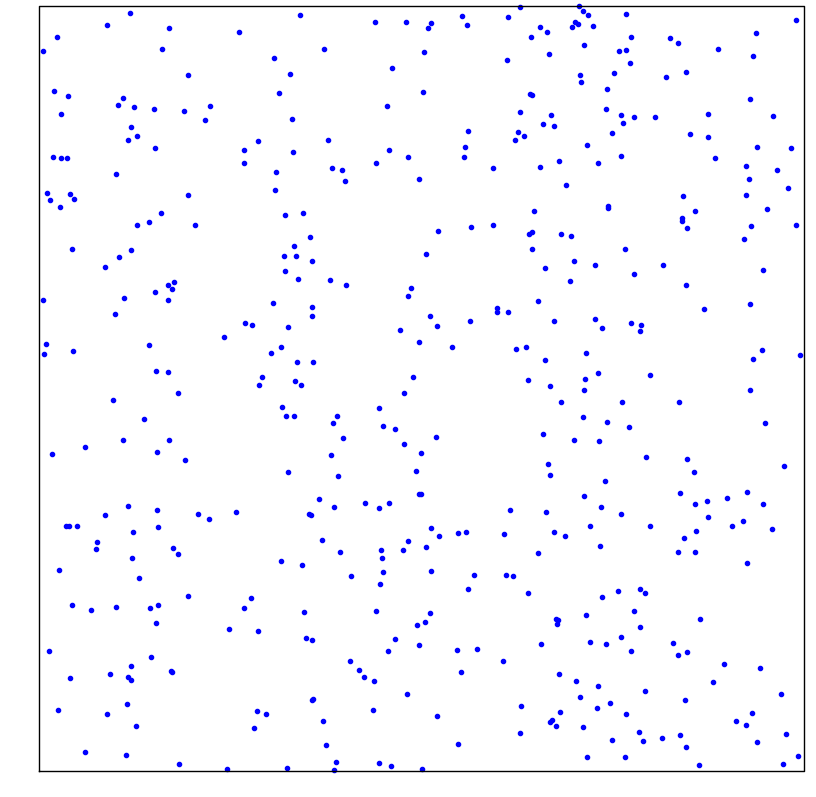} \\
    \includegraphics[width=0.45\linewidth]{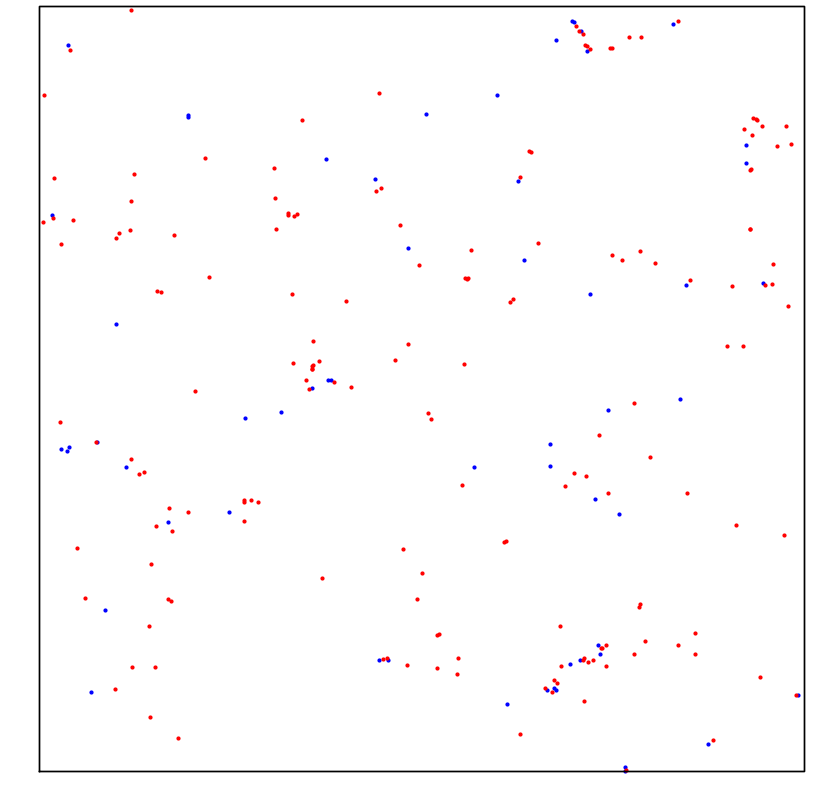} &
    \includegraphics[width=0.45\linewidth]{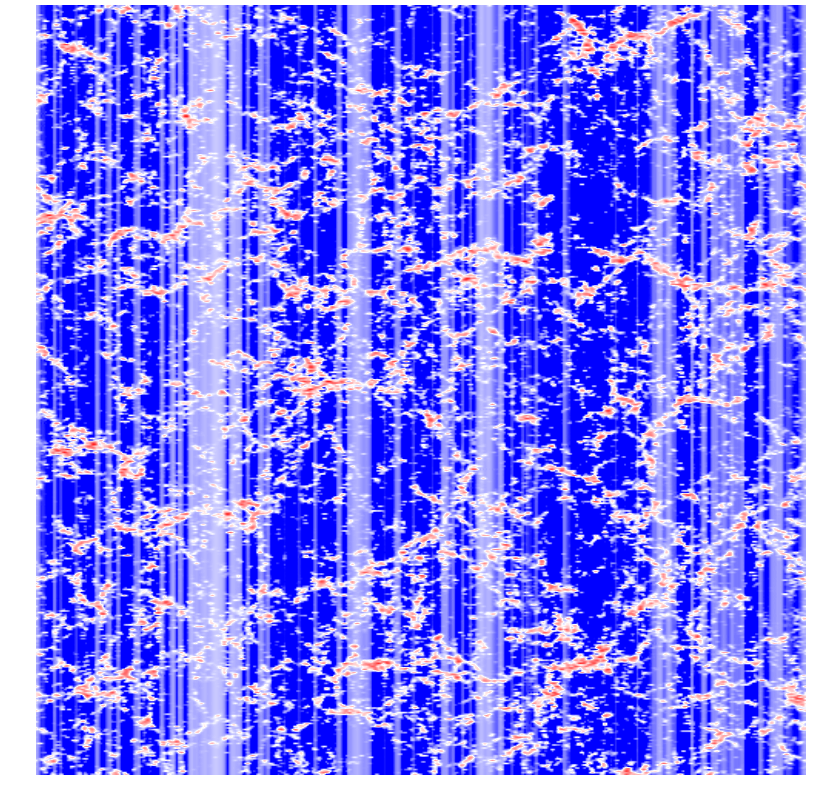} \\

  \end{tabular}
  \caption{This figure shows the same slice of the simulation at
    redshift $z=3$ as ``seen'' by different probes. We show a
    $300\times300\times4\,h^{-1}$Mpc slab of an approximate simulation
    with horizontal direction corresponding to transverse direction
    and vertical direction to radial direction in redshift
    space. The upper left plot shows the underlying dark matter
    density. The upper right plot shows the LSST sources, where
    structure is erased due to photometric redshift errors. The lower
    left shows a putative drop-out spectroscopic survey selecting
    $m_{\rm UV}<24.5$ (blue and red dots) and those going to just
    $m_{\rm UV}<24$ (blue dots). The lower right plot shows a raw
    image from a {\stagetwo}-like instrument. The vertical striping is
    due to foreground removal and there is a visible smoothing in the
    transverse direction. The plot uses logarithmic scaling with
    a non-linear color scale to make features more visible.  See
    Appendix \ref{app:panella} for discussion of assumptions that
    went into making of this figure and Figure \ref{fig:panella2}.}
  \label{fig:panella}
\end{figure}

\begin{figure}
  \centering
  \begin{tabular}{cc}
    \includegraphics[width=0.45\linewidth]{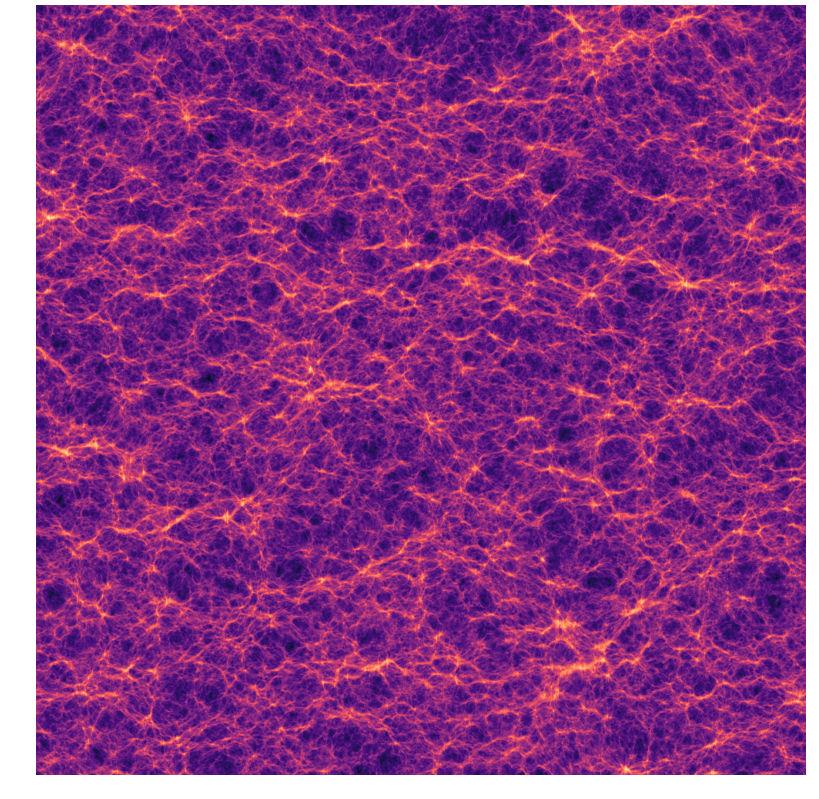} &
    \includegraphics[width=0.45\linewidth]{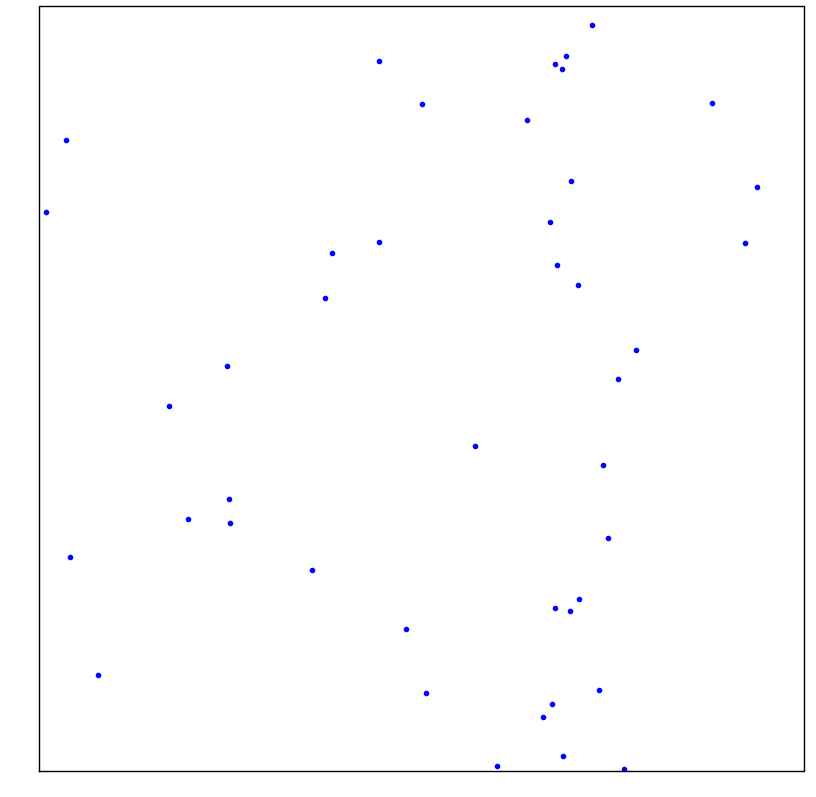} \\
    \includegraphics[width=0.45\linewidth]{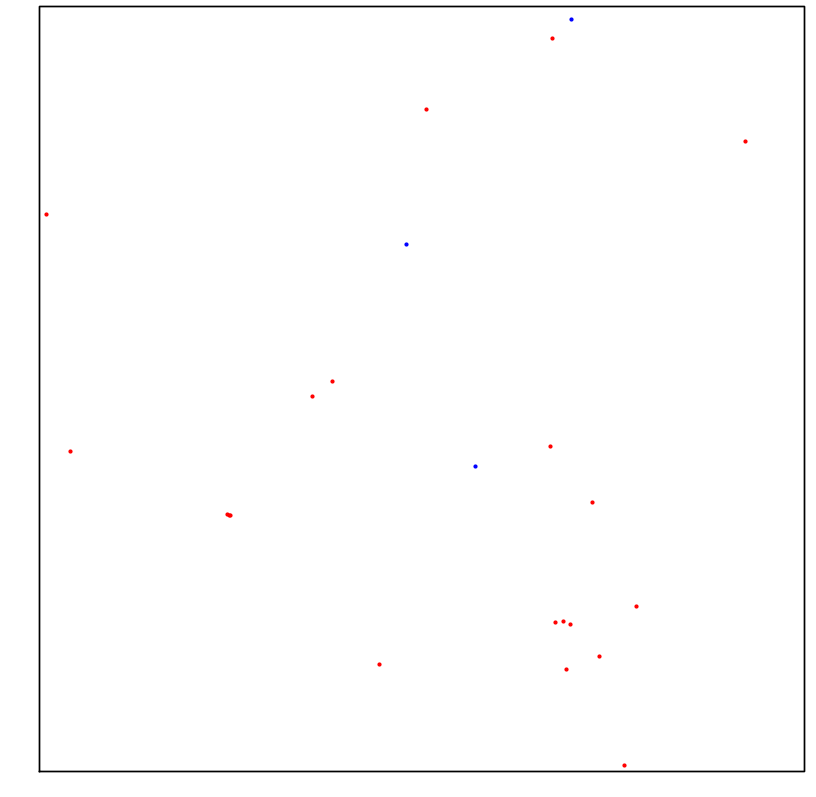} &
    \includegraphics[width=0.45\linewidth]{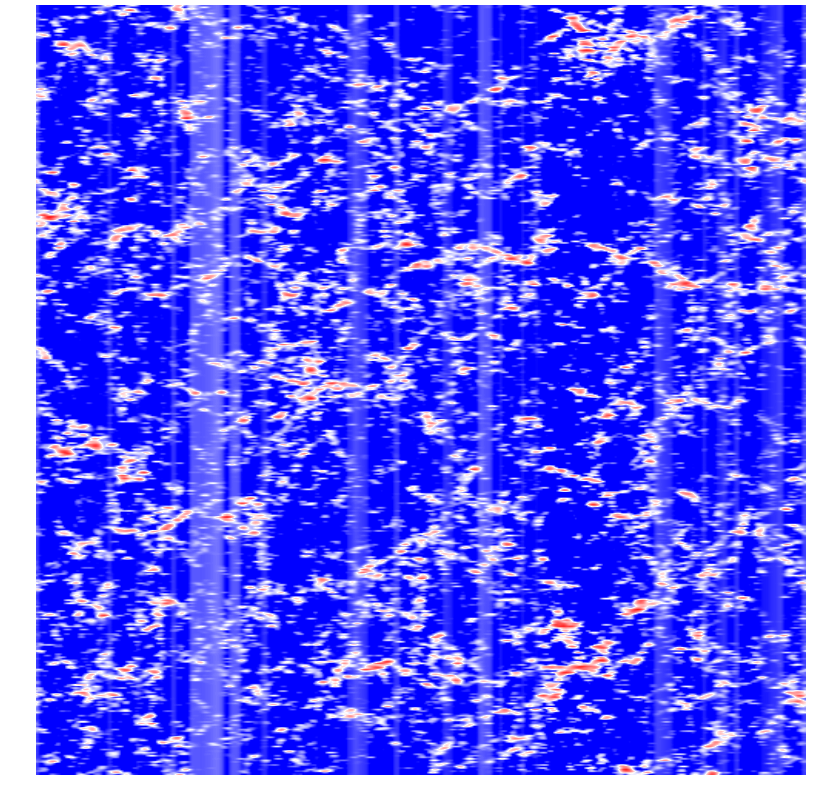} \\

  \end{tabular}
  \caption{Same as Figure \ref{fig:panella} but at $z=5$. Compared to
    lower redshift, the number of sources tracing the structure
    decreases further to become completely shot-noise dominated. The
    transverse smoothing for the \stagetwo\ experiment becomes more
    pronounced, but it nevertheless captures the richness of the
    underlying dark matter structures.}
  \label{fig:panella2}
\end{figure}

Nevertheless, with these caveats in mind, the figure of merit in Eq.~\eqref{eq:fom} is a
rough proxy of instrument capability and Figure~\ref{fig:stages} shows the improvement with
time of the current and proposed experiments. To visually demonstrate the capability of a \stagetwo\ experiment, we
refer reader to Figures \ref{fig:panella} and
\ref{fig:panella2}. These figures display how the proposed instrument
would faithfully measure the structures in the Universe up to very
high redshifts at the large scales relevant to cosmological analysis.

We again iterate that this section was focused on the
post-reionization experiments. There is a vibrant community of
epoch of reionization $21\,$cm experiments and ideas for even higher
redshift. These share many of the technical issues with the \stagetwo\
experiment even though the science is considerably different and are
discussed in Section \ref{sec:intro-highz}.

\subsection{Practical challenges}
\label{sec:intro-pract-limit}

There are several known issues for achieving $21\,$cm cosmology goals compared to traditional galaxy surveys. These call for a coherent
development plan that will allow this technique to reach its full
potential. We stress that the challenges are in the instrument and not fundamental to the signal: with sufficient care, we can build a calibrated system that
will be dominated by statistical rather than systematic errors. These complications and our suggested mitigation for a successful survey are:

\begin{itemize}
\item \textbf{Loss of small-$\kpar$ modes}. The foreground radiation
  is orders of magnitude brighter than the signal, but spectrally
  smooth (see Figure~\ref{fig:schem} for a schematic illustration). Thus, the signal can be isolated but only for modes
  whose frequency along the line of sight ($\kpar$) is sufficiently
  large. As a consequence, the low-$\kpar$ modes are lost and this
  precludes direct cross-correlation with tomographic tracers such as
  weak lensing. However, as we discuss, these modes can be
  reconstructed from their coupling to the measurable small-scale
  modes, with non-trivial precision for a sufficiently aggressive system.

\item \textbf{The foreground wedge}. Interferometers are naturally chromatic
instruments, since their fringe patterns---and therefore the cosmological lengthscales
that they probe---are dependent on frequency. This can cause extra spectral
features to be imprinted on the (in principle) spectrally smooth foregrounds. For
a power spectrum measurement, this results in a set of Fourier modes on the
$\kperp$-$\kpar$ plane (``the foreground wedge") that are heavily contaminated by foregrounds.
 This problem becomes more important at
  higher redshift and is acute for epoch of reionization experiments.
  We note that there is nothing fundamental about this problem: the mathematics
  behind the wedge are well-understood \cite{2012ApJ...756..165P,2014PhRvD..90b3018L}, and thus an appropriate analysis pipeline applied to a
  well-calibrated system with sufficient baseline coverage can
  in principle perfectly separate the foregrounds from the signal even
  inside the wedge \cite{2014PhRvD..90b3019L,Koopmans2017}. The problem is therefore primarily a technical challenge rather
  than a fundamental limitation.  We discuss our modeling of, and assumptions about, the foreground wedge in Appendix \ref{app:wedge}.
  
\item \textbf{The mean signal is not measured.} Because the mean
  signal is not measured, the redshift-space distortions in the linear regime are related
  to the growth parameter $f\sigma_8$ via an unknown
  constant. Cross-correlations with optical surveys \cite{Stephen18} and modeling the mildly-non linear regime of structure formationc\cite{CastorinaWhite19} are effective ways to break this degeneracy. 

\end{itemize}

These issues need to be studied in detail, both in theoretical terms
and through a vigorous experimental program. We argue that major US
agencies should support this research program in order to allow truly
competitive experiments to become reality in the coming decades.

\subsection{Roadmap}
\label{sec:intro-roadmap}

This white paper argues for a long-term development of the $21\,$cm cosmology
program in the USA, led by the Department of Energy but working in
conjuction with other agencies where shared science warrants
cooperation. In particular, a similar model to that of LSST is
envisioned, in which DOE takes up particular aspects of the
development which are well matched to its expertise and a
collaborating agency takes over some of the other aspects that might
not be an optimal fit for the DOE. To this end, we argue for a staged approach that
includes three nominal steps  leading to a Dark Ages experiment, as outlined in Table \ref{tab:roadmap}. 

\begin{table}
 \includegraphics[width=\linewidth, trim = 0 40 0 0]{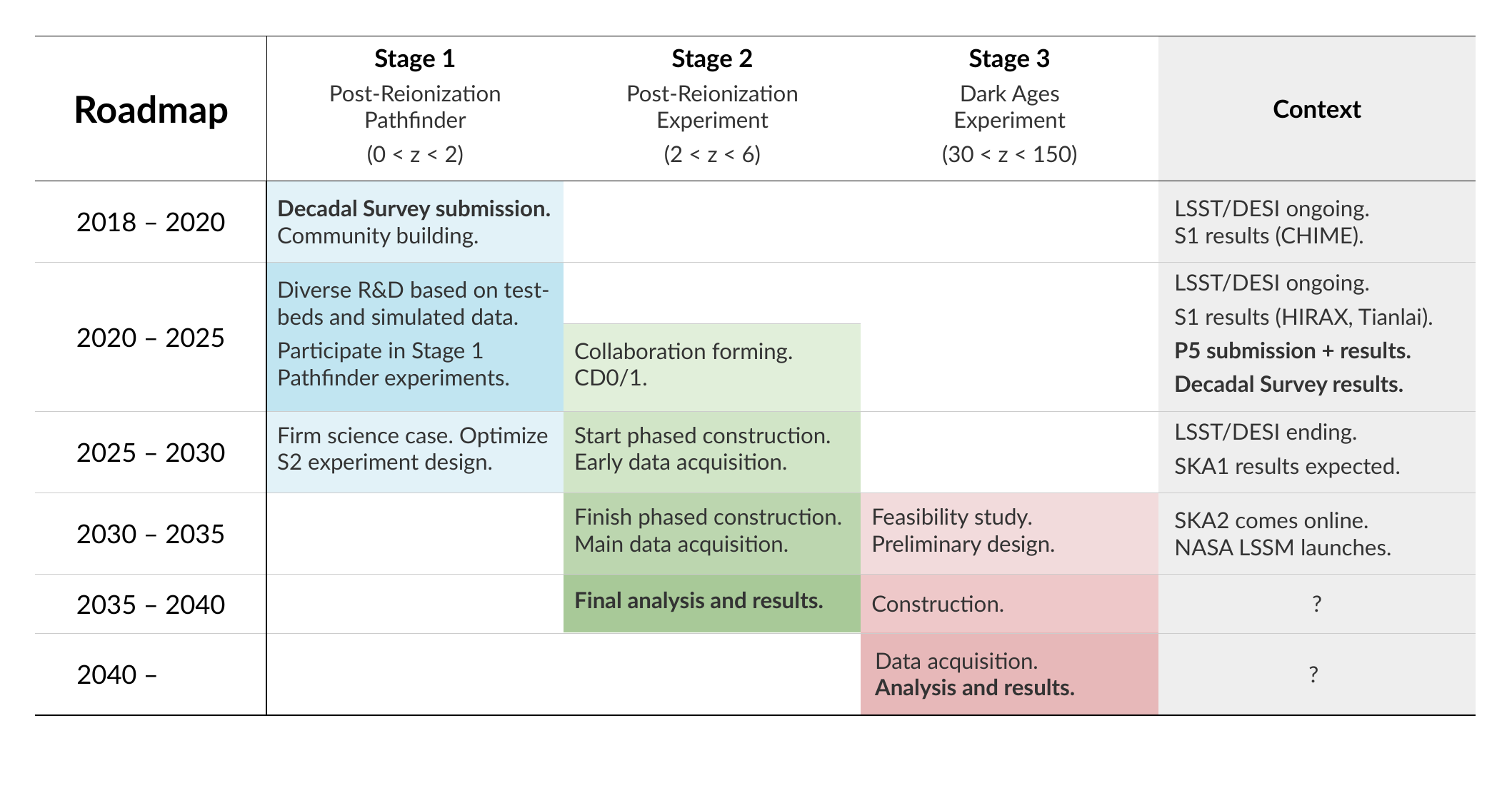}
 \caption{Notional roadmap of the proposed $21\,$cm cosmology program.}
  \label{tab:roadmap}
\end{table}

\begin{itemize}
\item Our first step in the roadmap is an era of vigorous research and
development, probably in conjuction with a small-scale test-bed
experiment. During this stage, the following should be accomplished:
\begin{itemize}
\item \textbf{Refine the scientific reach of a Stage {\sc ii} experiment.} In
  Chapter \ref{sec:sciencecase} we start this process by
  describing some of the exciting science that is achievable using
  a straw man design. The design of the instrument should be driven by
  science and not the other way round, but in practice one needs to
  start with a given design to see the ballpark science achivable and
  then iterate until a convincing science-driven experiment design
  emerges. Our Chapter \ref{sec:sciencecase} is the first step
  in this direction.

\item  \textbf{Advocate for support from major scientific commissions.} In
  particular, the 2020s \emph{Astronomy and Astrophysics Decadal
    Survey} and the next P5 report will need to strongly endorse this
  technique to keep it a viable option.

\item \textbf{Resolve technical challenges.} There are
  numerous technical challenges, particularly in terms of calibration
  and data analysis. We suggest a two-pronged approach: first to benefit from the experience of current-generation experiments in mitigating these challenges, and second to support instrumentation development and theoretical progress using a combination of computer
  simulations, lab experiments, and small, dedicated pathfinder
  instruments. We describe this program in greater detail in Section \ref{sec:challenges}.

\item \textbf{Optimize a \stagetwo\ instrument configuration.} Parameters
  like redshift range, number of elements and their optical designs,
  calibration schemes, etc.\ can crucially affect scientific
  outcome. We will refine and
  optimize the array parameters to both minimize the systematic
  effects and maximize possible science.

\item \textbf{Maintain flexibility in approach.} New exciting scientific developments obtained with optical surveys will be considered when designing the $21\,$cm array proposed here. For example, a sign of early dark energy might motivate a shift towards higher
redshift, while evidence for a non--cosmological-constant equation of state  parameter, $w\neq -1$, might favor lower redshift. Moreover, if fast radio bursts turn out to have useful cosmological applications, they might also affect various design choices. The most important point is that
sufficient resources must be available at this stage to develop the
technique and maximize its promise.

\end{itemize}

\item The next step is a post DESI/LSST experiment, which we call a Stage {\sc ii}
experiment in this document, becoming reality in the later part of the next
decade. To reach interesting cosmological constraints, the experiment will have to be an order of magnitude larger
than current experiments. In this document we consider a
particular fiducial Stage~{\sc ii} experiment operating at  redshifts
$z=0.3-6$, whose parameters we discuss in Section~\ref{sec:fiducial}. 
The main scientific output of this survey will come from surveying the
high-redshift universe, but because the majority of the cost is in the
infrastructure and metal, we have not sacrificed the low-redshift
component, which will offer ample cross-correlation opportunities and
moderate increase in total signal-to-noise.  However, this
particular aspect of the design, as any other, remains on the table to be changed
and optimized as we learn more about the most compelling scientific targets.

\item If successful, we expect this could be followed by a Dark Ages
  experiment. This is the most vaguely defined and forecasted
  instrument, and will require significant improvements and R\&D,
  pushing its timeframe to two or three decades from now. To motivate
  an experiment probing the high redshift $21\,$cm signal, we discuss
  some of the unique science opportunities in Section
  \ref{sec:dark-ages}.  The most important aspect is that there exists
  a long-term scientific opportunity which could be built on top of
  the \stagetwo\ experiment.

\end{itemize}

\subsection{Synergies with optical surveys}
\label{sec:intro-advantages}

Optical galaxy surveys are now a mature observational tool, having
gone from pioneering surveys of a few thousand galaxies, through
definitive detections of cosmological clustering signals like baryon
acoustic oscillations, to now routinely producing precision
cosmological constraints. This successive, multi-generational
development path continues, as next-generation experiments like DESI
are poised to improve over current experiments by an order of
magnitude in depth, and by pushing to significantly higher redshifts.

The $21\,$cm intensity mapping technique is much earlier along its
development track, and must yet pass through a series of milestones
before it can be considered truly competitive with optical surveys. We
can already discern some of the main synergies with the optical
surveys: \\[-0.8em] 

\setlength{\parindent}{0cm}
\paragraph*{\bf 3D information.} 

Optical galaxy surveys fall into two categories: either they survey a huge
sample of galaxies at low redshift resolution (photometric) or survey a subset
of selected galaxies at high redshift resolution (spectroscopic). However, in
both cases we have additional information about galaxies: from photometric
surveys the actual image of the galaxy can be used to infer not just galactic
morphology, but also gravitational lensing and the detailed optical spectra can
be used to infer physical properties of the galaxy, such as star formation. 
$21\,$cm surveys on the other hand provide an avenue that identifies galaxies
and at the same time recovers their redshifts (in an aggregate sense) allowing an
efficient mapping of the full 3D structure in our Universe. This inevitably
loses some information that can be present in the full optical survey, but
offers a complementary path towards a cost-effective survey at high redshift. \\[-0.8em]

\paragraph*{\bf Shot noise vs sample selection.} Any point tracer of
large-scale structure suffers from the fact that we are sampling a
continuous field using a finite number of objects. This Poisson
component, also known as shot noise, acts as a source of noise in any
statistics derived from the large-scale structure observable. To
reduce shot noise, one needs to take spectra of more objects, but most
often there simply are not enough objects up to a given flux, limiting
the ability to mitigate shot noise. In $21\,$cm observations, we are
measuring integrated intensity from all objects, even the very small
and faint ones, and so the shot noise is lower by several orders of
magnitude. In fact, all \stageone\ experiments will be limited by
continuous sources of noise (sky noise and thermal amplifier noise)
and only \stagetwo\ will start to be sensitive to the underlying shot
noise. On the other hand, optical surveys allow one to slice the
galaxy sample into individual sub-samples that can be selected to have
certain properties. Together, both techniques offer complementary
views of the
same underlying structure. \\[-0.8em]

\paragraph*{\bf Scaling with redshift.} Optical measurements excel at lower
redshifts, but they become increasingly difficult as the redshift range of
surveys is pushed towards the more distant universe. First, observations must be
performed in the infrared, where they suffer from brighter sky that has many
more sky-lines which are also more variable than in the optical. Second, the
infra-red detectors are more expensive and less efficient than optical
charge-coupled devices (CCDs).  Third, the objects themselves are fewer in
number and fainter, since we are observing a younger universe. In radio, the
primary limitation is from foreground emission; however, the same foreground
removal techniques vetted by previous generations of $21\,$cm experiments can be
applied because the foregrounds do not fundamentally change across the redshift
range of interest. In addition, at higher redshifts, the same bandwidth covers
more cosmic volume and requirements on things like reflector surface accuracy
become less demanding. In short, for the $z<1.5$ universe,  optical surveys offer
many advantages and offer an excellent tool for studying the universe down to the
smallest scales, but radio techniques scale better towards higher redshift. \\[-0.8em]

\setlength{\parindent}{1em}

\clearpage


\section{Science case for a  post-reionization \texorpdfstring{$21\,$cm}{21cm} experiment}
\label{sec:sciencecase}

This section focuses on preliminary science forecasts for a Stage~{\sc
  ii} $21\,$cm experiment to demonstrate the potential science reach
of such an instrument. A \textit{Stage~{\sc ii} experiment} refers to
an experiment that will build upon the current, non-US, Stage~{\sc i},
pathfinder telescopes such as CHIME and HIRAX\@. We focus on redshifts
after reionization ($z<6$) that will be mostly unexplored by optical
surveys. We design an array to probe these redshifts, based on what
would be possible with current technology at a price-point that is
consistent with a medium-size high-energy-physics experiment. In this
Chapter we envision a realistic experiment that is ``shovel-ready'',
assuming the technical challenges discussed in the next chapter are
feasible and Stage~{\sc i} experiments do not uncover any unexpected
significant issues.

We will describe the science potential that our proposed design could
achieve, briefly in Section~\ref{sec:fiducial} and then in more detail
in the following subsections. We conclude with a discussion of other
relevant science. In this version of the document, we assume PUMA
parameters \cite{PUMAAPC} as a concrete realization of the \stagetwo\
concept, because its design has been optimized in outline for the
science goals at hand.  In later stages of the planning process the
science goals and instrument parameters will be refined further with a
proper flowdown study, likely motivating various modifications or
improvements to the design choices we present here.

\subsection{Science drivers and the straw man experiment}
\label{sec:fiducial}

As outlined in the introduction, there are three  main science drivers for the
proposed experiments: measurement of the properties of dark energy in
the preacceleration era (goal A1), constraints or detection of 
inflationary relics in the shapes of features present in the primordial
power spectrum (goal A2) and constraints or detection of non-Gaussian
correlations in the primordial fluctuations (goal A3).

\begin{figure}[t]
  \centering
  \includegraphics[width=\linewidth]{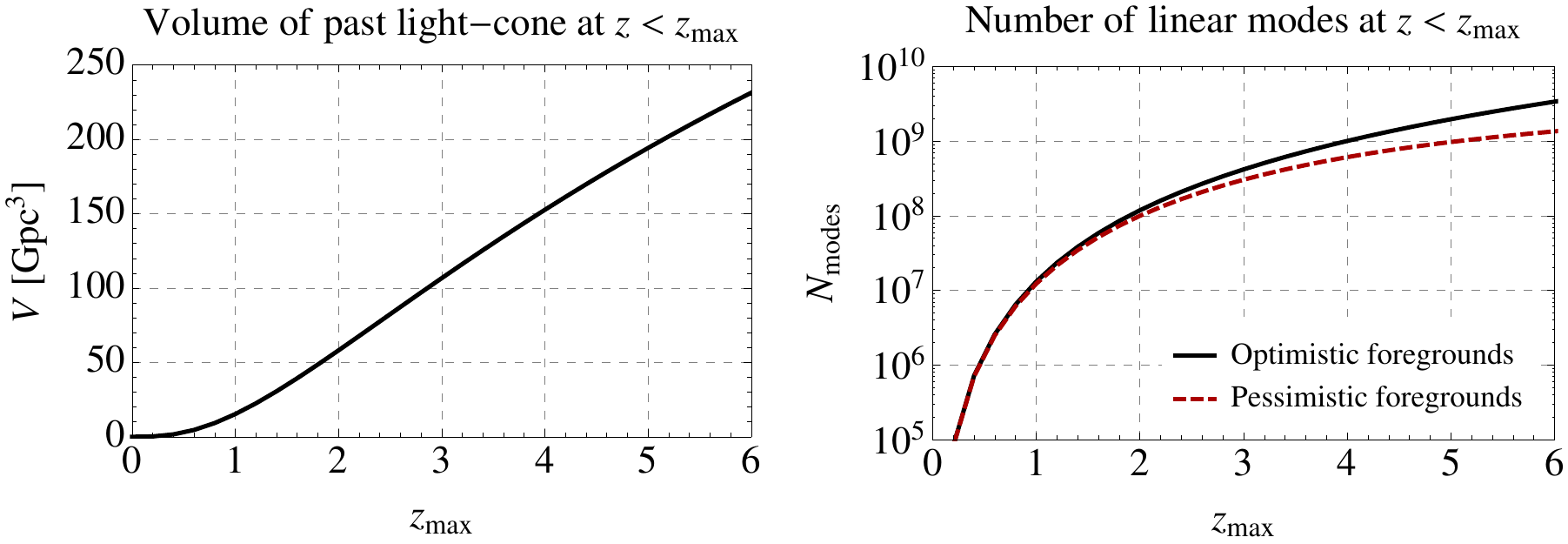}
  \caption{{\it Left}: Cumulative volume observable along our past
    light-cone up to maximum redshift $z_{\rm max}$. {\it Right}:
    Number of linear Fourier modes of the density field observable up
    to $z_{\rm max}$, where ``linear" refers to modes whose statistics
    can be predicted at the few-percent level (the precision required
    for many science cases in this section) by modern perturbation
    theories of large-scale clustering. A full-sky $21\,$cm survey
    over $2<z<6$ can in principle access $\sim 3$ times more volume
    and $\sim 30$ times more linear modes than a survey up to
    $z=2$. Even under the pessimistic assumptions about foreground
    contamination, a Stage~{\sc ii} $21\,$cm survey can still access
    $\sim 10$ times more modes than a $z<2$ survey.  }
  \label{fig:volModes}
\end{figure}

\begin{figure}
 \centering
   \includegraphics[width=\linewidth]{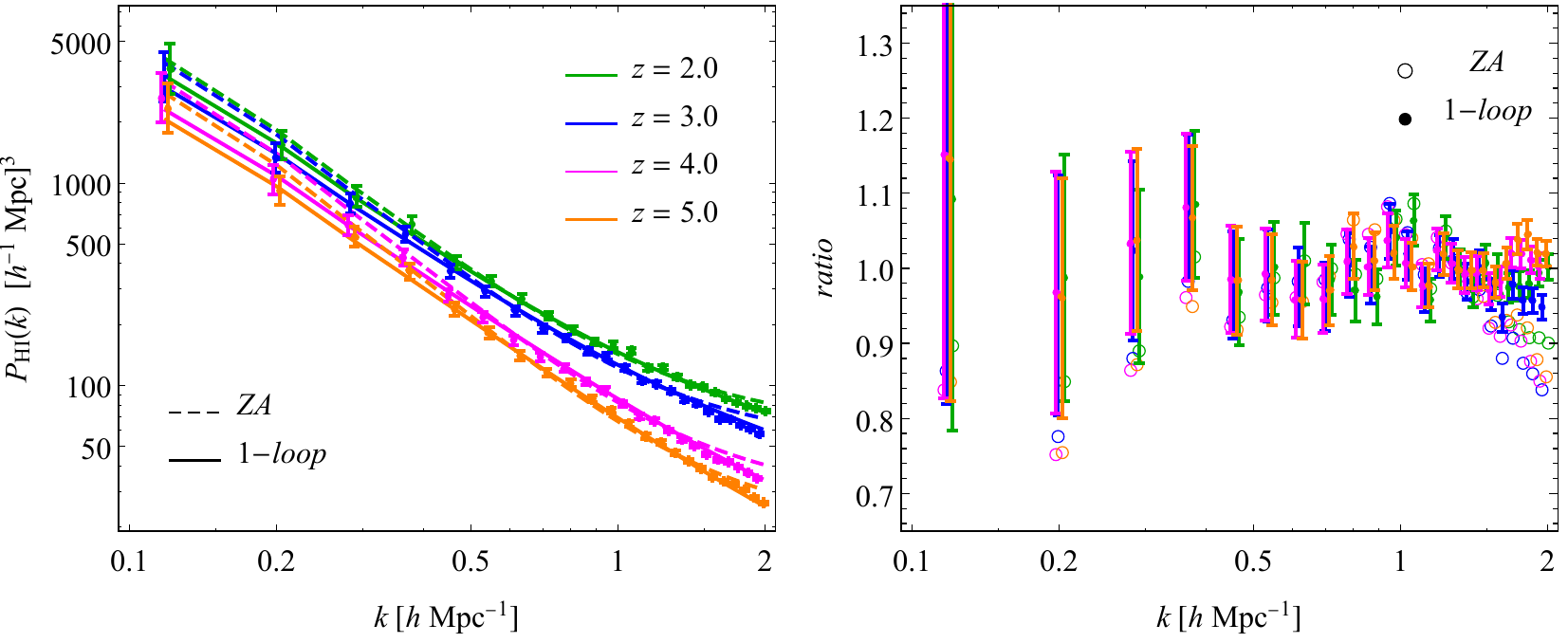}
  \caption{
   Comparison of 1-loop Eulerian perturbation theory and the Zeldovich approximation
   ($1^{\rm st}$ order Lagrangian perturbation theory) to the Illustris
   simulation (from Ref.~\cite{2018ApJ...866..135V}).
   This plot demonstrates that even simple, \textit{ab initio} theoretical models
   can be used to fit $21\,$cm data to very high $k_{\rm max}$, due to both the more
   linear universe at higher redshift and the greater linearity with which the
   neutral hydrogen gas traces these structures.
}
   \label{fig:castorina2}
\end{figure}

\begin{figure}
  \centering
  \resizebox{6.5in}{!}{\includegraphics[trim=0 20 0 0]{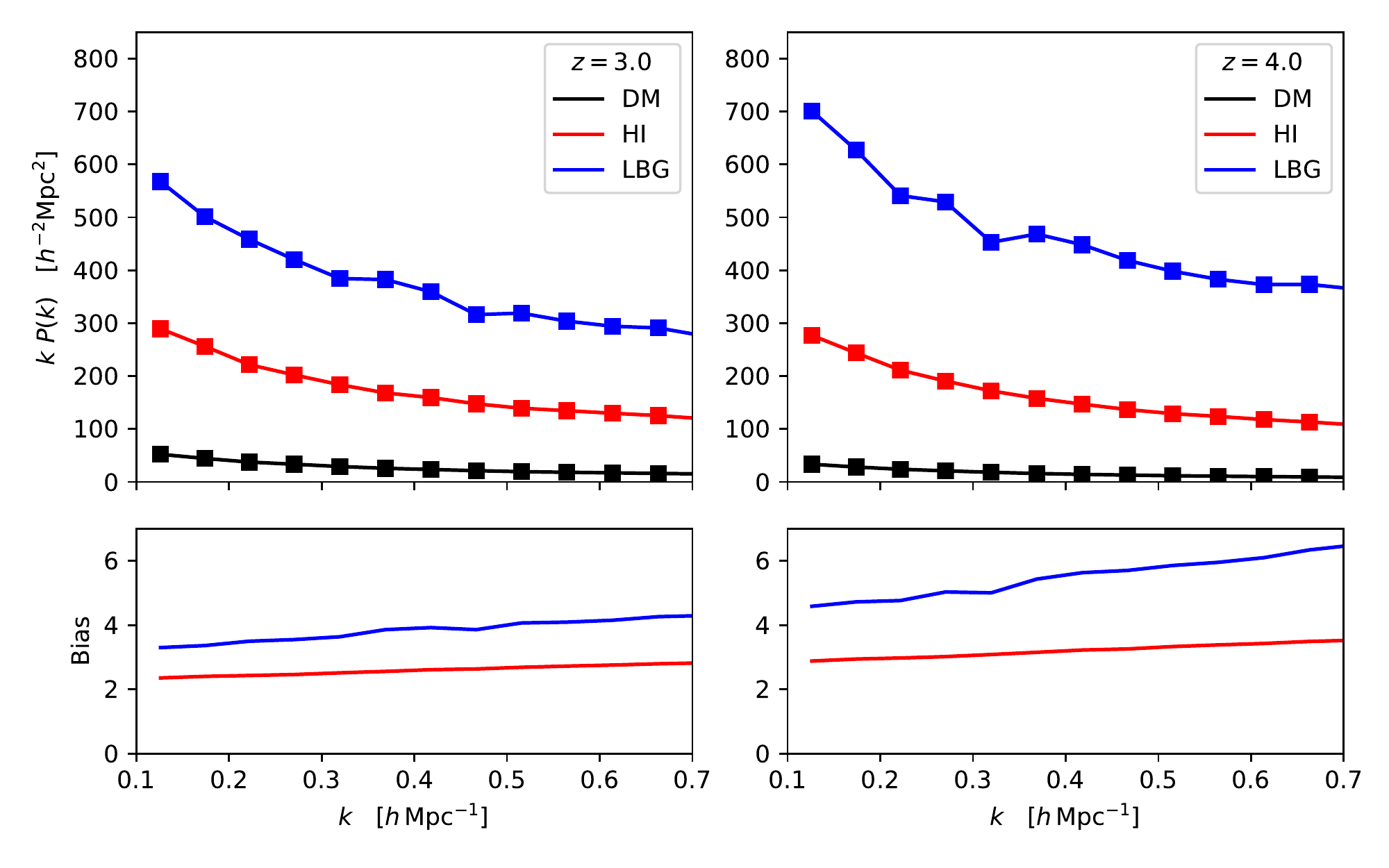}}
  \caption{(Upper) A comparison of the (real space) power spectra for dark matter, mock H{\sc i} and `dropout selected' Lyman Break Galaxies (LBGs) at $z=3$ and 4.  The power spectra are computed from an N-body simulation employing $2560^3$ particles in a $256\,h^{-1}$Mpc box \cite{2015MNRAS.453..311S,2015MNRAS.453.4311S}.  The H{\sc i} is painted into halos and subhalos of the simulation following Ref.~\cite{2018ApJ...866..135V} while the galaxies populate halos following Ref.~\cite{2018PASJ...70S..11H} for $m_{UV}=25$, close to the spectroscopic limit for large samples.  (Lower) The bias, defined as $b_i=\sqrt{P_i/P_m}$, as a function of scale for the H{\sc i} and LBGs.  Note the LBG bias is both larger and more scale dependent than the H{\sc i} bias, because LBGs populate higher mass halos.}
  \label{fig:cmp_bias}
\end{figure}

Goals A2 and A3 are best served by an experiment that has access to a large number of linear or quasi-linear modes.
Given a sufficient density of tracers, the total number of modes scales
as $V k_{\rm max}^3$, where $V$ is the survey volume and $k_{\rm max}$ is
the maximum wavenumber amenable to theoretical predictions.
Going to higher redshift helps both cases. First, there
is more volume per unit redshift at higher redshifts: as indicated in
the left panel of Figure~\ref{fig:volModes}, the total volume available
over $2<z<6$ is roughly triple the volume at $z<2$. The effect is even
more pronounced if one considers the amount of cosmic volume per unit
bandwidth of the radio signal. Second, at a given comoving wavenumber
$k$, the field is more linear at higher redshift, leading to an
increase of $k_{\rm max}$. This translates into a large increase in
the number of usable linear modes at higher redshift, as shown in the
right panel of Figure~\ref{fig:volModes} (see App.~\ref{app:forecasts-modes} for the details of our definition of ``linear modes.''). Figure~\ref{fig:castorina2} confirms that even low order perturbation theory calculations can accurately describe the results of hydrodynamical simulations out to a sufficiently high wavenumber.  Though it is not shown in Figure~\ref{fig:castorina2}, the cross-correlation between the observed and initial fields also remains higher to smaller scales for the $21\,$cm field.  Finally, the bias of the $21\,$cm field is less scale dependent, and easier to model, than a coeval population of galaxies because the neutral hydrogen traces lower mass halos (Figure~\ref{fig:cmp_bias}).  This effect becomes particularly pronounced at the highest redshifts.

By a fortunate coincidence, all three science drivers naturally lead
to a high-redshift experiment. The upper limit is set by the
requirement that the universe has reionized and thus astrophysics does
not limit our modeling, which requires $z<6$ or the low frequency edge
of 200MHz.  The lower limit is set by what we think are practical
considerations in terms of the maximum fraction bandwidth that we
believe is credibly obtainable. Based on \cite{Vanderlinde}, we set
the upper frequency limit to $1100$MHz, resulting in a lower redshift
limit to $z=0.3$. This gives the total bandwidth of $\sim$5.5,
somewhat lower than three octaves.

In \cite{PUMAAPC}, we have identified a 32000 array of 6-m dishes
operating at 200 -- 1100 MHz as a straw man configuration that would
achieve the three main scientific goals specified above.  The Science
Traceability Matrix developed for PUMA for the same goals as discussed
here, calls for hexagonally closed packed array with 50\% fill
factor. We adopt the same configuration in this revision of the
roadmap document, unless stated otherwise.  Such experiment is
$\sim$30 times larger than the partly funded HIRAX experiment,
currently under construction in South Africa.  The total collecting
area of such experiment would be around 0.9 square kilometers. While
this is more than SKA, we stress that the low frequencies and in
particular the non-actuating nature of the transit arrays makes such a
design orders of magnitude cheaper. We assumed a 5-year on-sky
integration, requiring a somewhat longer total duration of experiment,
but note that compared to optical experiments the achieved observing
efficiency can be considerably larger since radio telescopes can often
observe during the day and through cloudy weather.

\begin{figure}
 \centering
   \includegraphics[width=\linewidth, trim=0 10 0 0]{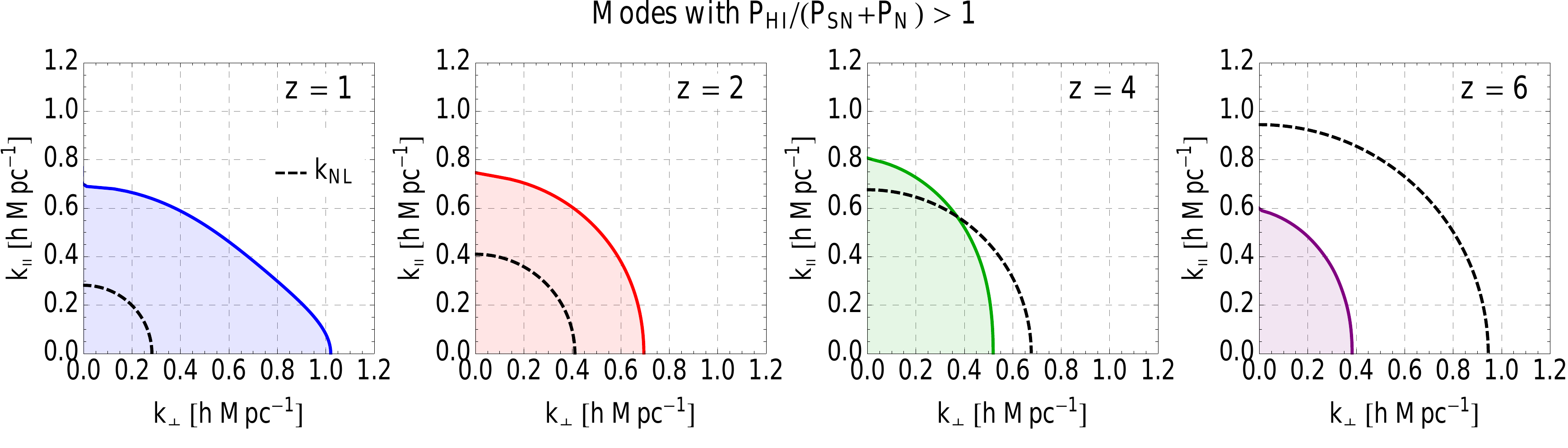}
  \caption{
The shaded regions indicate which Fourier modes of the 21$\,$cm temperature field have ${\rm S/N}>1$ at a few representative redshifts for a \stagetwo instrument, where the noise is a sum of thermal and shot noise. With such an instrument, ${\rm S/N}>1$ can be achieved for all linear modes at 
$z\lesssim 4$ and all modes with $k\lesssim 0.4h {\rm Mpc}^{-1}$ at $z\lesssim 6$. Foregrounds will of course reduce the number of accessible modes in practice, but will nevertheless leave a huge number of modes useful for cosmological and astrophysical studies.
}
   \label{fig:kmax-2d}
\end{figure}

In addition to the main science goals, such experiment would enable
a wide range of other science, both in the field of cosmology and
fundamental physics as well as in related astrophysical sciences that
could be of interest to a broad community. 
One can obtain intuition for the range of available science by asking
which modes of the 21$\,$cm temperature field will have signal
higher than the sum of thermal and shot noise. We show this 
at a few representative redshifts in Fig.~\ref{fig:kmax-2d}, finding
that ${\rm S/N}>1$ can be achieved for all linear modes at 
$z\lesssim 4$ and all modes with $k\lesssim 0.4h {\rm Mpc}^{-1}$ at $z\lesssim 6$. 
In the rest of this chapter,
we study a subset of the most interesting science that would come from
this experiment, with a focus on the cosmological arena.

In our forecasting we assume the existence of the DESI and LSST
experiments. When relevant we also discuss and compare with the CMB-S4
survey, but we note that its final design is less certain than that of
DESI and LSST.  In some sections, we impose additional 2\% or 5\%
priors on cosmic neutral hydrogen abundance, as motivated
by~\cite{2017MNRAS.471.1788C} or achievable using cross-correlation
with other tracers. The results presented in this chapter were derived
using several forecast codes. The common assumptions used to forecast
main results can be found in Appendices \ref{app:forecasts-signal},
\ref{app:wedge} and \ref{app:forecasts-noise}, but even when slightly
different assumptions are used the results are typically consistent to
around 20\% in accuracy over the relevant scales. We regard this as
sufficient at this early stage. Throughout this chapter we will
present forecasts for foreground optimistic and foreground pessimistic
case that are likely to bracket the true value of what level of
foreground cleaning is realistically achievable for the \stagetwo\
experiment.

\subsection{Early dark energy and modified gravity} 
\label{sec:ede}

A concerted, community-wide effort to explain the origin of cosmic
acceleration has uncovered a vast zoo
of dark energy and modified gravity models. These can be broadly
classified according to how they modify GR or replace the cosmological
constant, $\Lambda$ -- for example, by adding new scalar,
vector or tensor fields; adding extra spatial dimensions;
introducing higher-derivative or non-local operators in the action; or
introducing exotic mechanisms for mediating gravitational interactions
\cite{JaiKho10,Clifton:2011jh, Weinberg13, Joyce15, Joyce16, Amendola18}.
A summary of some possible new gravitational phenomena that can result from
these modifications is included in Table~\ref{tbl:newphys_summary}.

\begin{table}
  \centering
  \includegraphics[width=1.0\linewidth, trim = 30 50 30 0]{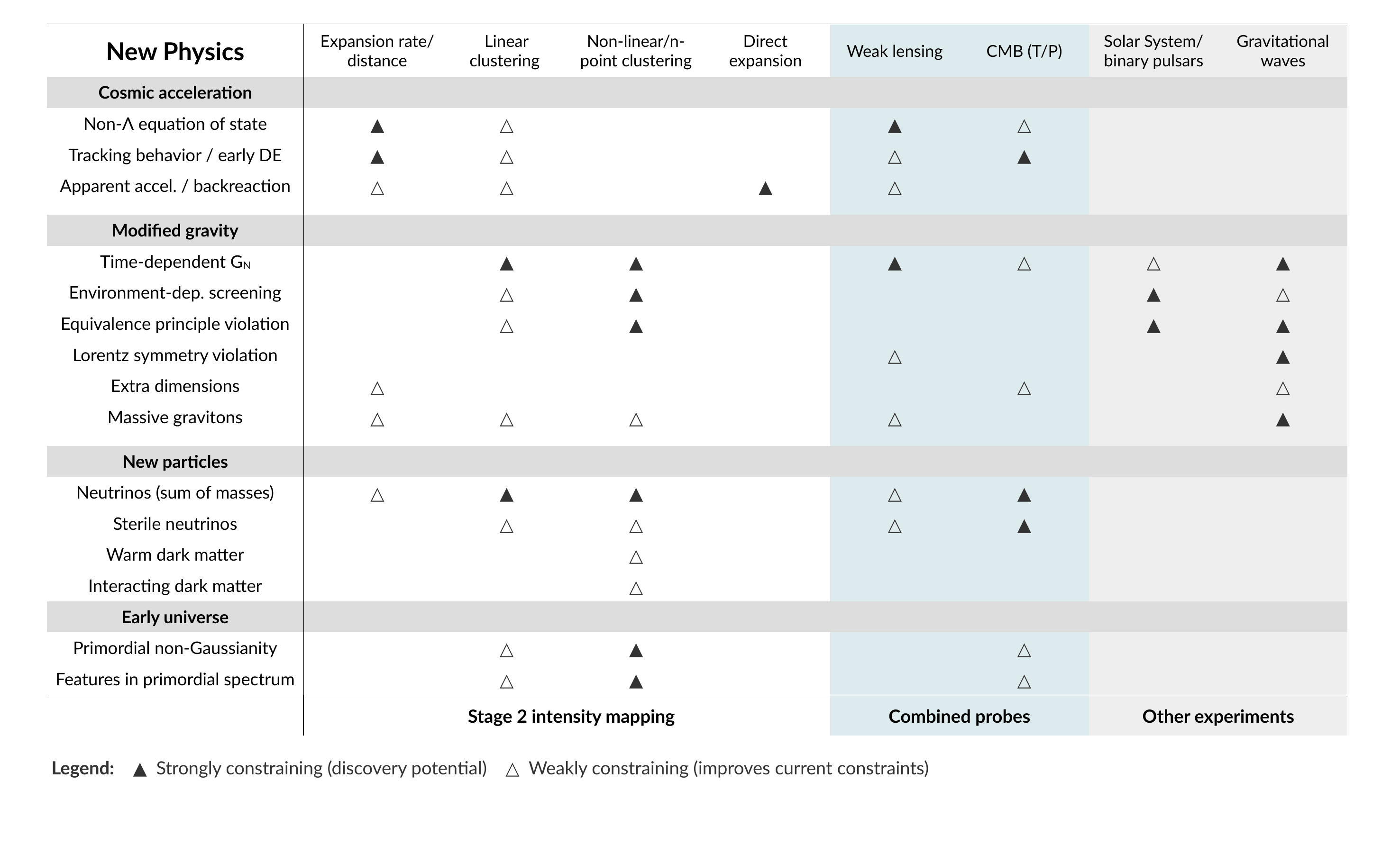}
  \caption{Summary of some possible signatures of new physics, along with a qualitative assessment of how well each of them can be constrained by \stagetwo\  and other experiments. Empty triangles denote observables that are only expected to offer mild improvements in constraining power, e.g. improvements of at most a factor of a few over current constraints. Filled triangles denote observables that have the possibility of yielding strong, possibly decisive constraints on new physics, e.g. improvements of an order of magnitude or more. Combined probes are observations from other experiments that can be combined synergistically with \stagetwo\  measurements.}
  \label{tbl:newphys_summary}
\end{table}

A systematic study of these models suggests a
number of new gravitational phenomena that can arise if
there are any deviations from the standard cosmological model. These
include the possibility of a time-varying equation of state for the
component that sources the cosmic acceleration; time- and
scale-dependent variations in the gravitational constant (leading to
modifications to the growth rate of large-scale structure and
gravitational lensing \citep{2013LRR....16....6A, 2013arXiv1309.5389J, 2013PhRvD..87b3501A, 2014PhRvD..90l4030B, 2015PhRvD..91h3504L}); and `screening' effects, where the strength of
gravity becomes dependent on the local environment \citep{Khoury:2003rn, 2010PhRvL.104w1301H, 2012PhRvD..86d4015B, 2013arXiv1309.5389J, 2016RPPh...79d6902K}. It is also the
case that current constraints on possible deviations from GR are quite
weak on cosmological scales, compared to the extremely precise
measurements that have been obtained on Solar System and binary pulsar
scales \citep{2015ApJ...802...63B, 2015CQGra..32x3001B}.
The application of GR to cosmology therefore represents an
extrapolation of the theory over many orders of magnitude in scale
from where is has been well tested. Constraints on GR on
cosmological scales are therefore a natural programmatic goal for
cosmology.

Observational constraints on possible deviations from GR+$\Lambda$ are
only now becoming sufficiently accurate to constrain a wide variety of
these scenarios. Recent theoretical work has significantly simplified
the task of testing dark energy and modified gravity theories, by
collecting many possibilities into a handful of broad classes,
such as the Horndeski class of scalar field theories, which can then
be studied in a general sense, instead of on an individual
`model-by-model' basis \cite{2013JCAP...02..032G,
  2013JCAP...08..010B, 2014JCAP...07..050B}. Although measurement of
the speed of propagation of gravitational waves based on the gravitational
wave event GW170817 and its electromagnetic counter-part GRB170817A
\cite{2017ApJ...848L..13A} has tightly constrained a large number of possible
modified gravity theories \cite{2016JCAP...03..031L,
  2017PhRvL.119y1302C, 2017PhRvL.119y1303S, 2017PhRvL.119y1304E,
  2017PhRvL.119y1301B, 2018PhRvL.120m1101A} (although see Ref.~\cite{deRham:2018red} for a critique that may mitigate this conclusion), large parts of parameter space remain unconstrained.

One can make predictions for observables within the context of these
general classes, to see where the possibility of detecting a
(potentially quite small) deviation from the standard cosmological
model might be maximized. This exercise has so far been performed for
a handful of theory classes and observables. In \cite{2017PhRvD..96h3509R},
for example, generic predictions were obtained for the behavior of the
equation of state of dark energy $w(z)$, within the full Horndeski
class. Interestingly, many of these theories predict a `tracking' type
behavior, where $w(z)$ scales along with the energy density of the
dominant fluid component at any given time. This leads to the
expectation that $w \simeq -1$ at low redshift, $z \lesssim 2$, where
dark energy begins to dominate, but $w \to 0$ at higher redshift, deep
within the matter dominated regime. This behavior is caused by
couplings between the scalar field and the matter sector that
generically arise in many branches of the Horndeski theories (although
tracking can also be realized in models without such couplings,
e.g.\ freezing quintessence models \citep{2006PhRvD..73f3010L}). The fact that this behavior is a
reasonably generic prediction of a large and important class of models
(most scalar field dark energy theories are included within the
Horndeski class) highlights the need for precision observations in the
intermediate redshift regime, $z \gtrsim 2$. If the equation of state
can be reconstructed at these redshifts, possible tracking behaviors
can be either definitively detected or thoroughly ruled out. Without
such direct observations however, it will be difficult to tell whether
a transition is occurring, or whether a possible disconnect between
observations at low and high redshifts is due to some other factor
(e.g.\ systematic effects).  In Section~\ref{sec:meas-expans-hist} we discuss how the \stagetwo\  experiment will
measure the expansion history at sufficiently high redshifts to constrain these models.

It is similarly important to test the growth rate of large scale
structure over a range of redshifts, to ensure that possible
deviations from GR on large scales have not been missed or absorbed
into constraints on other parameters at late times
\citep{2008PhRvD..77h3508D, 2010PhRvD..81d3512S, 2014PhRvD..89b4026B,
  Perenon:2015sla}. As with the equation of state, the $z \gtrsim 2$
range is currently lacking in direct observational probes of the
growth rate. In Section~\ref{sec:rsd} we will discuss ability of
\stagetwo\ experiment to measure the growth rate at high redshift.

Finally, we observe that 21\,cm is uniquely sensitive to very small
halos, where galaxies are usually too faint to be observed
directly. Some of the modified gravity theories that pass all
current observational tests predict that the abundance of those light halos
could be a sensitive probe of gravity modifications, rendering
\stagetwo\ a unique probe \cite{2019arXiv190702981L}.

\subsection{Measurements of the expansion history}
\label{sec:meas-expans-hist}

Baryonic Acoustic Oscillations have been a staple of survey science for
the past decade. They allow measurements of the expansion history
of the universe, whose relative calibration is naturally below percent
level and whose absolute calibration depends only on the well
understood plasma physics in the early universe.

In the early Universe, before hydrogen recombination, electrons, baryons and photons formed a tightly coupled plasma with a short mean-free path.  Perturbations in this plasma, seeded at much earlier times by inflation, propagated as acoustic waves until the photons decouple from the plasma at recombination.
The compressions and rarefactions in the plasma leave an imprint on the distribution of matter in the Universe at a characteristic scale of $r_d\simeq150\,\text{Mpc}$: the speed of sound in the primordial plasma times the age of the Universe at decoupling.  This scale is most commonly measured from the peak in the correlation function or, equivalently, the series of oscillations in the power spectrum known as baryon acoustic oscillations (BAOs; see Refs.~\cite{Weinberg13,PToday17,PhysRevD.98.030001} for recent reviews).

\begin{figure}[b!]
  \centering
  \includegraphics[width=0.7\linewidth,trim=0 20 0 0]{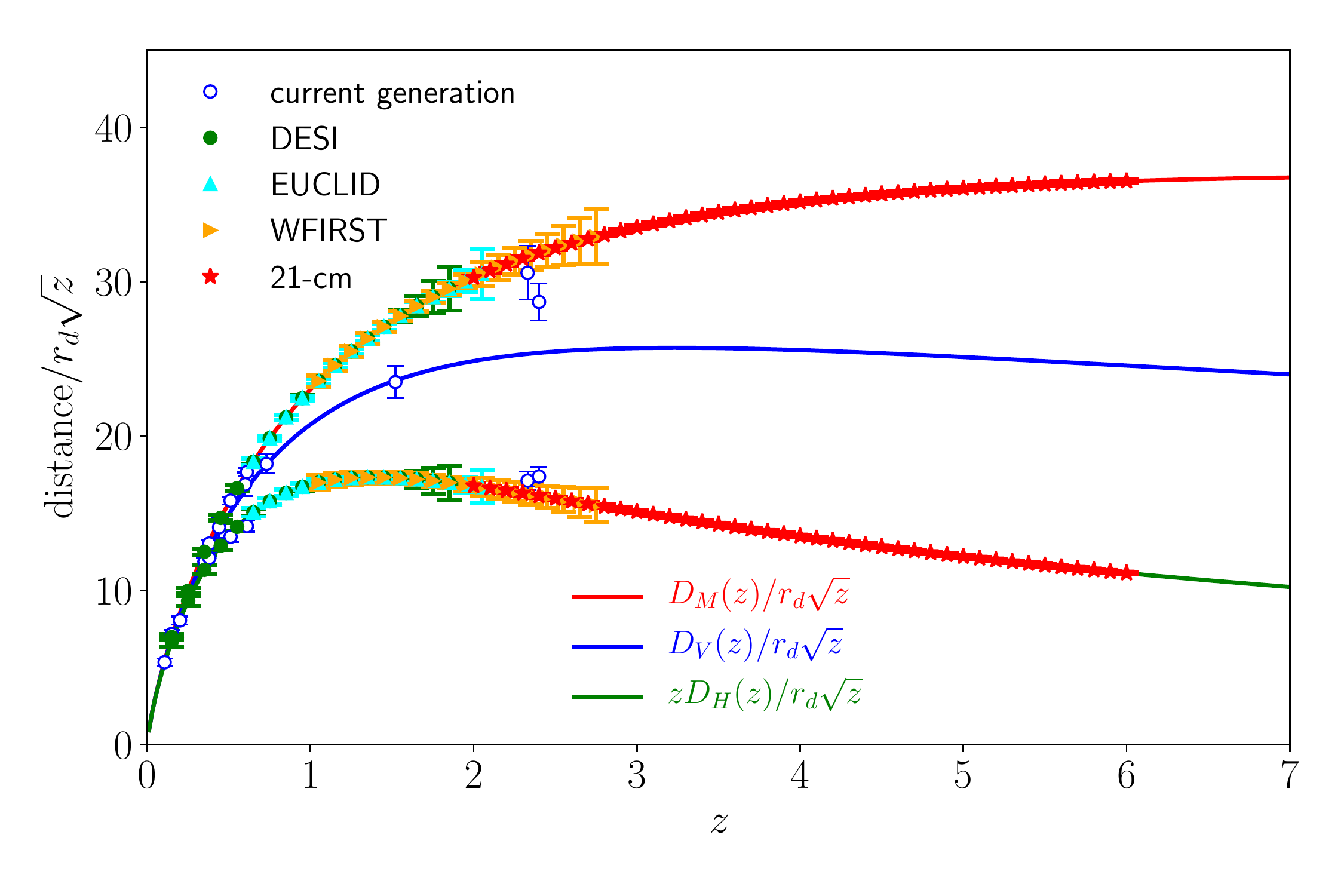}
  \caption{Constraints on the distance-redshift relation(s) achievable
    with the BAO technique for some current experiments (empty symbols),
    some up-coming experiments (based on \cite{2014JCAP...05..023F})
    and our Stage {\sc ii} experiment (based on \cite{2017arXiv170907893O}).
    This Figure is an adaptation of Figure 1 from
    \cite{2015PhRvD..92l3516A}. Lines from top to bottom correspond to
    transverse, spherically averaged and radial BAO for best-fit
    Planck $\Lambda$CDM model. $21\,$cm lines are for foreground
    optimistic case but with no reconstruction. 
    }
  \label{fig:bao}
\end{figure}
These correlations have been successfully detected using galaxies, quasars and the Lyman-$\alpha$ forest \cite{DR12BAO,DR12LyaBAO,SlosarBAO,DR14QSOBAO,WigglezBAO}. In fact, due to the large scales involved and the differential nature of the measurement (one or more peaks on top of a smooth background signal), BAOs are among the most robust measurements in cosmology.  Because the physics of early universe is well known, and highly constrained by CMB observations, the BAO method provides a well-calibrated standard ruler \cite{PCP18}.  With such a ruler BAOs can robustly measure the comoving angular diameter distance, $D_M(z)/r_d$, using transverse modes and the expansion rate, $1/H(z)r_d$, using radial modes; both as a function of redshift.
For this reason current and future spectroscopic surveys (e.g.~\cite{DR12BAO,DESI2016,Euclid2011} or Table \ref{tab:current}) have BAO as a major science driver.  A measurement of BAOs at $2<z<6$, complementary to the next generation of experiments, is one of the scientific opportunities in our proposed Stage {\sc ii} experiment.

In Figure~\ref{fig:bao} we estimate constraints on the distance scale
from a Stage {\sc ii} experiment. The forecasting was done using the
standard approach of Ref.~\cite{2003ApJ...598..720S}, adapted for
$21\,$cm measurements. In particular, at each redshift bin, we add the
shot-noise and thermal noise contribution at wavenumber 
$k=0.2\,h/$Mpc to power spectrum, and convert these back to an
effective number density of sources. The results are largely
independent of choice of fiducial $k$ at which we do this conversion. 
Figure \ref{fig:bao} shows that current and
next generation optical/IR experiments lose constraining power at
$z\simeq 2$, while we forecast a Stage {\sc ii} $21\,$cm experiment
can map the expansion history with high precision all the way to up to
the end of epoch of reionization ($z\simeq 6$).

\begin{figure}
  \centering
\begin{tabular}{cc}
\includegraphics[width=0.49\linewidth, trim=0 10 0 0]{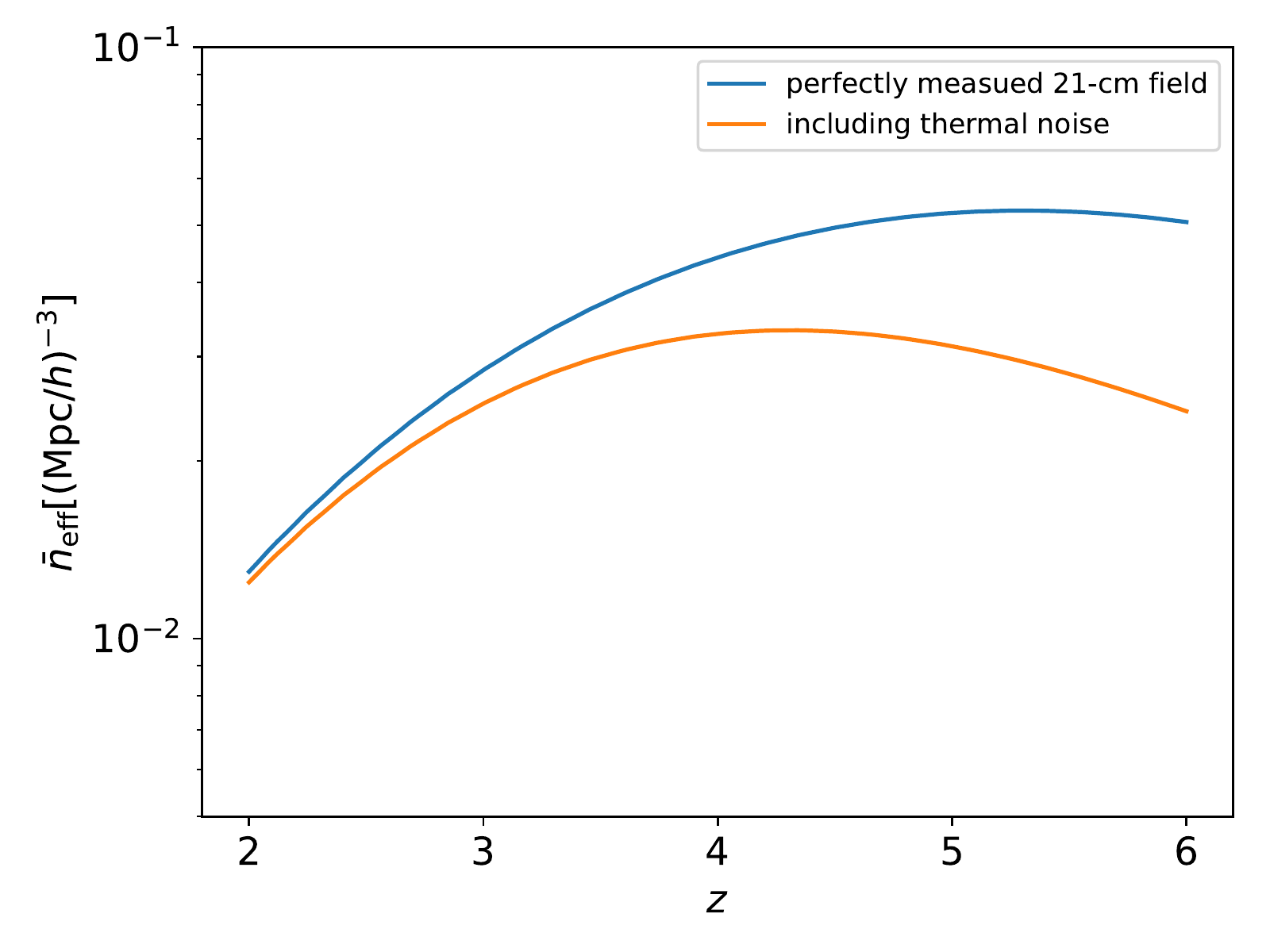} &
\includegraphics[width=0.49\linewidth, trim=0 10 0 0]{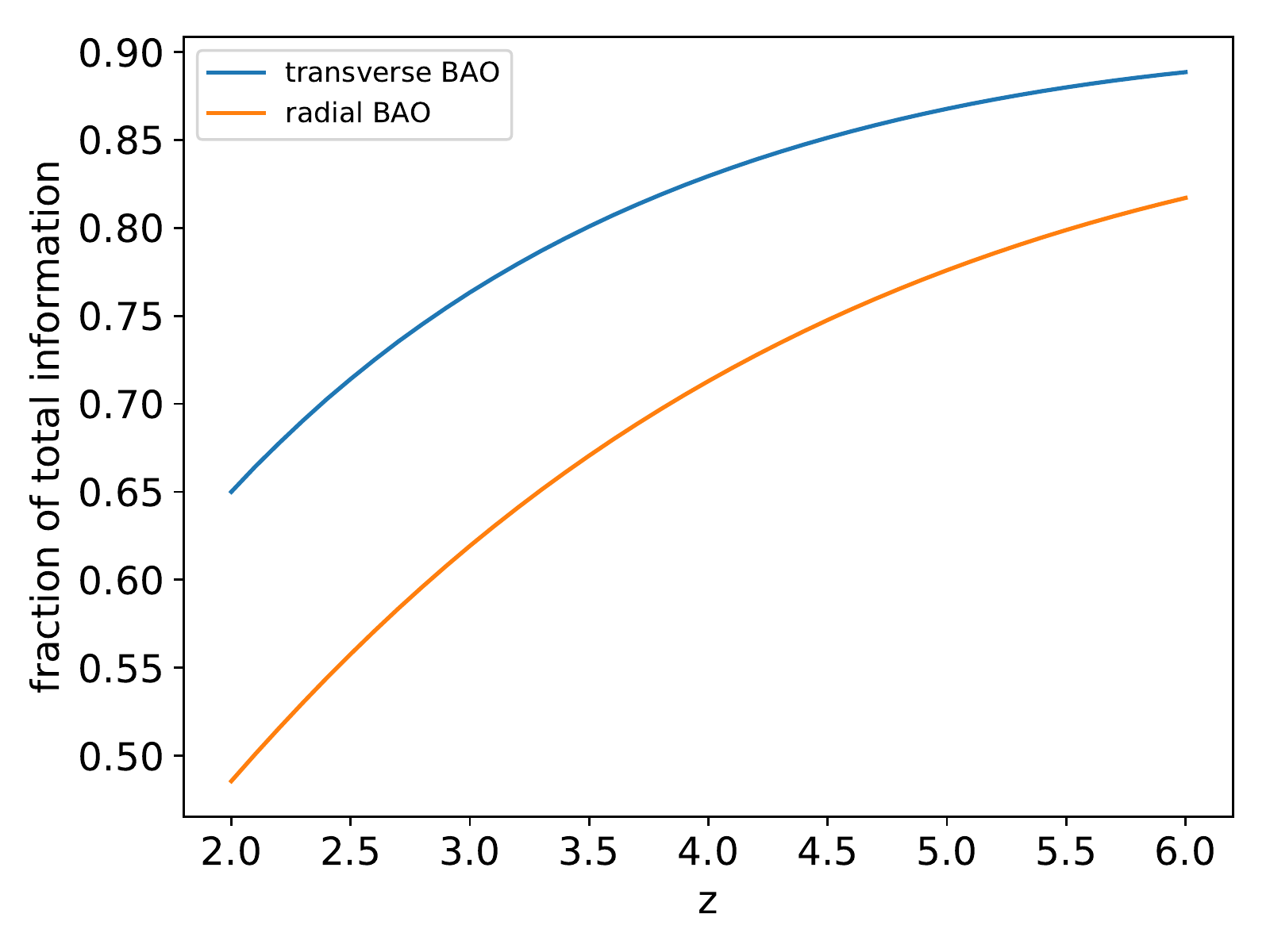}\\
\end{tabular}
\caption{Left: The effective number density of galaxies (i.e. the
  galaxy number density that would have an equivalent shot-noise
  contribution) for perfectly measured $21\,$cm field (blue) and
  effective number density for Stage {\sc ii} experiment after the
  effect of thermal noise has been accounted for (orange). Right: The
  fraction of total signal-to-noise obtained by Stage {\sc ii}
  experiment assuming no reconstruction compared to performing BAO
  measurement on perfectly measured and reconstructed field.}
\label{fig:bao2}
\end{figure}

The high precision achievable with a Stage {\sc ii} experiment is due
in part to the very high number density of $21\,$cm sources, which
provide sample-variance limited measurements of the relevant scales.
The $21\,$cm signal is dominated by numerous, small galaxies with
number densities greater than $10^{-2}\,h^3{\rm Mpc}^{-3}$.  This can
be compared to typical values for galaxy surveys which are around
$10^{-4}-5\times 10^{-3}\,h^3\,{\rm Mpc}^{-3}$ or less.  We plot these
numbers in the left panel of Figure \ref{fig:bao2}. The effect of
the thermal noise of the system (which is not present in optical
galaxy surveys) does lead to a decrease in the effective number
density of sources but for our Stage {\sc ii} survey this is a modest
change.  Provided foregrounds can be controlled, we are close to
saturating the information content in BAO that can be achieved over
half the sky -- no future BAO experiment could do significantly
better as illustrated in the right panel of \ref{fig:bao2}.

\subsection{Cosmic inventory in the pre-acceleration era}

The measurements of the cosmic expansion history and distance-redshift relation described above constrain the abundance and time evolution of the various components of the cosmic fluid.  Radial BAO directly probe the expansion history, $H(z)$, while the angular BAO are related to the angular diameter distance,
\begin{equation}
  D_M(z) = \frac{c}{1+z} \int \frac{1}{H(z)}dz \quad .
\end{equation}
Within GR, both are related to the evolution of the sum of the energy 
densities of components in the Universe
\begin{equation}
  H^2(z) = \frac{8\pi G}{3} \sum_{i} \rho_i(z) \quad .
\end{equation} 
Since the scaling of the energy density with time is known for matter, 
radiation, curvature, and neutrinos, the redshift dependence of $H(z)$ 
can be used to infer the time dependence of the dark energy density.  Assuming basic thermodynamics, this is in turn determined from the dark energy equation of state, $w=p/\rho$.
As discussed in previous sections, 
$w(z)$ is an extremely interesting quantity for studying dark energy models, 
and is being increasingly well constrained at relatively low redshifts, 
$z \lesssim 2$, where dark energy is a large fraction of the total cosmic 
energy density.
In Section~\ref{sec:ede}, we discussed a number of theoretical reasons why the equation of state might be near $-1$ at low redshift but transition to $w\approx 0$ at higher redshift, making it difficult to definitively distinguish dynamical dark energy from a Cosmological Constant using only low $z$ measurements. Indeed, 
some models only show large deviations from $w = -1$ at $z \gtrsim 2$, where dark energy is already a subdominant component of the cosmic energy density \cite{2006JCAP...06..026D, 2016PhRvD..94j3523K, 2017PhRvD..96h3509R}.
This makes these `early' dark energy scenarios relatively difficult to probe, as even quite large changes in equation of state only have a small effect on the total cosmic energy density \cite{2001PhRvD..64j3508B, 2009JCAP...04..002X, 2011PhRvD..83l3504C,2013PhRvD..87h3009P,PCP18}.
BAO measurements from a Stage {\sc ii} $21\,$cm experiment  will make it possible to measure the energy density with sufficient precision to put constraints on early dark energy scenarios however, allowing us to constrain this class of (scalar field) dark energy models.

\begin{figure*}
  \centering
  \includegraphics[width=0.49\linewidth, trim=0 10 0 0]{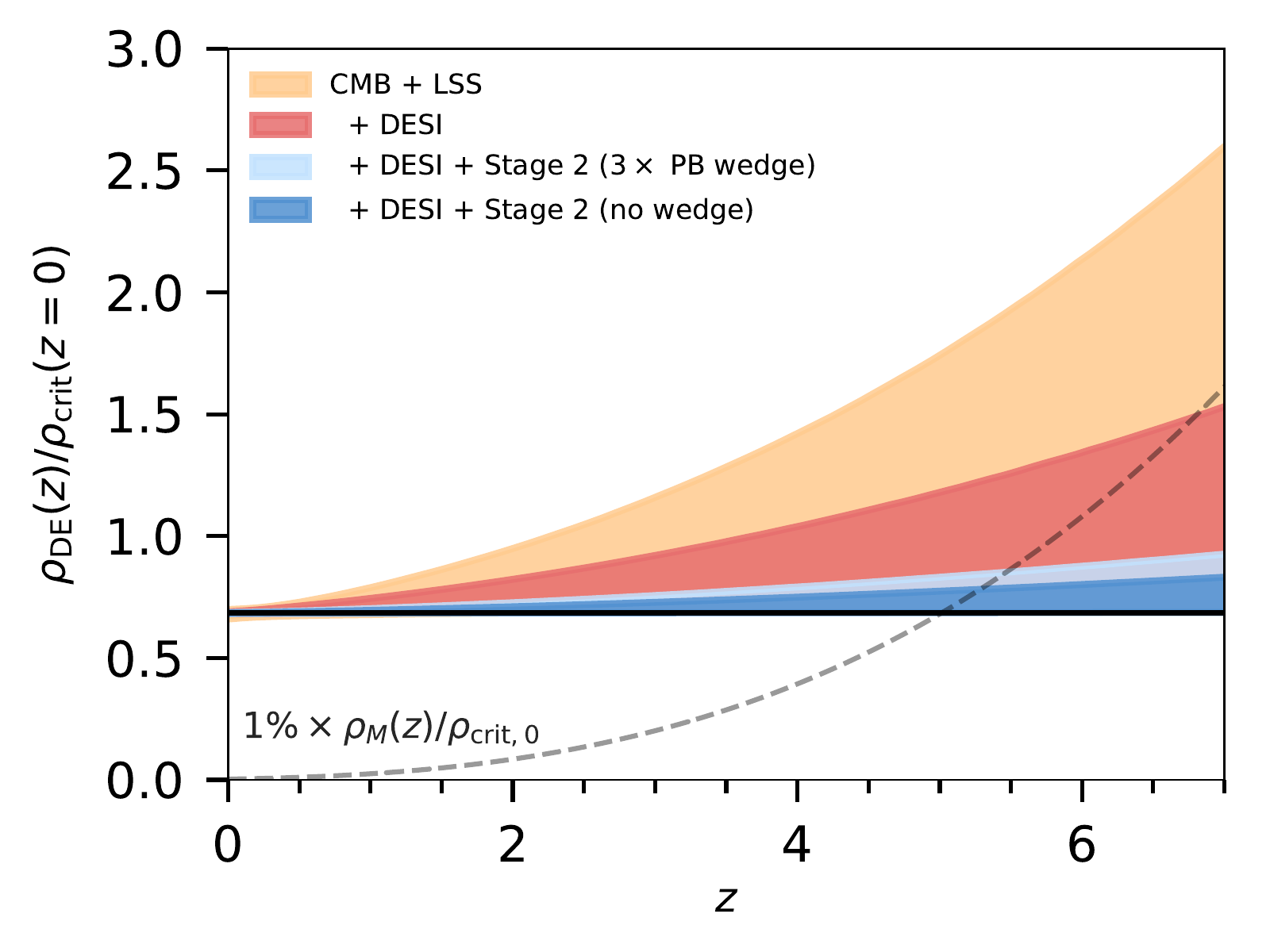}
  \includegraphics[width=0.49\linewidth, trim=0 10 0 0]{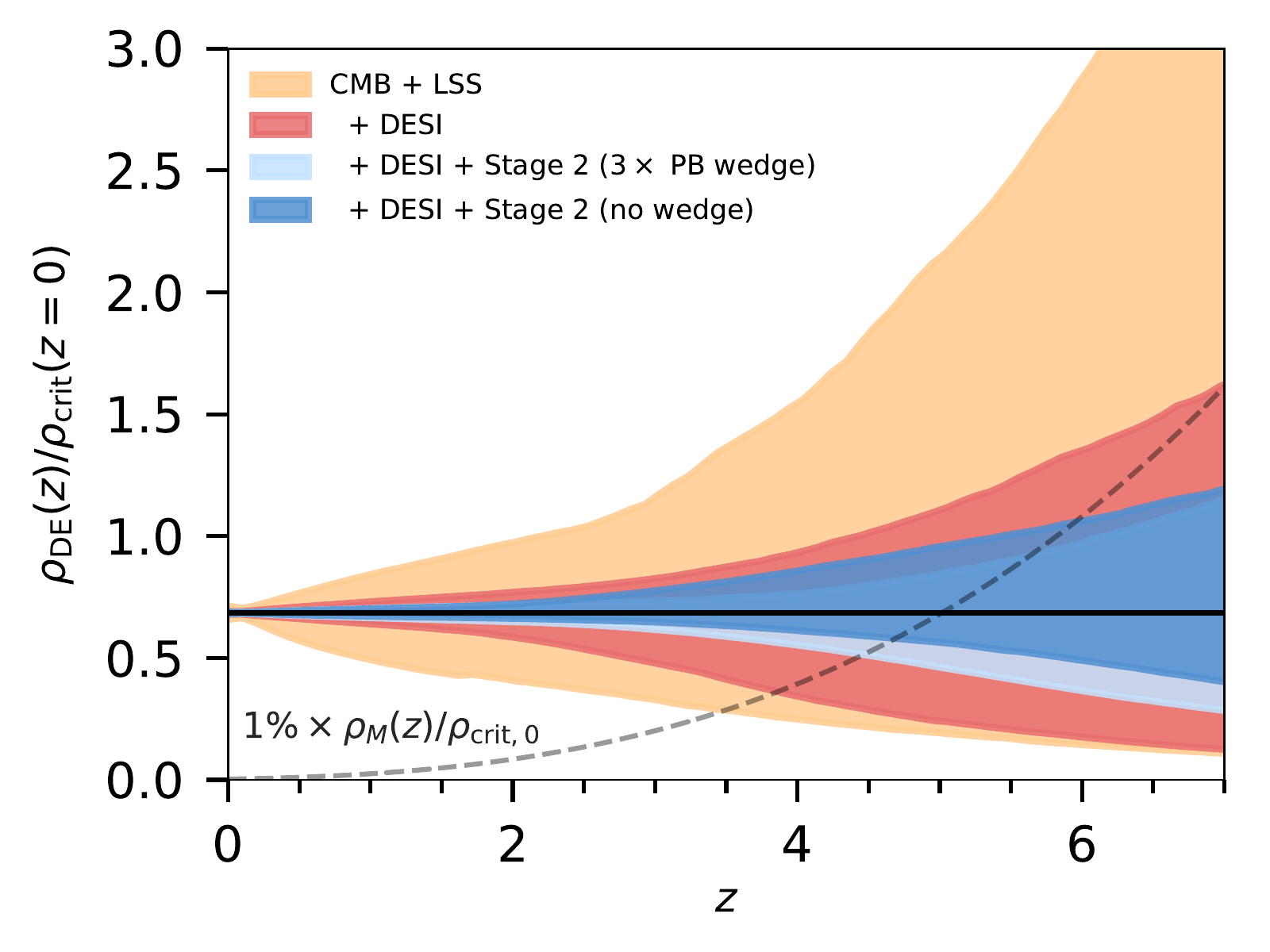}
  \caption{Current and forecast constraints on the redshift evolution
    of the dark energy density out to high redshift, allowing for two
    types of early dark energy -- `mocker' models (left), and
    Horndeski-inspired $\tanh$ models (right). The colored regions
    show the 95\% confidence intervals for several combinations of
    experiments: Planck CMB and existing BAO measurements from
    $0 < z < 0.6$; adding forecasts for DESI BAO at $0.7 < z < 1.6$;
    and adding forecasts for Stage {\sc ii} intensity mapping BAO at
    $2 < z < 6$ with the foreground pessimist and optimistic
    cases. The gray dashed line tracks 1\% of the matter energy
    density, to give some indication of how subdominant the dark
    energy component is.}
  \label{fig:earlyde}
\end{figure*}

To illustrate this, Figure~\ref{fig:earlyde} shows current and forecast 
constraints on the energy density of dark energy as a function of redshift. 
We compare two models that allow early dark energy behaviors, while also 
admitting a fiducial flat $\Lambda$CDM case -- `mocker' models \citep{2006PhRvD..73f3010L, 2006APh....26...16L}, which are a particular class of quintessence models with a smooth transition to a matter-like equation of state at high redshift; and `tracker' models, which are phenomenological models with a smooth step-like transition in the equation of state, motivated by the Horndeski model priors discussed in Section~\ref{sec:ede}. The mocker models are minimally-coupled, and so are constrained to not cross the phantom divide (i.e. go from $w \ge -1$ to $w < -1$), while the tracker models are not subject to this restriction.

In both cases, it can be seen that current data (CMB plus BAO at
$z < 0.6$) constrain any early dark energy component to be less than
about 3\% of the cosmic energy density at $z = 6$, with significant
growth (or decay) in the energy density allowed. Adding the DESI
constraints at $0.7 < z < 1.6$ would improve the upper limit to around
1\% at $z=6$, while still allowing considerable deviations from a
cosmological constant -- e.g. by a factor of 2 in energy density at
$z=6$ for the Mocker models. Adding a Stage {\sc ii} $21\,$cm
experiment, covering $2 < z < 6$, improves the constraints by at least
another factor of two, depending on the model, even in the foreground
pessimistic case. This is a significant improvement
considering that the dark energy density is strongly subdominant at
these high redshifts.

\subsection{Growth-rate measurement in the pre-acceleration era}
\label{sec:rsd}

Redshift-space distortions are an anisotropy of the power spectrum
along the line of sight caused by the peculiar velocities of sources
that add to the cosmic redshift. Since these velocities are sourced by
the same fluctuations in the universe, the result is a particular
distortion of the power spectrum. To lowest order, these distortions
multiply the standard power spectrum by $[b+f\mu^2]^2$, where $b$ is
the large-scale bias, $\mu$ is
the cosine of the angle to the line of sight and $f=d\log D/d\log a$
is the logarithmic derivative of the growth factor.  Given that the
shape of power spectrum is known to a good degree, redshift-space
distortions in traditional radio surveys measure $f\sigma_8$, where
$\sigma_8$ is the linear-theory value of the rms fractional
fluctuations in density averaged spheres of $8\,h^{-1}$Mpc radius at
$z = 0$.  The $\Lambda$CDM model, constrained by current CMB
observations \cite{PL18,PCP18}, predicts both $\sigma_8(z)$ and
$f\sigma_8(z)$ at $2<z<6$ to better than 0.5\% (or about 1.1\% if we
allow neutrino masses to vary).  This provides a firm prediction which
can be tested using precise observations at high $z$.

\begin{figure}
  \centering
  \includegraphics[width=1\linewidth, trim=0 20 0 0]{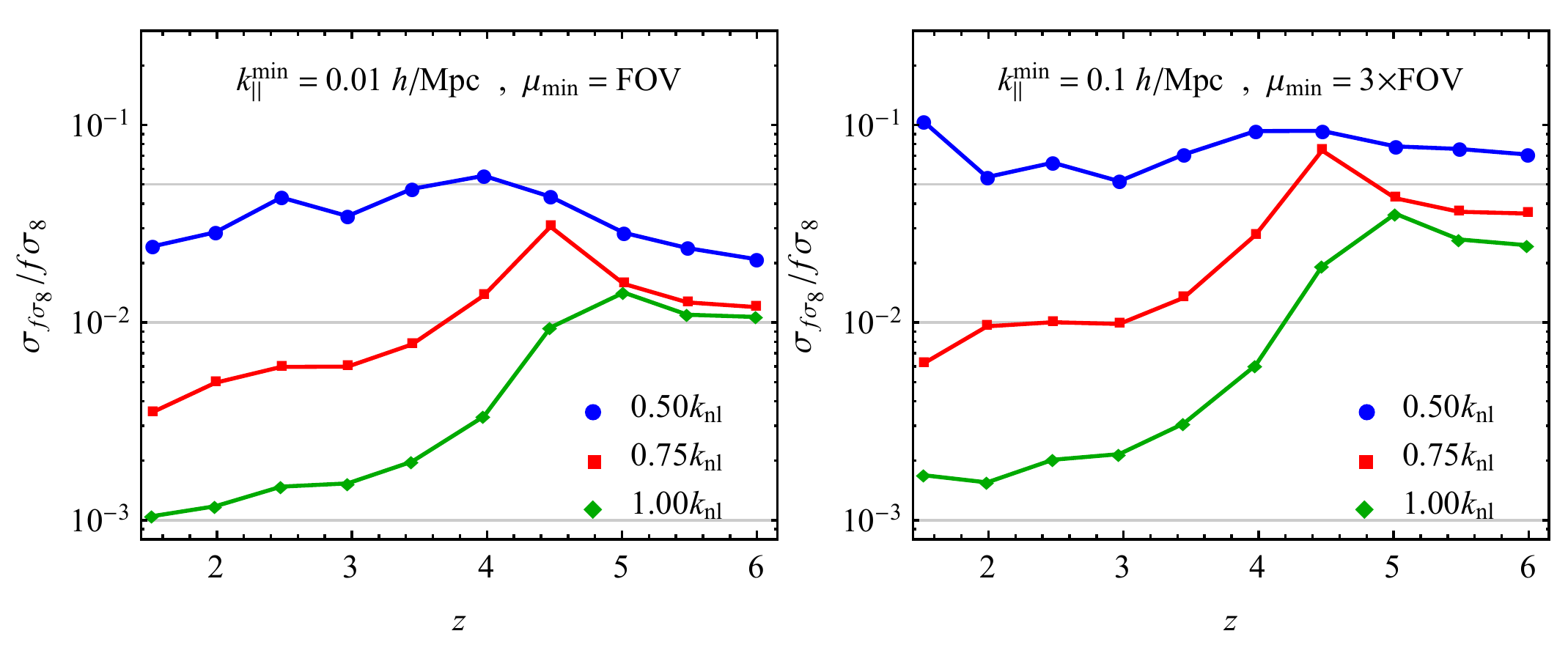}
  \caption{Constraints on the growth rate of structure, $f\sigma_8$,
    for the Stage {\sc ii} experiment assuming no priors on
    $\Omega_{\rm HI}$ from external data but modeling of the power spectrum in the mildly non linear regime using perturbation theory.  Left panel: An optimistic foreground removal scenario where only modes with $k_{||}<0.01 \,h/\text{Mpc}$ are lost and the wedge extends to the size of the primary beam. Different colors show different choices for the smallest scales included in the forecast. Right Panel: A pessimistic case with only $k_{||}>0.1 \,h/\text{Mpc}$ modes available and a wedge extending to three times the primary beam.}
  \label{fig:fsig8}
\end{figure}
\begin{figure}
  \centering
  \includegraphics[width=0.95\linewidth]{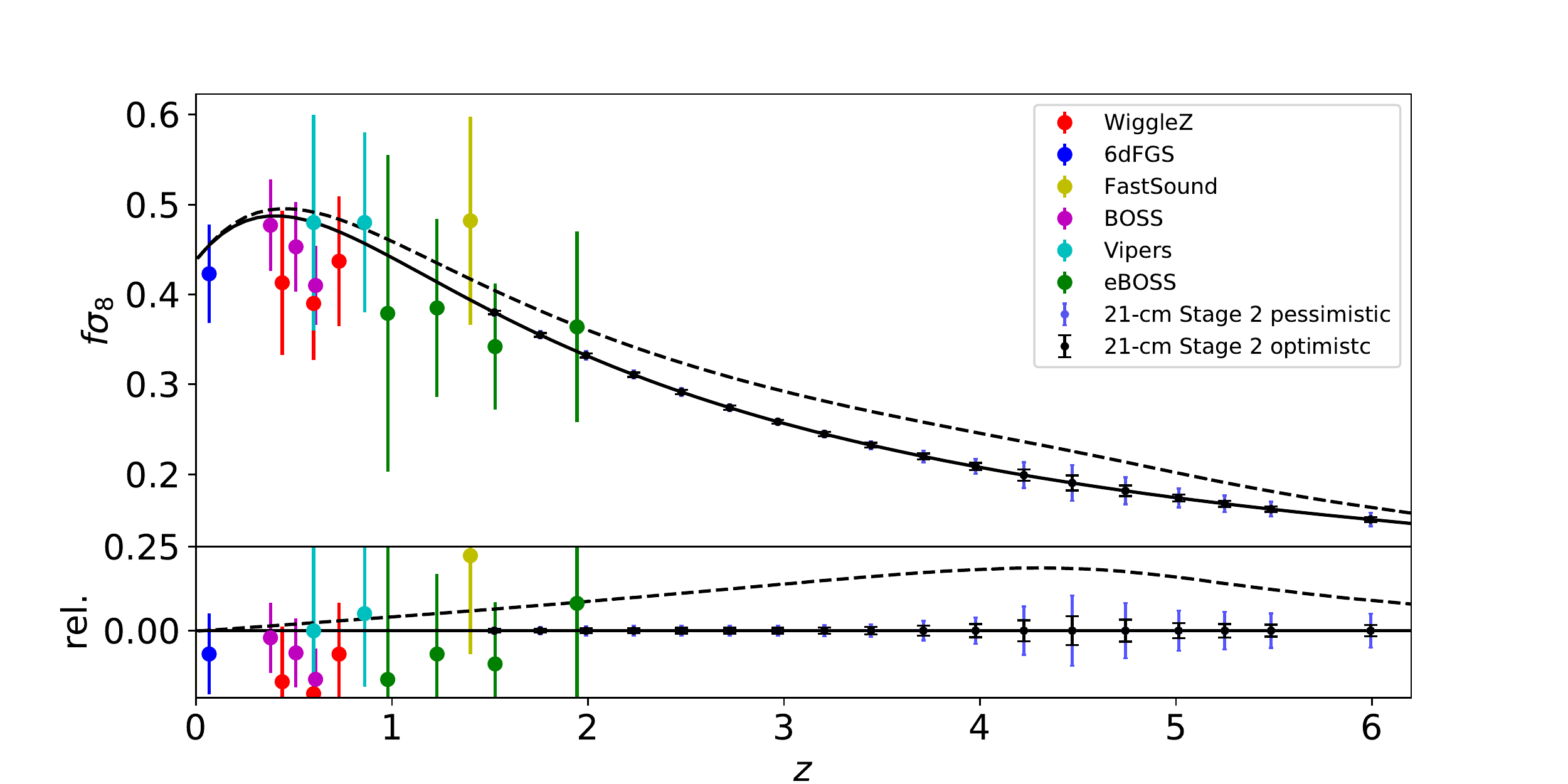}
  \caption{Same data as in in Figure \ref{fig:fsig8} for
    $k<0.75k_{nl}$, but now plotted together with compendium of
    current constraints on $f\sigma_8$ (points; see text).  Lines
    are theoretical models: $\Lambda$CDM is plotted with solid line
    while dashed is the modified gravity model described in the text
    with vanishing effects at high redshift and an expansion history
    equal to that of $\Lambda$CDM.  }
  \label{fig:fsig8mg}
\end{figure}
In $21\,$cm, the mean signal is unknown, so in effect linear redshift-space
distortions instead measure the product $\omegahi f \sigma_8$, with
$\omegahi$ being a nuisance parameter. However, there are three main
ways to go around this limitation.  The first is to use the method of
Ref.~\cite{2017arXiv170907893O}, namely measure the bias and brightness temperature from
complementary data such as the Lyman-$\alpha$ forest, where the
sources relevant for $21\,$cm emission appear as individually detected
hydrogen systems (for a summary of our current understanding of the uncertainties in neutral hydrogen abundance, see refs.~\cite{2015MNRAS.452..217C, Padmanabhan2015,2018MNRAS.473.1879R}). Assuming the foreground contamination can be brought under control, the
resulting constraints are dominated by this prior if it is weaker than
$\sim 1\%$ \cite{Obuljen:2018kdy}. Alternatively, it is possible to cross-correlate with
other tracers at the same redshift as we discuss in
Section~\ref{sec:cross-corr-stud} and Figure~\ref{fig:desiqso}.  Finally, one can use beyond-linear effects to break the degeneracy between $\omegahi$ and $f\sigma_8$ \cite{CastorinaWhite19,2019arXiv190411923M}.  All
methods allow redshift-space distortions to be measured with the
precision of a few percent. This also happens to be close to the current level
of theoretical uncertainty in the modeling of redshift-space
distortions.
Figure~\ref{fig:fsig8} shows the RSD constraints between $1.5<z<6$ achievable by Stage {\sc ii} $21\,$cm for different foreground removal assumptions, in the left panel an optimistic case and in the right panel a more pessimistic one.
Different colors show the smallest scales, largest wave number $k$, included in the forecast in units of the non linear scale $k_{\rm NL}$. We consider somewhere between the red and green line a realistic scenario, for which Stage {\sc ii} $21\,$cm will be able to measure RSD at a few \% precision even in the most pessimistic cases.

We replot the same data in Figure~\ref{fig:fsig8mg} for the red curve together with a selection of current constraints for comparison
\cite{2012MNRAS.425..405B, 2012MNRAS.423.3430B, 2016PASJ...68...38O,
  2017MNRAS.466.2242B, 2017A&A...608A..44D, 2018arXiv180103043Z}.
The theoretical models are the fiducial $\Lambda$CDM model (plotted as a
solid black line) and a moderately tuned modified gravity model (plotted
as a dashed black line) chosen so that the expansion is unaffected at
$z>6$ and the effects are small at low redshift. In particular, we
use the Horndeski formalism of Ref.~\cite{2014JCAP...07..050B}, with the expansion
history fixed to mimic $\Lambda$CDM, $\alpha_{\rm T}=0$ (motivated by
LIGO results) and other parameters proportional to
$\alpha_i(a) \propto (a/a_t)^r/[(a/a_t)^r+1]^2$ with $a_t=1/8$ and
$r=4$.  The theoretical models are generated using the
\texttt{hi\_class} package \citep{2017JCAP...08..019Z,
  2011JCAP...07..034B}. It is clear from the plot that the Stage {\sc ii}
will be extremely powerful in telling departures from $\Lambda$CDM
growth of fluctuations over significant portions of the evolution of
the universe.

\subsection{Features in the primordial power spectrum}
\label{sec:feat-prim-power}

\begin{figure}[b]
  \centering
  \includegraphics[width=0.8\linewidth, trim=0 20 0 0]{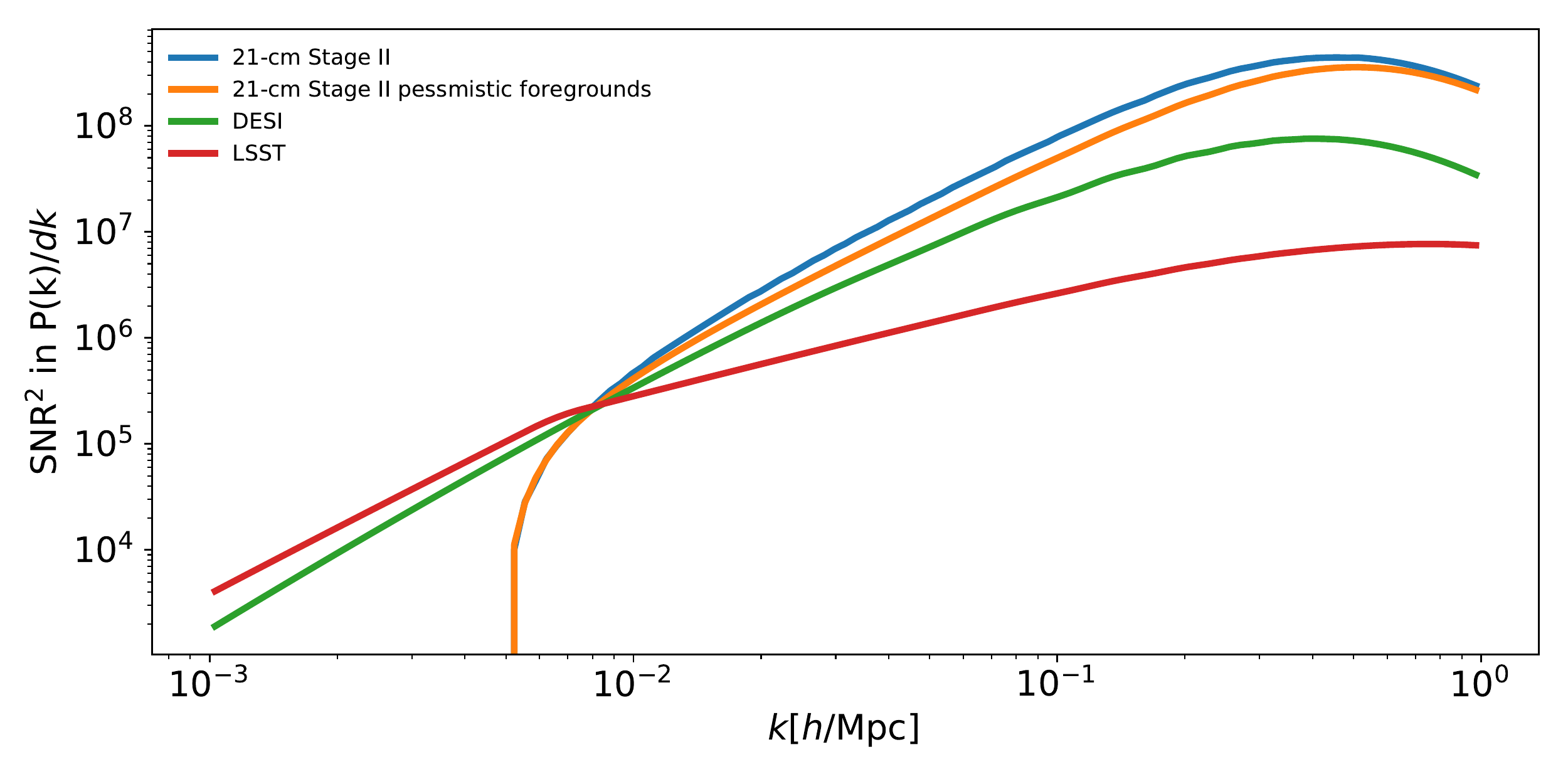} 
  \caption{Total signal-to-noise ratio (squared) in power spectrum  measurements as a function of wavenumber~$k$ for different experiments.  The knee in the results for~LSST~(red) corresponds to the scale at which LSST transitions from being a three-dimensional survey to a two-dimensional survey due to photometric redshift smearing. On the other hand, the turn-over for DESI~(green) at high~$k$ is driven by the observed number density of galaxies. Stage {\sc ii} $21\,$cm has the largest volume and the most accurate sampling, but it looses sensitivity at small wavenumbers due to foreground contamination. Note that we normalized the signal-to-noise ratio per linear step~$\Delta k$, which implies that changing this to logarithmic steps~$\Delta\log k$ (which might be a more natural basis) would further boost the relative usefulness of $21\,$cm by a factor of~$k$.}
  \label{fig:pksnr}
\end{figure}
The baryon acoustic oscillations are well-understood features in the matter power spectrum that are introduced during the evolution of the universe. In addition, there might be other oscillatory features of various origins in the power spectrum that we can search for with a Stage {\sc ii} experiment. In general, the matter power spectrum~$P_m$ at a wavenumber~$k$ and redshift~$z$ is in linear theory given by
\begin{equation}
  P_m(k,z) = T^2(k,z)\, P_\zeta(k)\, ,
\end{equation}
where~$T(k,z)$ is the transfer function and $P_\zeta(k)$ is the dimensionful primordial power spectrum. Assuming standard slow-roll inflation, the power spectrum of curvature perturbations~$P_\zeta(k)$ is well approximated by a power law ($A_\mathrm{s} k^{n_\mathrm{s}-4}$ with $n_\mathrm{s} \approx 0.96$~\cite{PL18, PCP18}). However, numerous mechanisms could have imprinted oscillations around this power law in the primordial spectrum (see e.g.~\cite{Chluba:2015bqa, Slosar:2019gvt} for recent reviews), while exotic physics in the dark sector can add additional features to the transfer function (see e.g.~\cite{Cyr-Racine:2013fsa}).

Detecting a deviation from a featureless power spectrum of primordial fluctuations would provide unique insights into the physics of the primordial universe. These features can provide evidence for particular inflationary scenarios, or identify the existence of new particles and forces during inflation or in the thermal plasma. In most cases, the feature amplitude is a free parameter, which could be unobservably small, and the precise characteristics of the feature can have a great impact on its detectability. While there exist two major classes of models (broadly defined as harmonic in~$k$ or~$\log k$), the details can still vary significantly, with possible runnings of the frequency~\cite{Flauger:2014ana}, locality of the feature~\cite{Achucarro:2013cva} and multiple features~\cite{Chen:2014joa, Chen:2014cwa} all possible within the vast landscape of models. Having said this, the cosmic microwave background~(CMB) puts stringent constraints on the amplitude of features over a wide range of their frequency, but no significant evidence has been found for such signals~\cite{Meerburg:2011gd, Peiris:2013opa, Meerburg:2013dla, Easther:2013kla, Fergusson:2014tza, Ade:2015lrj, Akrami:2018odb, PL18}. In addition, competitive limits have been derived in~\cite{Beutler:2019ojk} from current BOSS~data alone demonstrating the feasibility of this measurement in the (galaxy) power spectrum.

The $21\,$cm signal could significantly extend the search for features to much smaller scales, but also provide (significantly) improved constraints on scales already constrained by the~CMB~\cite{Chen:2016zuu, Xu:2016kwz} (cf.\ also~\cite{Beutler:2019ojk}). This is related to the fact that primordial features are in principle easier to find in the matter power spectrum since the intrinsic signal in the observables is suppressed in the~CMB. The reason for this is that the shape of the LSS~transfer function is smoother resulting in a larger intrinsic signal compared to the~CMB. We refer to~\cite{Beutler:2019ojk} for a detailed discussion.

Moreover, we show the total signal-to-noise ratio in the power spectrum measurement as a function of wavenumber in Figure~\ref{fig:pksnr}. This signal-to-noise can be thought of as the most model-independent proxy for comparing different surveys in their ability to constrain these models because of the broad prior model and parameter space. We see that Stage {\sc ii} $21\,$cm covers a very large $k$-range with exquisite signal to noise. In fact, due to the scaling of the accessible number of modes, it is always preferable to use smaller scales unless there is a theoretical prior to favor looking for such signals at large scales.

\begin{figure}
	\centering
	\includegraphics{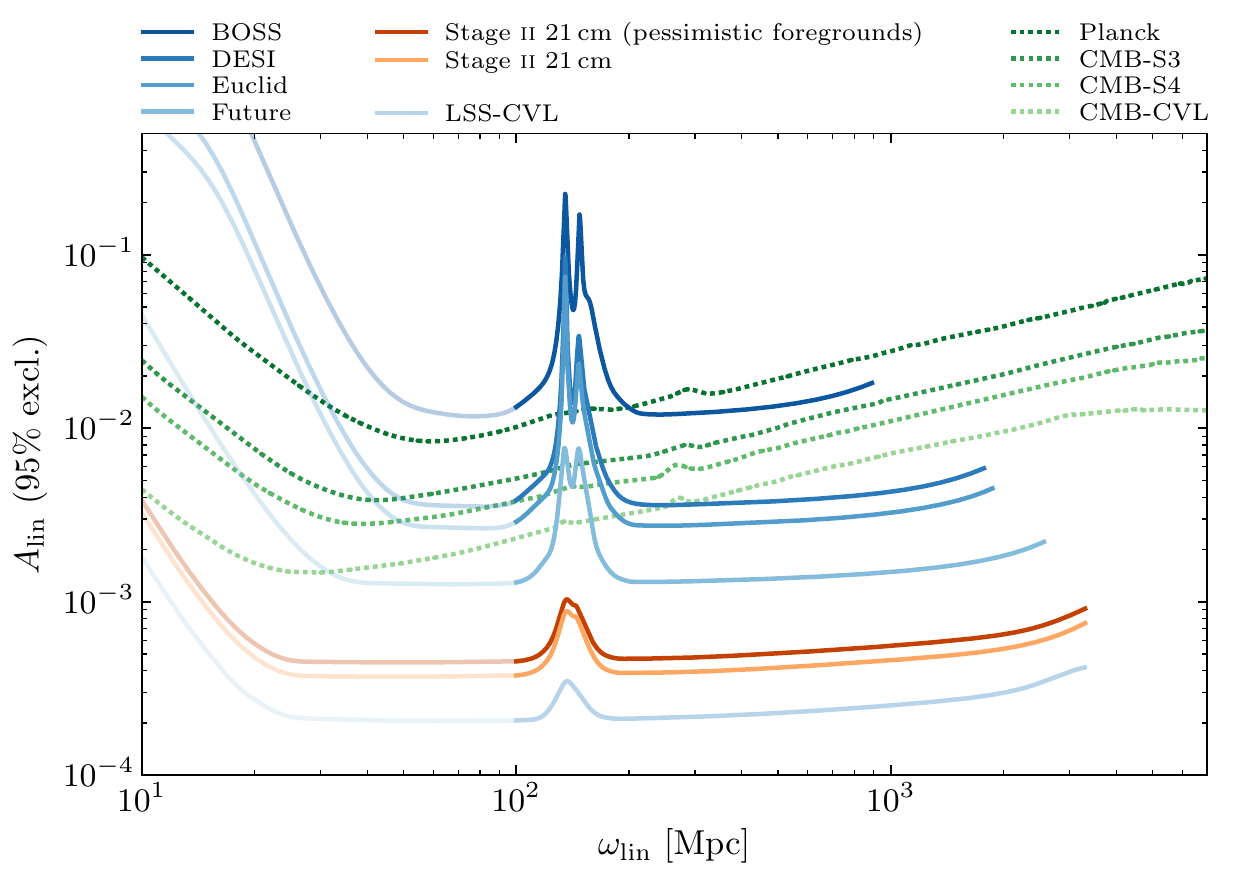}
	\caption{Forecasted upper limits at 95\% c.l.\ on the amplitude~$A_\mathrm{lin}$ of linear primordial feature oscillations as a function of their frequency~$\omega_\mathrm{lin}$ for different cosmological surveys (based on~\cite{Beutler:2019ojk}). We compare the sensitivity of current and future galaxy surveys~(blue) and CMB~experiments~(green) to the reach of a Stage~{\sc ii} $21\,$cm intensity mapping survey (orange). While the loss in constraining power at small frequencies $\omega_w$ is due to the degeneracy with the broadband power spectrum, the peak around $150\,\mathrm{Mpc}$ is due to the inference with the standard BAO~signal. Intensity mapping could significantly improve the constraints achievable in future galaxy surveys and even those of a cosmic variance-limited CMB~experiment (`CMB-CVL') over essentially all accessible frequencies.}
	\label{fig:features}
\end{figure}
In Figure~\ref{fig:features}, we provide forecasts for the sensitivity of a suite of cosmological surveys, including Stage {\sc ii} $21\,$cm, to primordial features. Since a wide range of feature models can be decomposed into a basis of oscillations which are linear in the wavenumber~$k$, we display the estimates of the 95\% upper limits for these linear feature models as a function of their frequency~$\omega_\mathrm{lin}$. To this end, we added an oscillatory feature with amplitude~$A_\mathrm{lin}$ to the primordial power spectrum of fluctuations,
\begin{equation}
  P_\zeta(k) = \frac{2\pi\,A_\mathrm{s}}{k^3} \left(\frac{k}{k_\star}\right)^{\!n_\mathrm{s}-1} \left[1 + A_\mathrm{lin} \sin(\omega_\mathrm{lin} k + \varphi_\mathrm{lin}) \right] ,
\end{equation}
with pivot scale~$k_\star$. We derived the future limits on these inflationary wiggles from the relative power spectrum following~\cite{Beutler:2019ojk}. In particular, we modeled the suppression of primordial power from nonlinear evolution in the Zeldovich approximation based on~\cite{2003ApJ...598..720S} with 50\% reconstruction efficiency. Moreover, we separately marginalized over the standard BAO~signal and six additional additive and multiplicative polynomial (`broadband') terms in redshift bins of width $\Delta z = 0.1$. We also take the effects of the survey window function into account. Due to the robustness of the signal given the analytic insights, we employed wavenumbers up to $k_\mathrm{max}= 0.75\,h\hskip1pt\mathrm{Mpc}^{-1}$. (We refer to~\cite{Baumann:2017gkg, Beutler:2019ojk} for a detailed description of these forecasts, the treatment of gravitational nonlinearities and further discussion of the estimated sensitivity of future surveys.) Figure~\ref{fig:features} shows that a Stage {\sc ii} $21\,$cm intensity mapping survey could significantly improve the limits on primordial feature models over currently planned galaxy surveys. In fact, it has the potential to be noticeably more sensitive than a future galaxy survey mapping about $10^8$~objects up to $z_\mathrm{max}=3$ and lead to constraints (or detections) within a factor of a few of a half-sky cosmic variance-limited experiment covering $z \leq 6$.

\begin{figure}
	\centering
	\includegraphics{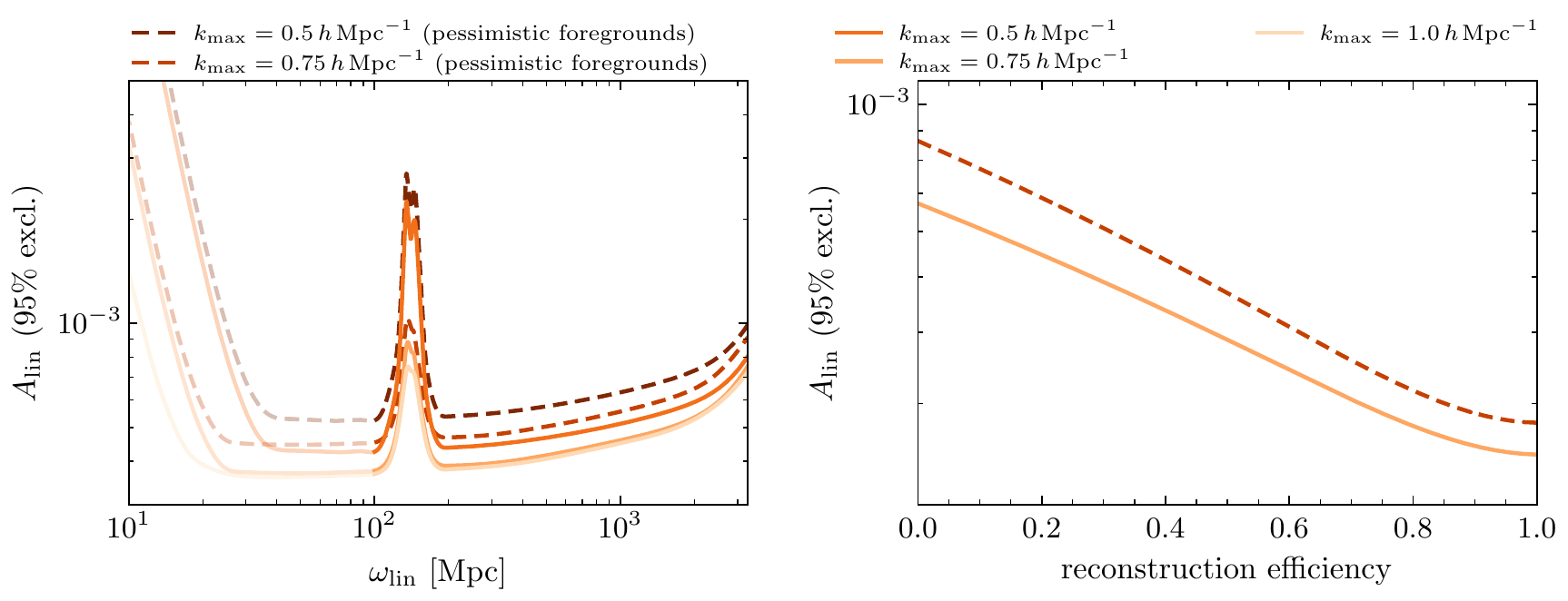}
	\caption{Impact of the maximum wavenumber~$k_\mathrm{max}$ and the reconstruction efficiency on the sensitivity to the amplitude~$A_\mathrm{lin}$ of primordial features. \textit{Left}:~The estimated 95\%~upper limit on~$A_\mathrm{lin}$ is shown as a function of feature frequency~$\omega_\mathrm{lin}$ for several values of~$k_\mathrm{max}$ and a reconstruction efficiency of~50\%. \textit{Right}:~Constraints on~$A_\mathrm{lin}$ as a function of reconstruction efficiency for $k_\mathrm{max} = 0.75\,h\hskip1pt\mathrm{Mpc}^{-1}$ and $\omega_\mathrm{lin}=200\,\mathrm{Mpc}$. Given the naturally more linear high-redshift universe observed by \stagetwo, this dependence is not very strong.}
	\label{fig:wiggs}
\end{figure}

The achievable constraints (or detection limits) on these oscillatory features depend not only on the foregrounds, but also our ability to undo the nonlinear `smearing' of structure (due to gravitational evolution in the late universe) by means of reconstruction. Our fiducial choice for the reconstruction efficiency is 50\%, which is almost certainly a conservative choice in a foreground-optimist scenario (cf.~\cite{Karacayli:2019iyd, Modi:2019hnu}. The mild sensitivity to this assumption is shown in the right panel of Figure~\ref{fig:wiggs}, while the left panel indicates the weak dependence on the maximum wavenumber~$k_\mathrm{max}$. We observe that we can expect measurements that are considerably better than $10^{-3}$ at 95\%~c.l.\ over the majority of the relevant parameter space.

\subsection{Primordial non-Gaussianity}
\label{sec:prim-non-gauss}

One of the exciting targets for future large-scale structure experiments is to obtain evidence for primordial non-Gaussianity (see e.g.~\cite{2010AdAst2010E..72C,Alvarez:2014vva,Meerburg:2019qqi} for reviews).
In the minimal model of slow-roll, single-field inflation, the primordial density field is perfectly
Gaussian. The detection of non-Gaussianity in the primordial field would therefore be immediately informative about the details of the inflationary process.

Observable deviations from Gaussian statistics in the density field are a direct measurement of the particle spectrum and interactions relevant to the inflationary sector.  As such, either a detection or an upper limit is testing particle physics at inflationary energy scales, which could be as high as $10^{14}\,$GeV. These energies are unlikely to be probed in collider experiments and thus are the unique domain of cosmological surveys. Furthermore, self-interactions of the inflaton that lead to non-Gaussian signatures are often tied to the fundamental mechanism for inflation itself.

These interactions often lead to a non-zero 3-point function of fluctuations in the primordial
curvature perturbation $\zeta_\vk$,
\begin{equation}
\left<\zeta_\vk \zeta_{\vk'} \zeta_{\vk''}\right> =  \delta^{(D)}(\vk+\vk'+\vk'')\ B(\vk, \vk', \vk''),
\end{equation}
where the Dirac delta-function is imposed by translational
invariance and $B$ is the bispectrum, which is a function of the
triangle configuration of the wavevector arguments. While a Gaussian field
has $B=0$, deviations from Gaussianity lead to non-zero bispectra, whose amplitude
is proportional to parameters traditionally denoted as $\fnl$, normalized so that
$\fnl\sim 10^5$ would correspond to $\mathcal{O}(1)$ non-Gaussianity~\cite{2010AdAst2010E..72C,Alvarez:2014vva}.\footnote{The numerical value
  of $10^5$ comes from the fact that primordial curvature fluctuations
  have a root mean square of $\sim 10^{-5}$.}

\begin{table}[b!t]
  \centering
  \renewcommand{\arraystretch}{1.4}
  \begin{tabular}{|c|c|c|}
  \hline
    & $\fnl^{\rm loc}\lesssim 1$ &  $\fnl^{\rm loc}\gtrsim 1$ \\
\hline
   $\fnl^{\rm eq,orth}\lesssim 1$ &  Single-field slow-roll & Multi-field \\
   $\fnl^{\rm eq,orth}\gtrsim 1$ &  Single-field non-slow-roll & Multi-field\\
   \hline
  \end{tabular}
\caption{Physical implications for qualitatively different measurements of the shapes of primordial non-Gaussianity (adapted from \cite{Alvarez:2014vva}). For $\fnl^{\rm loc}$, the bispectrum peaks where $k \ll k', k''$.  By contrast, the bispectrum for $\fnl^{\rm eq}$ and $\fnl^{\rm orth}$ peaks at $k\sim k' \sim k''$.
\label{tab:fnlped}
}
\end{table}

While the amplitude of $f_{\rm NL}$ reflects the strength of an interaction, the shape $B$ carries a wealth of additional information about the nature of inflation.  The local
bispectrum, parameterized by $\fnl^{\rm loc}$, is a shape for which the signal to noise is dominated by the limit of one of
the $k$-modes being soft (i.e.\ $k \ll k', k''$). This shape is of particular interest since it cannot arise in single-field inflation and would point directly to multiple light fields~\cite{Maldacena:2002vr,Creminelli:2004yq}. In contrast, equilateral and orthogonal shapes, with amplitudes $\fnl^{\rm eq}$ and $\fnl^{\rm orth}$, peak in configurations where $k\sim k' \sim k''$ and are typical of non-minimal interactions of the inflaton with itself~\cite{2009astro2010S.158K}. The target thresholds are $\fnl\simeq 1$ (see Table~\ref{tab:fnlped}). With sufficient signal to
noise, further information can be extracted either by considering
correlation functions beyond the bispectrum or by carefully exploring
the scale dependence of the bispectrum~\cite{Alvarez:2014vva}. In principle, it is possible to extract the spectrum of particles including their masses~\cite{Chen:2009zp,Baumann:2011nk,Noumi:2012vr} and spins~\cite{Arkani-Hamed:2015bza,Lee:2016vti}, which inspired the name {\it cosmological collider} physics~\cite{Arkani-Hamed:2015bza}.  

The best current constraints come from the CMB~\cite{2014A&A...571A..24P,2016A&A...594A..17P}
and indicate no statistically significant deviations from Gaussianity.
However, the error bars are too large to draw any meaningful conclusions about the
primordial dynamics. This motivates us to explore non-Gaussianity in
large-scale structure. While future constraints from the~CMB are limited by the number of 
available modes~\cite{Abazajian:2016yjj} (although large improvements can still be
achieved when considering bispectra involving tensors~\cite{2016PhRvD..93l3511M}),
we have access to a 3D~volume of modes with large-scale structure surveys. This is why 
it is expected that constraints from large-scale structure will
eventually become better than those derived from the CMB~\cite{Alvarez:2014vva}.

In this respect, the $21\,$cm signal has been identified as unique
because it is present throughout the Universe and could provide us
with an enormous volume (the entire sky between redshift 0 and
$z<150$). While non-Gaussianity from $21\,$cm has been studied in the
dark ages~\cite{2006PhRvL..97z1301C,Pillepich:2006fj,2006PhRvL..97z1301C,2017JCAP...03..050M}
and the epoch of reionization~\cite{2013PhRvD..88b3534L,2013PhRvD..88h1303M,2009MNRAS.394..133C},
our focus in this section will be on the low-redshift universe~\cite{2013MNRAS.431.2017T,2013MNRAS.433.2900D,2015MNRAS.448.1035C,Xu:2016kwz,2017PhRvD..96f3525L},
specifically constraints coming from the proposed Stage {\sc ii}
experiment in the redshift range $2\leq z \leq 6$.  We follow the
forecasting methodology of~\cite{2018arXiv180109280K}. In
particular, our forecasts include a comprehensive list of effects,
including the bias expansion for non-Gaussian initial conditions up to
second order, redshift space distortions, theoretical errors and
trispectrum contributions to the bispectrum. We have expanded the
codes used for galaxy forecasting to take into account instrumental
noise and propagating beam size effects into an effective noise in power spectrum measurements as
described in Appendix~\ref{app:forecasts-noise}. We have further
implemented various cuts to simulate the effect of a low-$\kpar$ cut and
the foreground wedge to simulate our foreground pessimistic
scenario. 

Our results are summarized in Table~\ref{tab:fnl}, where we
see that even with conservative assumptions a Stage {\sc ii} $21\,$cm
experiment could reach $\fnl^{\rm local}=1$ at $2\,\sigma$. This
target defines a typical level of non-linearity that is inherent in
many multi-field models~\cite{Alvarez:2014vva}. Such a
measurement would provide valuable insight into the degrees of
freedom that actively produce initial density fluctuations. For
equilateral or orthogonal non-Gaussianity, the reach is likely
somewhere between the optimistic and pessimistic foreground scenarios. Including
the wedge increases the noise, but might not be necessary for
constraints on these bispectra.  Any measurement of
$\fnl^{\rm eq,ortho} > 1$ would be incompatible with single-field slow-roll 
inflation~\cite{2003JCAP...10..003C,2015JCAP...01..016B,2016PhRvD..93b3523B}.
While our forecasts cannot exclude all such possibilities, they would
cut out a large fraction of the currently viable parameter space and
thus represent an opportunity for a major discovery. It is also
possible that Stage {\sc ii} data could be combined with data from a
mission such as SphereX\footnote{http://spherex.caltech.edu/}, leading
to further improvements.

\begin{table}
  \centering
  \renewcommand{\arraystretch}{1.3}
  \begin{tabular}{|c|C{3cm}|C{3cm}|C{3cm}|C{3cm}|}
    \hline
    \multirow{2}{*}{$\fnl$} & \multicolumn{2}{c|}{CMB error}
    & \multicolumn{2}{c|}{Stage {\sc ii} $21\,$cm error} \\
    &Planck (current) & CMB-S4 (forecast)& FG pessimistic & FG optimistic \\
    \hline
    Squeezed (local) & 5.0 & 2.0  & 0.7 & 0.2    \\
    Equilateral &43 & 21 & 27 & 4.5\\
    Orthogonal & 21 & 9.0 &  7.7 & 2.5\\
    \hline
  \end{tabular}
  \caption{$1\sigma$ constraints on various types of $\fnl$ parameters
    (see text) for Stage {\sc ii} $21\,$cm as compared with the
    results from the~CMB, both currently achieved from Planck and forecast
    for CMB-S4. We see that even with the foreground pessimistic case
    results are competitive with other experiments and would be a
    significant step towards a characterization of the inflationary
    mechanism (cf.\ Table~\ref{tab:fnlped}).  \label{tab:fnl}}
\end{table}

\subsection{Weak lensing and tidal reconstruction}
\label{sec:lensing}

Gravitational lensing affects any map we make of the universe, with 
the gravitational fields of large scale structure deflecting photons and therefore
``re-mapping" the angular coordinates we associate with a given location
on the sky. This re-mapping probes the Weyl potential and is thus directly
related to the projected distribution of mass between the observer and the
source of the photons being measured.  Therefore, a reconstruction
of this re-mapping, either in terms of a deflection field or a
decomposition into magnification and shearing effects, can help to
address many of the science goals stated earlier, such as constraining
the behavior of dark energy, measuring deviations from general relativity, or
determining the masses of light neutrinos which suppress the power spectrum on
small scales. A lensing map can further be cross-correlated
with other maps of structure, adding redshift resolution and contributing
additional constraining power by breaking degeneracies present in individual
maps.

Lensing of the CMB has been detected at high significance (e.g.~\cite{PLens18}), and will be one of the main science deliverables of upcoming CMB projects such as the Simons Observatory \cite{Simons} and CMB-S4~\cite{Abazajian:2016yjj}.
The joint effect of lensing on both CMB temperature and polarization allows for a robust detection in several channels, but since the CMB is effectively a single screen, it only offers access to a single projection of all matter
between the observer and the surface of last scattering.  Redshift information can be obtained through cross-correlation with other tracers, but this introduces additional populations have associated modeling uncertainties.
Lensing can also be measured from the correlations between observed galaxy shapes
in a large optical survey (see Ref.~\cite{Troxel:2017xyo} for the
current state of the art).  By binning the galaxies in redshift,
one can access multiple projections with different redshift weightings.
However, there are several pernicious systematics that must be dealt with, ranging from the impact of the telescope's point-spread function on inferred galaxy ellipticities, to control over the uncertainties in photometric redshifts, to the ``intrinsic alignments" of galaxies with their nearby environments (e.g.~\cite{Mandelbaum:2017jpr}).

In some sense, lensing of $21\,$cm fluctuations represents the ``best of
both worlds.''  $21\,$cm intensity maps have angular resolution and other
properties that place them in roughly the same regime as CMB maps, 
so there is promise that the well-developed estimators and pipelines
for reconstruction of CMB lensing can be adapted to $21\,$cm
observations. However, since $21\,$cm maps will be intrinsically
three-dimensional, they will also enable the same ``tomographic" lensing
studies as in galaxy lensing, but free of many of the galaxy-specific systematics mentioned above.
The promise of $21\,$cm lensing has long been recognized in the literature
(e.g.~\cite{Cooray:2003ar,2004NewA....9..417P,Zahn:2005ap,Metcalf:2006ji,Pourtsidou:2013hea}), and work from both the simulation~\cite{Romeo:2017zwt} and analytical~\cite{Foreman:2018gnv} sides continues.

Of course, that is not to say that $21\,$cm lensing analyses will not have
their own systematics to account for, and these are starting to be investigated.
For example, the quadratic lensing estimators that are standard in CMB analyses rely on the Gaussianity and translation-invariance of the intrinsic statistics of the CMB, whereas $21\,$cm maps will have more complicated statistics that will
affect any reconstruction of the lensing map.
Refs.~\cite{Lu:2007pk,Lu:2009je,Foreman:2018gnv} have shown that
these effects will be significant at the redshifts relevant
here. Ref.~\cite{Foreman:2018gnv} has also presented a technique to
mitigate a portion of this impact, which will reduce the additive bias
on the power spectrum of a reconstructed lensing map, but will
generally increase the noise on the same quantity. In cross-correlations between lensing and other
tracers, the additional bias will not be present, but the noise will
remain, and this must be taken into account when performing forecasts.

However, the bias on the lensing estimator caused by nonlinear
clustering is an interesting signal in its own right, being sensitive
to the power spectrum of the long density modes that gravity couples
to shorter modes within the $21\,$cm map. (Note that the long modes
referred to here are in the {\em same}\ redshift range as the map;
lensing also couples long density modes to short modes within the map,
but those long modes are at strictly {\em lower}\ redshifts than those
being directly observed.) These modes can be reconstructed in the same
way as for lensing, a process often referred to as ``tidal
reconstruction'' because it relies mainly on tidal
effects~\cite{Pen:2012ft,Schmidt:2013gwa,Zhu:2015zlh,Zhu:2016esh,Karacayli:2019iyd}. This
method can be used to reconstruct modes with low $\kpar$, which would be obscured by foregrounds if attempts were made to measure them directly. These modes can then be cross-correlated with the CMB to constrain possible integrated Sachs-Wolfe signatures of early dark energy or modified gravity, or cross-correlated with other measurements of lensing to probe structure growth or neutrino mass.

In Table~\ref{tab:lensing}, we present forecasts for the total signal
to noise on the various auto or cross power spectra related to lensing
and tidal reconstruction, applying the forecasting strategy of
Ref.~\cite{Foreman:2018gnv} to the fiducial $21\,$cm instrument
described in Sec.~\ref{sec:fiducial}. The displayed signal to noise is
combined over lensing reconstruction from 20 redshift bins spanning
$1<z<6$, while, for simplicity, we treat LSST galaxies and shear
(i.e.\ galaxy shape correlations) non-tomographically. We also show
equivalent values for CMB-S4 lensing, assuming a $1'$ beam, noise of
$2\,\mu$K-arcmin, and $f_{\rm sky}=0.4$. Even in the case of pessimistic
foregrounds, we
expect that cross-correlations of $21\,$cm lensing with LSST can be
measured at a precision approaching that of CMB-S4; recall that these
cross-correlations will include much more tomographic information than
CMB lensing.

For the $21\,$cm lensing auto spectrum, the
``bias-hardening'' method mentioned above leads to so much noise that
this measurement is not competitive with CMB-S4, even if the
foreground wedge can be completely cleaned. However, the power spectra
of long density modes in each redshift bin can likely be accessed with
very high precision, with a total signal to noise on transverse density modes alone of several hundred
regardless of the foreground treatment. 
In terms of scales, for optimistic assumptions about foregrounds, modes transverse to the line of sight (i.e.~with $k_\parallel=0$) can be reconstructed with signal to noise greater than unity for $k_\perp \lesssim 0.1h {\rm Mpc}^{-1}$ for most of the range between $z\sim 1.5$ and $6$. With more pessimistic foregrounds, the signal to noise degrades to less than unity above $z\sim 3$ on all scales, although at $z\lesssim 3$, signal-dominated reconstruction of modes with $k_\perp \lesssim 0.04h {\rm Mpc}^{-1}$ will still be possible.

\begin{table}
  \centering
  \renewcommand{\arraystretch}{1.4}
    \begin{tabular}{|c|c|c|c|}
    \hline
\multirow{2}{*}{Quantity / experiment} & \multirow{2}{*}{~CMB-S4~} & \stagetwo & \stagetwo \\
& & FG pessimistic & FG optimistic \\
\hline
Lensing $\times$ LSST galaxies & 367 & 208 & 400  \\
Lensing $\times$ LSST shear & 178 & 129 & 225  \\
Lensing auto & 353 & 5 & 75  \\
\hline
Tidal reconstruction auto & --- & 480 & 1281  \\
\hline
  \end{tabular}

\caption{\label{tab:lensing} 
Total signal to noise on measurements of auto or cross power spectra related to gravitational lensing of $21\,$cm maps. We expect cross-correlations of $21\,$cm lensing with LSST galaxy clustering or cosmic shear (galaxy lensing) to be measured at a precision approaching that of cross-correlations with CMB-S4 lensing, with the advantage that the former will contain much more (tomographic) information about the growth of low-redshift structure. The lensing auto spectrum will be more challenging, due to confounding effects from nonlinear clustering in the $21\,$cm maps~\cite{Foreman:2018gnv}. However, these same effects are sensitive to the power spectrum of long density modes at the source redshift, which can be ``tidally reconstructed" using similar estimators~\cite{Pen:2012ft,Schmidt:2013gwa,Zhu:2015zlh,Zhu:2016esh,Foreman:2018gnv,Karacayli:2019iyd}. These measurements can be made very precisely with our fiducial $21\,$cm instrument, even in the presence of foregrounds. The numbers shown above correspond to reconstructed modes transverse to the line of sight (i.e.~with $k_\parallel=0$), which would ordinarily be inaccessible due to foreground contamination.
}
\end{table}

Overall, the signal-to-noise in these measurements is impressive. Following these predictions all the way to their implications for cosmological parameters or specific models of new physics goes beyond the scope of this document,
because its main strength will come in particular through interaction
of cross-correlations which require assumptions about the existence of
other experiments. However, this is a very promising direction to pursue, 
and warrants further investigation.

\subsection{Forward model reconstruction}
\label{sec:forward-model}

One advantage of the $21\,$cm field on the scales relevant to an intensity mapping interferometer is that it is well described by a quadratic Lagrangian bias model \cite{Modi:2019hnu}.  This allows a forward-model approach to reconstructing modes missing from the observations.  Such a reconstruction maximizes the likelihood of a forward simulation to match the observations, under given modeling error and a data noise model, and in simulations it has residual errors lower than shot noise.   For redshifts $z=2$ and $4$, ref.~\cite{Modi:2019hnu} are able to reconstruct the $21\,$cm field in simulations with cross correlations $r_c > 0.8$ on all scales for both optimistic and pessimistic assumptions about foreground contamination and for different levels of thermal noise.

The ability to perform such reconstructions opens up several science opportunities which would otherwise be hindered by foegrounds.  First, it lessens the impact of the foreground wedge on BAO reconstruction, tightening the constraints on the cosmic distance ladder by a factor of nearly 2 \cite{Modi:2019hnu}.  Perhaps more interesting is the opportunities it opens up for correlating with other surveys.
The utility of large 21-cm intensity mapping arrays is the highest in the high redshift regime, where there is a lot of cosmic volume to be explored and which is the most difficult to be accessed using other methods. While there are some fields that will cross-correlate straight-forwardly with the $21\,$cm data, notably the Lyman-$\alpha$ forest and sparse galaxy and quasars samples which entail true three-dimensional correlations, any field that is projected in radial direction occupies the region of the Fourier space that is the most contaminated with the foregrounds. Reconstruction techniques allow us to re-enable these cross-correlations and thus significantly broadens the appeal of high-redshift $21\,$cm experimentation.  Two areas which deserve special mention are the cross-correlation of reconstructed $21\,$cm fluctuations with photometric surveys to measure $dN/dz$ of the photometric sample and cross-correlation with CMB lensing to constrain the growth rate of structure \cite{Modi:2019hnu}.

\begin{figure}
  \centering
  \includegraphics[width=\linewidth]{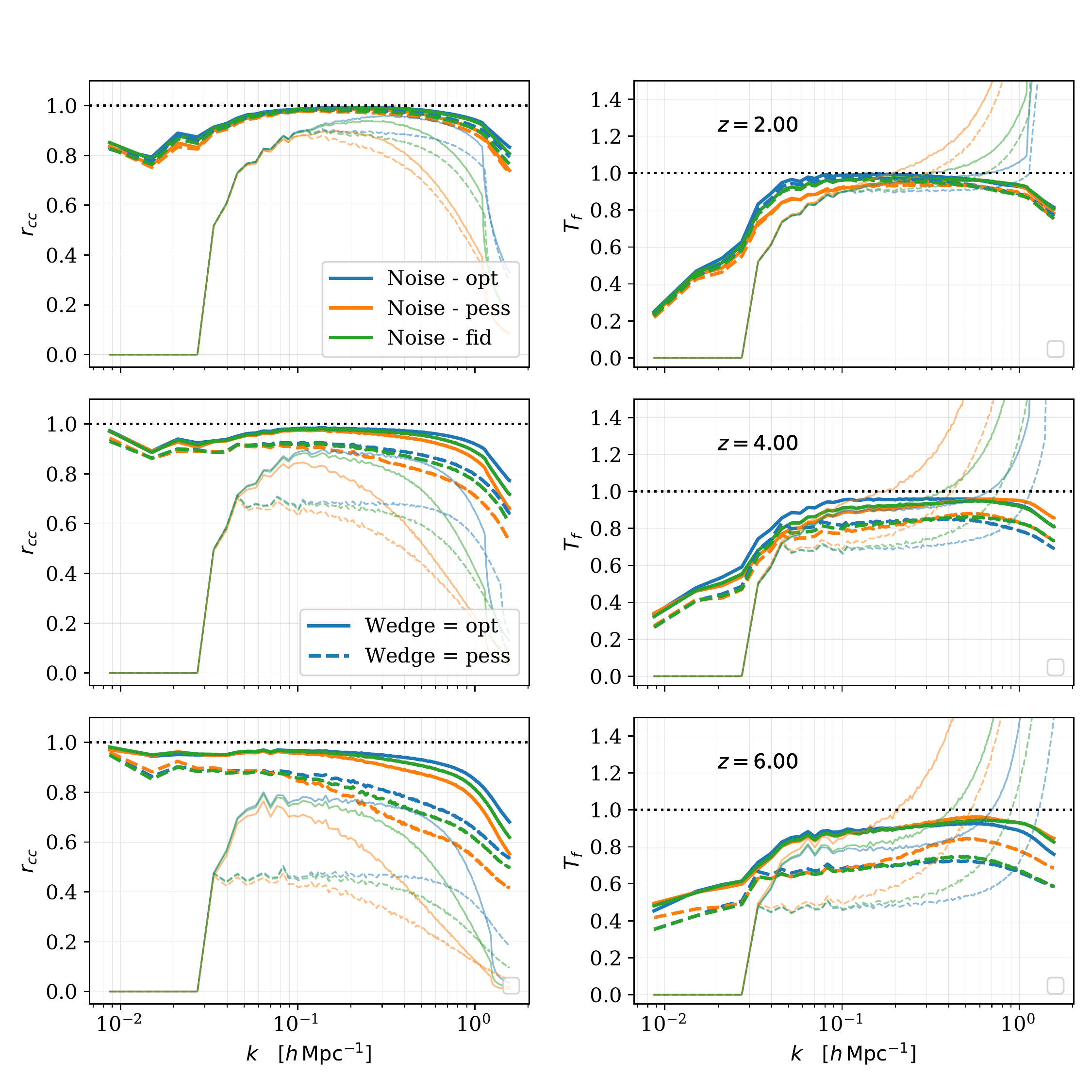}

  \caption{Forward model reconstruction of the HI field, from ref.~\cite{Modi:2019hnu}. Thick lines are for the reconstructed field and thin lines for the unreconstructed one. Solid and dashed line correspond to optimistic and pesimistic noise assumptions respectively. The three sets of panels correspond to different redshift: $z=2$ (top), $z=4$ (middle) and $z=6$ (bottom). The cross-correlation coefficient $r_{\rm cc}(k)=P_{FT}(k)/\sqrt{P_{FF}(k)P_{TT}(k)}$ for the input or reconstructed field $F$ and true field $T$, measuring the amount of total information recovered, is plotted in the left column. The transfer function $T_f=P_{FF}(k)/P_{TT}(k)$, giving the relative amplitude of modes in the reconstructed field, is plotted on the right. }
  \label{fig:recon}
\end{figure}

In Figure \ref{fig:recon} we show the performance of such reconstruction. The ability of the forward modeling to reconstruct the information that is completely lost to foreground is remarkable. The exceptional effectiveness of this reconstruction relies partly on the particular recipe that has been used to paint the neutral hydrogen onto dark-matter halos, but realistic complications are unlikely to drastically change these conclusions (see discussion in ref.~\cite{Modi:2019hnu}).

\subsection{Basic cosmological parameters: neutrino mass, radiation
  density, dark energy equations of state}

As a natural by-product of measuring the expansion history and
shape of the power spectrum, we can improve constraints on many
interesting cosmological parameters. While the expansion history is
directly sensitive to any of the parameters discussed below, improvements
in such measurements often additionally break degeneracies with other parameters
so that results in combination with standard datasets such as Planck often improve
considerably.
The shape of the power spectrum depends coarsely on the matter density
$\Omega_m$ and the epoch of the matter-radiation equality through
their impact on the transfer function $T(k)$. Additionally, distances in the universe
affect the conversion between the observed power spectrum (measured in
angles and redshifts) and the comoving power spectrum (measured in inverse
comoving distance units), an effect known as the Alcock-Paczynski test
\cite{1979Natur.281..358A}. In practice, redshift-space distortion
produce similar effects, which means that they must be modeled simultaneously.

In particular, a \stagetwo\ experiment would provide
valuable additional information on:

\textbf{Neutrino mass.} Cosmology is sensitive to the sum of neutrino mass eigenstates $m_\nu=\sum m_i$ (see e.g.~\cite{Dvorkin:2019jgs}). We know from neutrino oscillation experiments that $m_\nu\ge 0.06\,$eV in the normal hierarchy and $m_\nu\ge 0.1\,$eV in the inverted hierarchy \cite{Les13,Pat15,Arc17,LatGer17}.
Massive neutrinos affect the expansion history of the universe and they free-stream out of small scales density perturbations, making the field slightly smoother on scales smaller than the free-streaming length.
Their effect can be detected through a particular scale-dependence of the power spectrum between large and small scales, although this usually takes the form of comparing the fluctuation power measured by the CMB with that measured at low redshifts.
The general expectation is that the neutrino masses will be detected in the coming years using a number of related methods. The combination of CMB lensing with BAO, the broad-band power measurements in galaxy surveys and weak gravitational lensing of galaxies all have sufficient sensitivity. We expect a Stage {\sc ii} $21\,$cm experiment could improve the signal to noise of all these measurements.

\textbf{Energy density of radiation.} The amount of radiation in the early universe is usually parameterized by the effective number\footnote{This nomenclature can be misleading because any component with an equation of state like radiation ($w=1/3$) that is coupled gravitationally will contribute to this
quantity.} of massless neutrinos $\neff$ (cf.\ e.g.~\cite{Green:2019glg}).  Measuring $\neff$ is an important discovery channel for new physics, since any light particle that was in thermal equilibrium with the Standard Model will contribute an additional $\Delta \neff \geq 0.027$ unless its contribution is diluted by other decays.
At the high temperatures thought to be present in the early universe, even very weak interactions are sufficient for thermalization.  As a result, percent-level measurements of $\neff$ can be an extremely sensitive and broad probe of new physics (see e.g.~\cite{Baumann:2016wac,Abazajian:2016yjj}).
Currently, the best measurements arise from a combination of CMB and BAO data, but future $21\,$cm measurements of the matter power spectrum could help push the CMB measurement to $\Delta\neff=0.027$ at more than $1\sigma$~\cite{Baumann:2017gkg}.

\textbf{Dark energy equation of state.} While we stress that the main
strength of the Stage {\sc ii} experiment lies in directly measuring the
properties of dark energy at high redshifts, it is also capable of
determining low-redshift dark energy properties since these change the
expansion-history and hence mapping between angles and scales to
redshift $z\sim 2$. As an example, we measure the standard dark energy
equation of state parameter $w$, but any model with dynamical dark
energy at low-redshift will benefit from these observations of the
universe in the pre-acceleration era (cf.\ e.g.~\cite{Slosar:2019flp}).

These effects are studied though general Fisher matrix formalism,
following the methodology of \cite{2017arXiv170907893O}.
In Table~\ref{tab:params}, we summarize these forecasts alone and in
combination with some standard cosmological probes that will be
available towards the end of the next decade. These parameters are the
focus of the most important DOE-sponsored upcoming surveys and as
such warrant further examination.

\begin{table}
  \centering
  \renewcommand{\arraystretch}{1.4}
  \begin{tabular}{|c|p{2.1cm}|p{1.9cm}|p{2.2cm}|p{2.3cm}|p{2.5cm}|p{2.3cm}|}
  \hline
\multirow{2}{*}{Parameter / combination}  & LSST + DESI + Planck & \multirow{2}{*}{CMB S4} & \multirow{2}{*}{\stagetwo + Planck} & LSST~+~DESI + \stagetwo~+~Planck & \multirow{2}{*}{CMB-S4 + \stagetwo} & \multirow{2}{*}{Everything bagel} \\
\hline
$\sum m_\nu$ [meV] & 38 & 59 & 31~/~27 & 25~/~22 & 24~/~21 & 15~/~14 \\
$\sum m_\nu$ + 3\% $\tau$ prior [meV] & --- & 15 & --- & --- & 14~/~13 & 10.4~/~10.2 \\
$\sum m_\nu$ [meV] (free $w$) & 50 & --- & 33~/~29 & 26~/~23 & --- & --- \\
\hline
$N_{\rm eff}$ & 0.050 & 0.026 & 0.043~/~0.037 & 0.033~/~0.030 & 0.014~/~0.013 & 0.012~/~0.011 \\
\hline 
$w$ (free $\sum m_\nu$) & 0.017 & --- & 0.006~/~0.005 & 0.005~/~0.004 & --- & --- \\
\hline
\end{tabular}

\caption{\label{tab:params} Combination of parameter forecasts for a compendium of future DOE experiments.
  All combinations include a Planck 2015 CMB prior to promote stability of the Fisher matrix, and are for a $\Lambda$CDM
  cosmology unless stated otherwise. For combinations
involving $21\,$cm we state both the pessimistic and
optimistic foreground removal cases respectively, separated by a slash.}
\end{table}

These constraints were derived assuming
$k_{\rm max}=0.4\,h\,{\rm Mpc^{-1}}$ for 21cm and
$k_{\rm max}=0.2\,h\,{\rm Mpc^{-1}}$ for DESI (LRGs+ELGs only), and
for simplicity and fair comparison no BAO damping in both cases. Note
that the higher $k_{\rm max}$ for $21\,$cm is justified given its higher
redshift and less complex biasing arising from probing less massive halos.
Since the non-linear damping of BAO increases with time, our neglect of BAO
damping for all surveys overestimates the power of lower-redshift probes.
The LSST Fisher matrices were based on the updated work of \cite{2014JCAP...05..023F},
while CMB-S4 Fisher matrices were provided through private communication
\cite{CMBS4FM}. Following \cite{2017arXiv170907893O} we used 5\%
priors on both $b_{\rm HI}$ and $\Omega_{\rm HI}$.

\subsection{Cross-correlation studies}
\label{sec:cross-corr-stud}

In the next decade we will see many different probes measure the same
volume of space using different tracers and different techniques and
cosmology should enter a golden era of cross-correlations. In
general cross-correlations are extremely useful for three reasons:
(1) any contaminating signal that is not present in both probes
will not affect the signal; (2) the value of cross-correlations
grows as the number of pairs, i.e.\ with the square of the number of probes,
while the total signal-to-noise in auto-correlations grows only linearly;
(3) cross-correlations allow the possibility of sample-variance-free
measurements of some quantities, and more generally allow breaking of
degeneracies.

In our fiducial experiment, we have assumed a wide redshift range,
covering both low redshift $0.3<z<2$ and high redshift $2<z<6$. These
offer different cross-correlation opportunities. At low-redshift the
universe will be well sampled by other spectroscopic surveys, allowing
us to directly cross-correlate with:

\begin{itemize}
\item \textbf {Spectroscopic galaxy samples}. For example, \stagetwo\ 
has redshift overlap with the DESI BGS (Bright Galaxy Survey) sample at
low redshift and with the DESI ELG (Emission Line Galaxy) and
LRG (Luminous Red Galaxy) samples in the intermediate redshift range.
These cross-correlations will measure the relative bias
factors of the samples which would have implications for our
understanding of galaxy evolution at these redshfits. Moreover, the
large number densities of BGS might allow efficient sample-variance
cancellation and perhaps provide a new avenue for constraining
non-Gaussianity \cite {2009PhRvL.102b1302S} or growth
\cite{2009JCAP...10..007M}.
This is in addition to kSZ cross-correlations discussed
in Section \ref{sec:kinet-suny-zeld}. 
\end{itemize}

\begin{figure}
  \centering
  \includegraphics[width=0.7\linewidth]{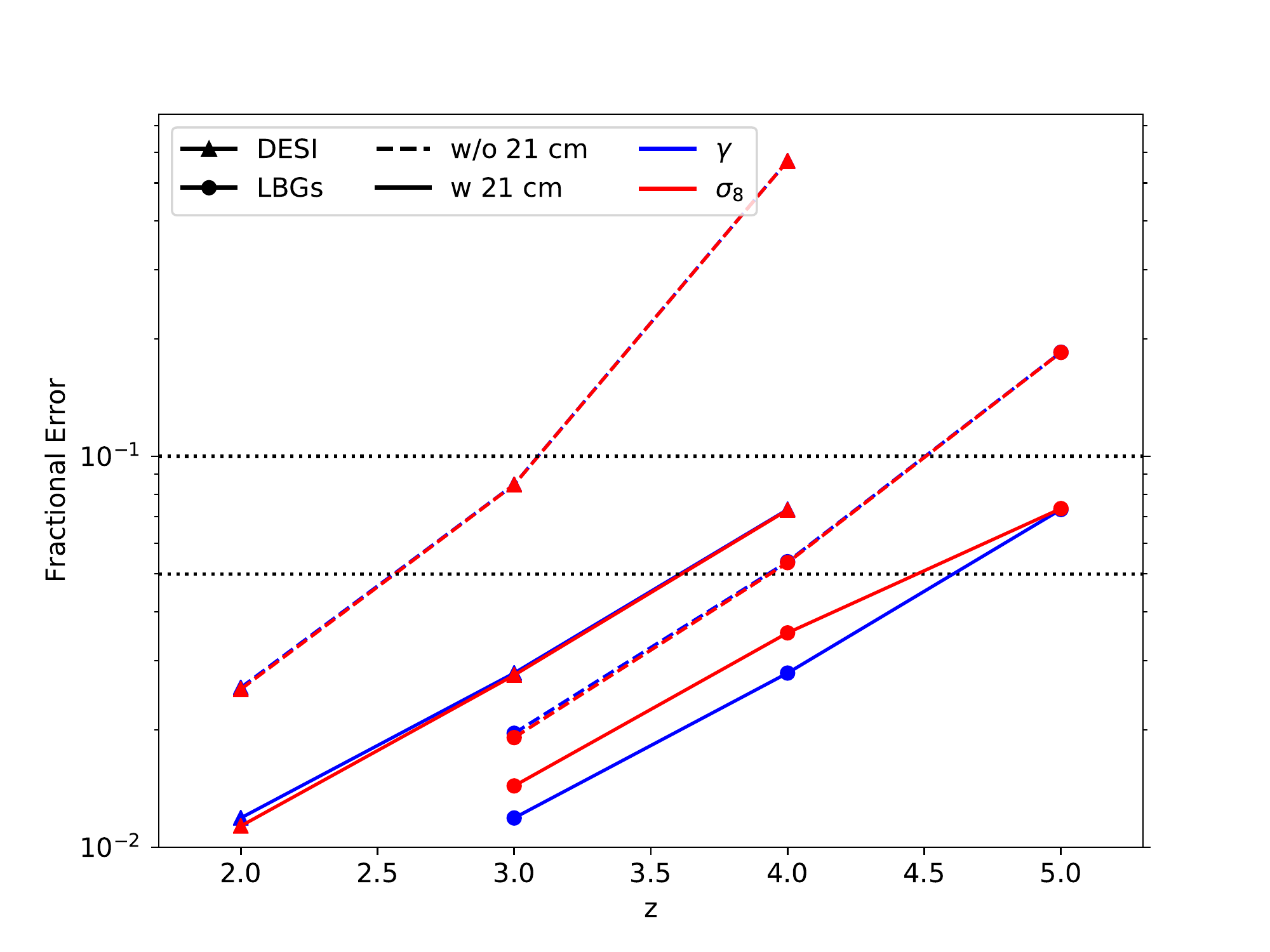}

  \caption{Forecasted constraints on the gravitational slip parameter, $\gamma$, and the power spectrum
normalization, $\sigma_8$ , using a combination of RSD and CMB-lensing $\times$LSST cross-correlations assuming
$f (z)$ is fixed. RSD data assume DESI QSOs, a LBG survey at $m_{\rm UV}^{\rm th}= 24.5$ over 1000 square degrees, or
either cross correlated with \stagetwo\ data. Adding 21-cm data to the RSD side significantly improves
constraints on $\gamma$, yielding sub-4 percent constraints at z = 2, 3 and 4 from DESI QSOs and our
hypothetical LBG survey at z = 3 and 4, respectively. At z = 5 adding 21-cm data improves
constraints by more than a factor of two, to below ten percent. Adapted from \cite{Stephen18}}.
  \label{fig:desiqso}
\end{figure}

On the higher redshift end ($z>2$) our fiducial experiment has been designed
to probe volumes not well sampled by other tracers of large scale
structure. Nevertheless, there will be avenues for
direct cross-correlation, in particular with:

\begin{itemize}
\item \textbf{High-redshift quasars}. QSOs have been measured in
  large numbers by BOSS/eBOSS, but the size of the dataset will gain
  another considerable boost with DESI\@. This information will give
  extra BAO and RSD signal and help calibrate both $21\,$cm and quasar
  bias parameters (in conjunction with auto-correlation measurements).
  
  As an example of the science return enabled by the presence of 21 cm
  data, Figure \ref{fig:desiqso} shows the forecasted 1-$\sigma$ error
  on $\sigma_8(z)$ and the gravitational slip parameter $\gamma$ from
  a combination of spectroscopic data, CMB weak lensing and 21\,cm
  intensity mapping data.  While this method is not competitive with
  methods discussed in Section \ref{sec:rsd}, it is arguably less
  dependent on theoretical assumptions.

\item \textbf{Lyman-$\alpha$ forest}. The Lyman-$\alpha$ forest will have been
  probed by BOSS, eBOSS and DESI\@. This cross-correlation will go down
  to very small scales in the radial direction. Since both probes
  measure the neutral hydrogen this cross-correlation will help both
  probes achieve their full potential \cite{2017JCAP...04..001C}. In
  particular, it will help with measuring the contamination of the
  Lyman-$\alpha$ forest by damped Lyman-$\alpha$ (DLA) and high column
  density (HCD) systems and thus enhance the potential of the
  Lyman-$\alpha$ forest as a probe of a small-scale physics.

\item \textbf{High-redshift forests of other metals}. In addition to
  Lyman-$\alpha$, the high redshift universe also contains other
  metal forests, like the Si{\sc iii}, Si{\sc iv} and C{\sc iv} forests,
  whose physics and bias parameters can again be constrained by cross-correlation
  with $21\,$cm.

\item \textbf{Lyman-$\alpha$ emitters} will be detected in large numbers in
  surveys like HETDEX \cite{2008ASPC..399..115H}. Cross-correlations
  with $21\,$cm will allow determination of their physical parameters as
  well as constrain interlopers.
\end{itemize}

However, more indirect cross-correlations are also possible. The
reconstruction method discussed in Section \ref{sec:forward-model}
will enable cross-correlations with sparse galaxy samples from
photometric redshifts and reconstruction of CMB lensing. The lensing
field reconstruction discussed in Section~\ref{sec:lensing} will
enable cross-correlation with both lower redshift galaxy samples and
higher-redshift lensing screens (from CMB or EoR intensity mapping
\cite{2004NewA....9..417P}).

In short, it is clear from this discussion that the number of
cross-correlations grows quadratically with the number of tracers and
hence with the increased number of samples we are entering the era of
cross-correlations.

\subsection{Kinetic Sunyaev Zel'dovich Tomography with \stagetwo\ \texorpdfstring{$21\,$cm}{21cm} and CMB-S4}
\label{sec:kinet-suny-zeld}

Future CMB experiments such as CMB-S4 will have the resolution and sensitivity to produce highly significant measurements of the kinetic Sunyaev Zel'dovich (kSZ) effect, temperature anisotropies induced by the scattering of CMB photons from free electrons with a non-zero CMB dipole in their rest frame. The kSZ temperature anisotropies are correlated with the high-resolution maps of large scale structure produced by a Stage II 21cm intensity mapping survey. Using the technique of Sunyaev Zel'dovich (SZ) tomography, this correlation can be exploited to obtain a tomographic reconstruction of the remote CMB dipole field~\cite{Terrana:2016xvc,Deutsch:2017ybc,Cayuso:2018lhv,Smith:2018bpn}: the CMB dipole observed at different locations in the Universe, projected along the line of sight.~\footnote{The primary contribution to the dipole field at any spacetime location is the radial component of the peculiar velocity, however the Sachs-Wolfe and Integrated Sachs-Wolfe contributions are important when considering large-angle or large-redshift correlations. These contributions can be relevant for large volume surveys, as contemplated here.} Previous forecasts for kSZ tomography considered galaxy surveys as a tracer of structure, and demonstrated that future surveys such as LSST, in correlation with CMB-S4, can be used to obtain high fidelity tomographic reconstructions of the remote dipole field out to $z \sim 3$. The reconstructed dipole field can be used to place competitive constraints on primordial non-Gaussianity~\cite{Munchmeyer:2018eey,Contreras:2019bxy}, modifications of GR~\cite{Pan:2019dax,Contreras:2019bxy}, the growth function~\cite{Madhavacheril:2019buy}, and various early-Universe scenarios~\cite{Cayuso:2019hen,Zhang:2015uta}. It can be shown \cite{Padmanabhan2019} that using a large redshift coverage has excellent prospects to alleviate astrophysical systematic uncertainties in the resultant forecasts, by combining independent information from multiple tomographic bins.  An important property of kSZ tomography is that the measurement of large-scale modes using the reconstructed dipole field can be of higher fidelity than measuring these modes directly from the tracer itself~\cite{Smith:2018bpn}. This feature is particularly relevant in this context, since kSZ tomography probes radial modes which are poorly constrained in 21cm experiments due to foreground contamination.

\begin{figure}
  \centering
  \includegraphics[width=0.45\linewidth]{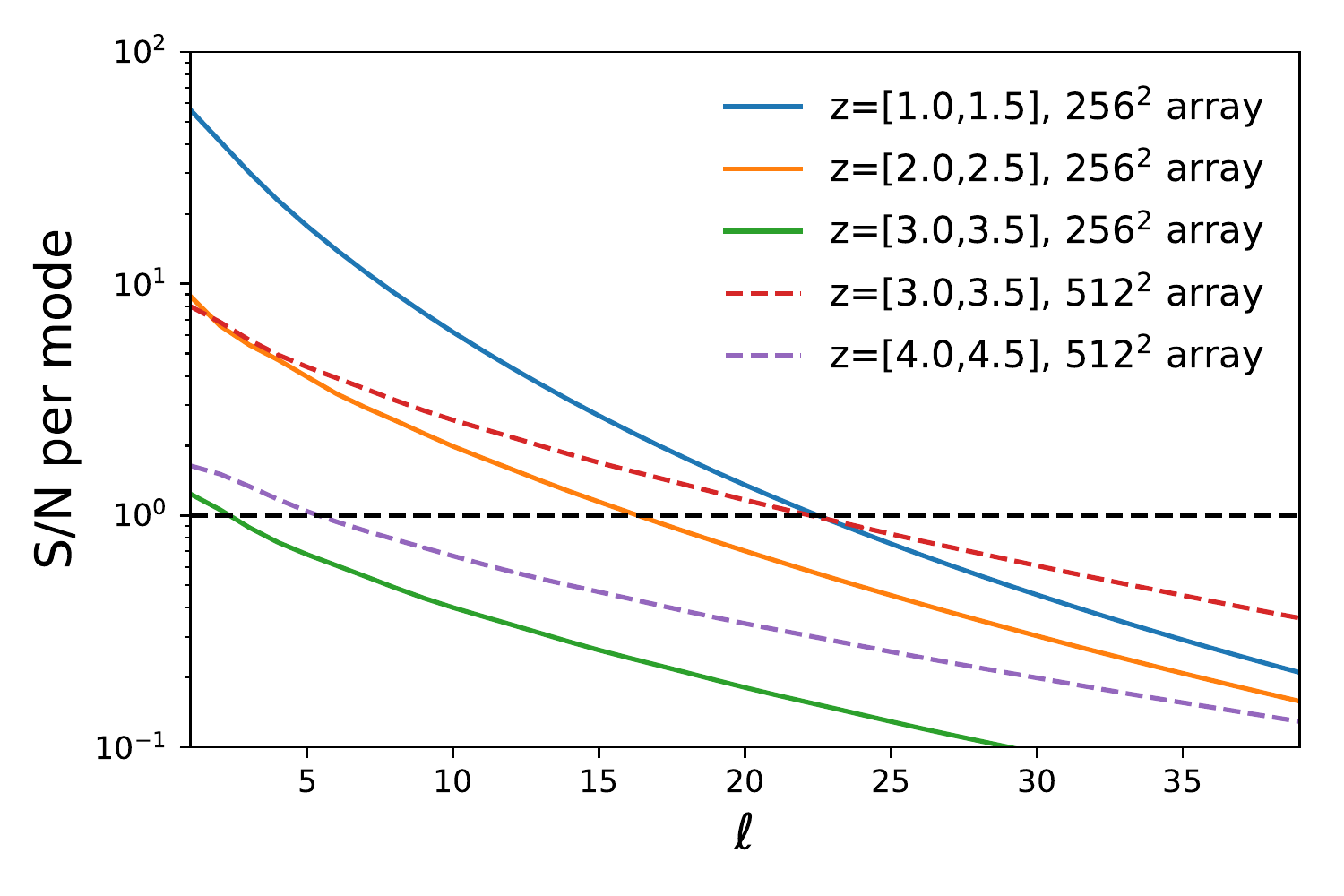}

  \caption{The forecasted signal-to-noise per mode of the reconstructed dipole field for some example redshift bins using a Stage II 21cm survey and CMB-S4. We consider both the default Stage II 21cm configuration (solid lines), as well as a configuration where the size of the array is doubled (dashed lines). Based on this result, we conclude that a Stage II 21cm survey will be a useful tool for kSZ tomography, with the reconstruction improving with increased resolution of the intensity maps.}
  \label{fig:ksz_tomography}
\end{figure}

In Fig.~\ref{fig:ksz_tomography}, we show the expected signal-to-noise per mode of the reconstructed dipole field using a Stage II 21cm survey and CMB-S4, for different redshift bins. We present two scenarios: the default specifications of the Stage II survey (solid lines), and a configuration where we doubled the size of the array (dashed lines). To produce this forecast, we have used the quadratic estimator of Ref.~\cite{Deutsch:2017ybc}, incorporating the halo model of Ref.~\cite{Obuljen:2018kdy} for the distribution of HI and the model of Ref.~\cite{Smith:2018bpn} for the distribution of free electrons. We assume the noise, beam, and sky coverage for CMB-S4 is $1 \mu K \ {\rm arcmin}$ / $1.5 \ {\rm arcmin}$ / $60 \%$, consistent with the current reference design~\cite{Abazajian:2019eic}; we have neglected foregrounds and systematics in making our forecast. 

The signal-to-noise per mode using the default configuration of Stage II 21cm and CMB-S4 is appreciable over a range of angular scales out to $z \sim 3$. The dipole field reconstruction can be improved by increasing the angular resolution of the intensity maps. In particular, doubling the size of the array (dashed lines) dramatically improves the reconstruction at $z>3$. This would allow one to take advantage of the very large volumes probed by a 21 cm survey, as compared to what is possible even with ambitious galaxy redshift surveys. We conclude that a Stage II 21cm survey can be useful for kSZ tomography. Cosmological constraints derived from the reconstructed dipole field would contribute to many of the science goals of a Stage II 21cm survey. 

\subsection{Direct measurement of cosmic expansion}

The measurement of the Universe's expansion in real time would be a
unique confirmation of the standard cosmological model. Cosmological
sources drift in redshift with the characterizing time-scale of a
Hubble time. Over a 10 year time-span, this results in the redshift
change of around $\delta z=10^{-9}$. This is challenging both
statistically and systematically.  However, if measured, it would be
one of the very few dimensional quantities that one can measure
directly in cosmology\footnote{The other prominent examples include
  time-delays in gravitational lenses that allow us to measure the
  Hubble rate and the temperature of the CMB.}. Controlling absolute
redshift calibration at the required level over a decade is extremely
difficult, but possible in optical \cite{2015APh....62..195K}. In
radio, however, it should be considerably easier, since clock
generators with sufficient accuracy are available off-the-shelf. Since
in radio systems the clock-generator sets the absolute time-scale and
thus frequency calibration, this (typically dominant) part of the systematic error budget 
is absent. There are additional subtleties to do with accurate clock
transport, or subtle changes in the beam due to changes in the physical
state of the reflecting material over 10 years, but while these can
produce anomalous changes in the measured signal, they are unlikely to
produce systematic shifts.

\begin{figure}
  \centering
  \includegraphics[width=0.7\linewidth]{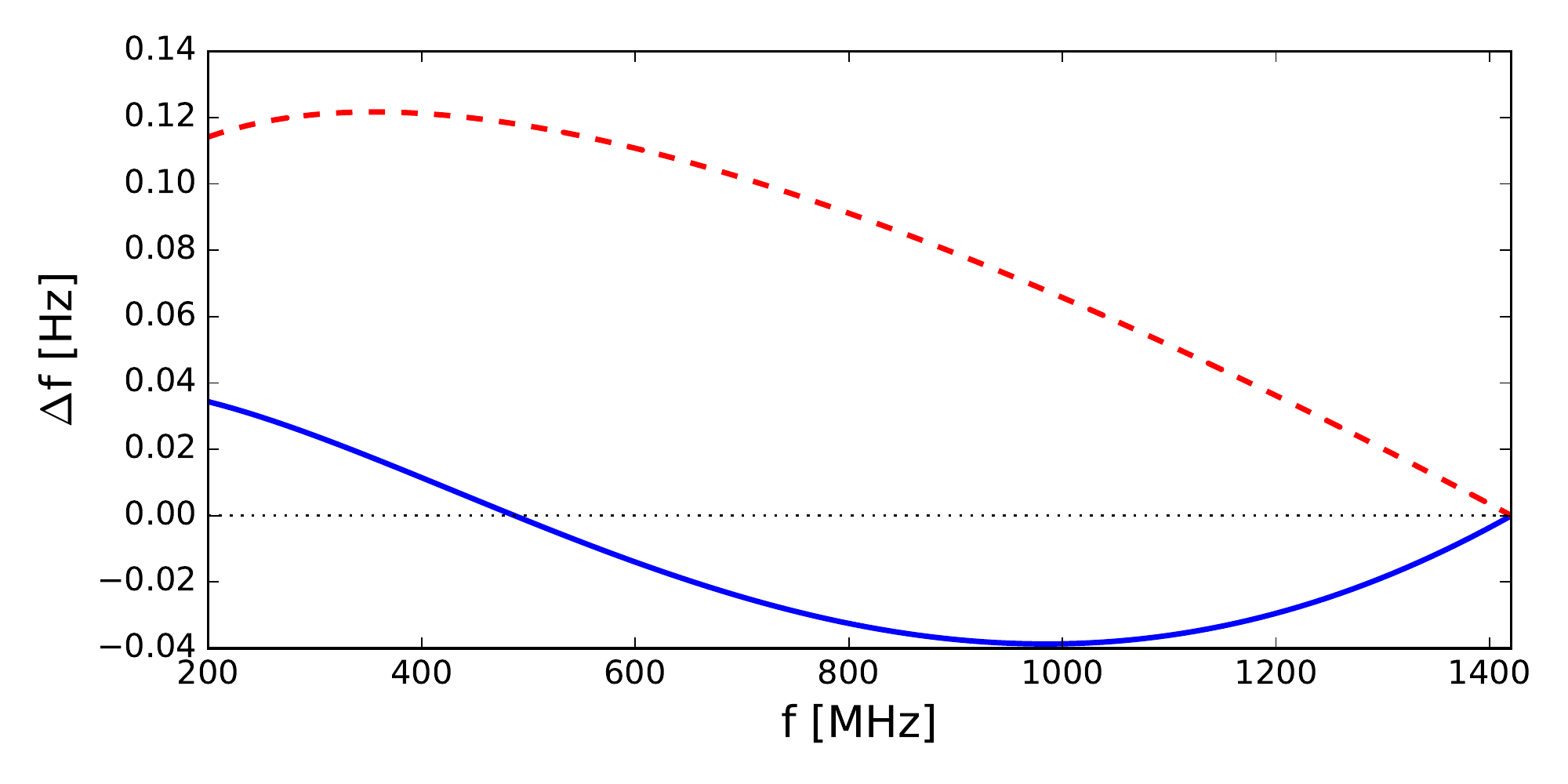}

  \caption{Predicted drift in frequency as a function of frequency for
  a standard cosmological $\Lambda$CDM model (blue, solid) and flat
  matter dominated model ($\Lambda=0$, red, dashed) over 5 years.
  We see that the required frequency precision is $\mathcal{O}(10^{-2})\,$Hz
  in order to distinguish these two scenarios.}
  \label{fig:drift}
\end{figure}

The basic formula for the redshift drift is given by
\begin{equation}
  \frac{dz}{dt} = (1+z) H(0) - H(z),
\end{equation}
where $H(z)$ is the Hubble parameter at redshift $z$.
In Figure~\ref{fig:drift}, we show a typical prediction for a total
drift as a function of frequency for a 5-year experiment. We see that,
in principle, the required accuracy is of the order of $10^{-2}$ Hz. If
there existed lines whose natural width would be this small, this
would have been a trivial measurement. In practice, however, the $21\,$cm
line is velocity smeared to a few 100 km/s giving the natural
smoothness of the cosmic signal of around $10^5$ Hz. Thus, one really
needs to rely on very precise measurements of the overall structure.
On the upside, we see that there is a very definite structure to the
shape of this function, so tracing the shift as a function of redshift
gives another leverage on systematic control.

There are two basic approaches to this measurement. The first is to 
rely on the apparent radial motion of the entire field of density
fluctuations. It can be shown that sensitivity of this method is given
by
\begin{equation}
  \sigma(\dot{z}) = \frac{1}{H(z)} \left( \frac{Vt^{3}}{48\pi^3} \int
    k_r^2 \frac{P_S(\vk)}{dP_N/d(t^{-1})(\vk)}d^3k \right)^{-1/2},
\end{equation}
where $V$ is the volume of the survey and $P_S$ and $P_N$ are the
signal power and noise power (per inverse year of integration) in
comoving space respectively and $k_r$ is the radial wave-vector. This
expression is correct even when field is non-Gaussian. We see that the
majority of the signal is coming from the fine, high frequency radial
modes that are suppressed by velocity dispersions in realistic
cosmologies. In our numerical work we have found that this technique
is not statistically promising for our straw-man configuration, but
that it could be for a low redshift $z<1$ array.

An alternative approach is to consider a finite number of cold systems
observed in absorption when backlit by distant sources. This has been
considered in~\cite{2014PhRvL.113d1303Y}, which finds a possibility of
a $5\sigma$ detection over 10 years. The forecasting is highly
uncertain due to poorly known redshift distribution of radio sources,
which is even more poorly known in our redshift range and therefore we
do not attempt it in this paper, but a simple extrapolation based on
CHIME numbers shows that this measurement would most likely be
possible with our \stagetwo\ proposed experiment.  However, both methods
would require saving data at a radial resolution that is beyond what
is necessary for the standard cosmological analysis, and might increase
the overall data-storage requirements by a factor of a few.

Finally, we note that in both cases, the scaling is $t^{3/2}$. This
very unusually favorable scaling comes from the fact that signal
increases linearly with time while noise falls as $1/\sqrt{t}$.

\begin{figure}
  \centering
  \includegraphics[width=0.7\linewidth]{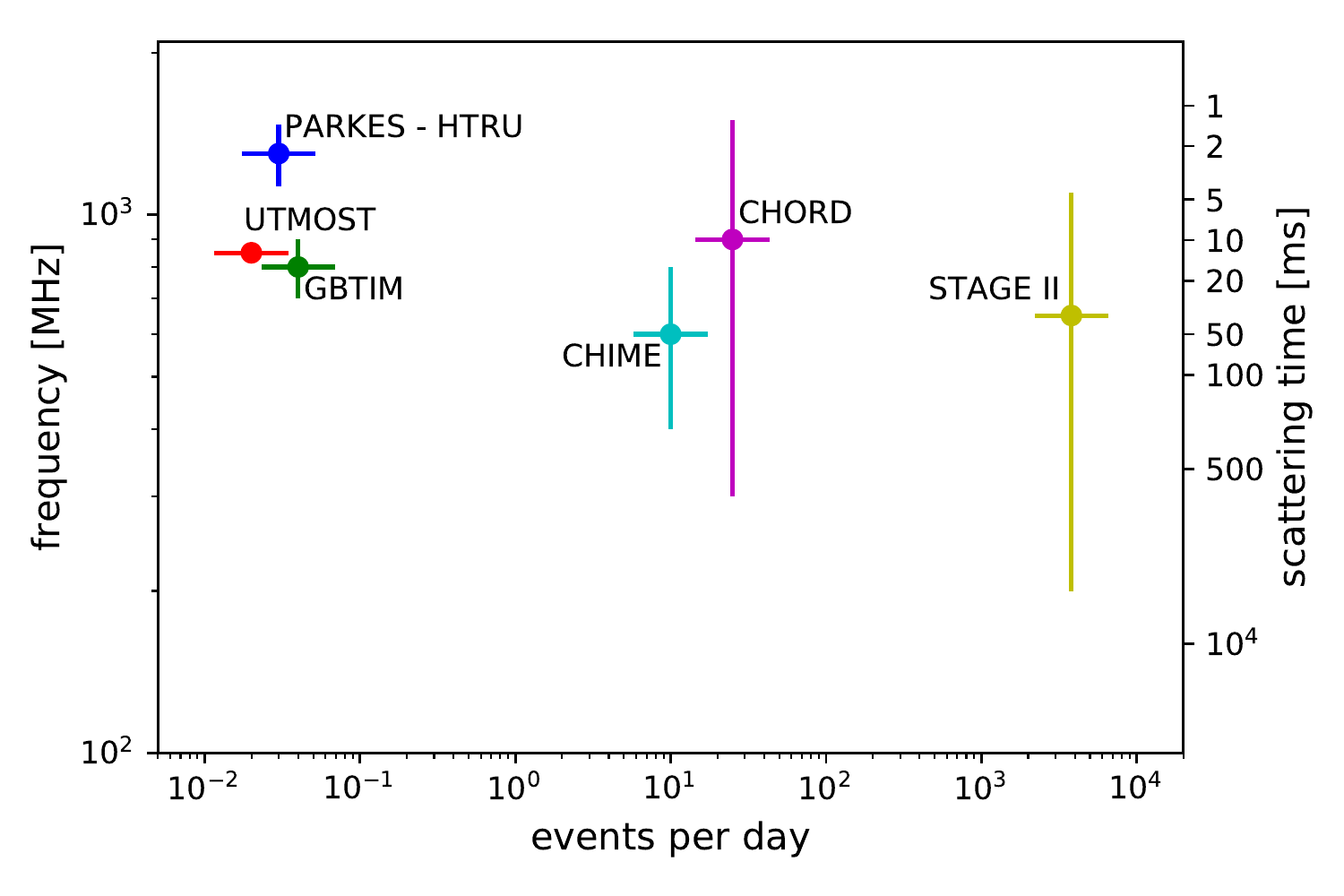}
  \caption{The expect total rates of FRBs for a few experiments
    currently operating or under construction.}
\end{figure}

\subsection{Ancillary science: Time-domain radio astronomy}
\label{sec:ancill-scienc-time}
\subsubsection{Fast Radio Bursts: A new cosmological probe}
\label{sec:fast-radio-bursts}
The extremely high mapping speed that makes transit interferometers sensitive
to large-scale structure also allows them to detect transients at very high
rates \cite{2016SPIE.9906E..5XN,2017PASA...34...45B,2018arXiv180311235T}.
Of particular interest are fast radio bursts (FRBs), a recently discovered
and poorly understood class of radio transient 
\cite{2007Sci...318..777L, 2013Sci...341...53T}.
FRBs are bright, broadband, millisecond flashes, which have now been
established to be of extragalactic and cosmological origin
\cite{2015Natur.528..523M, 2017Natur.541...58C, 2017ApJ...834L...7T}.

A defining feature of FRBs is that they are highly dispersed: their arrival
times depend on spectral frequency due to the frequency-dependent refractive
index of free electrons in astrophysical plasmas. This dispersion gives a precise measure of the
column density of electrons to the burst source, presenting opportunities to study
the distribution of plasma on cosmological scales. The large-scale distribution
of plasma is poorly understood since it mostly resides at densities and
temperatures where it does not significantly emit or absorb radiation. These
so-called ``missing baryons'' have
only recently been detected for the first time through stacking analyses of the
thermal Sunyaev-Zel'dovich effect \cite{2017arXiv170905024T, 2017arXiv170910378D}.
Beyond providing a better understanding of structure formation, a precise
measurement of the electron distribution would aid in the interpretation of
the kinetic Sunyaev-Zel'dovich (kSZ) effect. The kSZ effect measures a
degenerate combination of the electron power spectrum and of large-scale
velocity flows. Independent information about the electron distribution would
permit the velocity flows to be disentangled, providing a check on the theory
of dark-matter structure formation, a probe of the nature of gravity on large
scales, and constraints on modified gravity models.

\citet{2014ApJ...780L..33M} proposed measuring the plasma distribution
from a sample of FRBs by stacking their dispersion measures on
foreground optically-detected galaxies. The contribution to the
dispersion measure from the FRB hosts, as well as the
redshift-dependent contributions from interloping plasma, can be
separated from the signal using its dependence on impact
parameter. Such an analysis requires relatively precise sky
localizations to significantly better than an arcminute for the
FRBs. This could be achieved by adding a number of low-cost outriggers
to the array providing $\sim10$\,km baselines.

A second, related method is to measure the 3D clustering of FRBs directly using
dispersion, and thus electron column density, as a proxy for radial distance
and redshift \citep{2015PhRvL.115l1301M}. FRBs themselves are likely to be
biased tracers of the large-scale structure, however, their measured clustering
will be distorted by systematic errors in their radial distance measurements
from structure in the line-of-sight plasma. These dispersion-space distortions
can then be exploited to precisely measure the plasma distribution.

The proposed experiment operates at a factor of three lower frequency
than most FRB discoveries to date, despite some moderately sensitive searches in this band
\citep{2017ApJ...844..140C}. Only very recently have FRB discoveries at 400\,MHz
been announced \cite{2018ATel11901....1B}, and as such the rates and detectability
at low frequencies are highly uncertain. At these frequencies, the effects of scattering of
the burst signals by inhomogeneous plasma are expected to make them more
difficult to detect (although the presence of this scattering helps in
interpreting discovered bursts \cite{2015Natur.528..523M}).

To get a back-of-the-envelope event rate, we proceed as follows. We
assume that of the currently measured FRBs, the total signal bandwidth
is usually around $1/4$. Hence we take the \stagetwo\ experiment to be
effectively three independent experiments operating between
200~--~400 MHz, 400~--~700MHz and 700~--~1100 MHz.  We then assume the same
instrument properties as described in Appendix~\ref{app:forecasts-noise} (including sky-noise, etc) and assume that
the FRB brightness distribution is Euclidean calibrated to observed
CHIME event rate of $\sim$5/day. This gives approximately {\bf 3500}
events per day. Over the five year span of the experiment, this would
produce around 6 million FRB detections. While this number is
uncertain at a factor of a few, the reality can be better as well as
worse. More information is given in Table \ref{tab:FRB}.

\begin{table}\centering
  
\begin{tabular}{cccc}
  & 200~--~400 MHz & 400~--~700MHz & 700~--~1100 MHz \\
  \hline
  Bandwidth  & 200 MHz & 300 Mhz & 400 MHz\\
  Total System Temperature & 120K & 75K & 65K \\
  Instantaneous FOV & 91$^\circ$ & 27$^\circ$ & 10$^\circ$  \\
  10-$\sigma$ threshold fluence & 26 mJy\,ms & 13 mJy\,ms & 6 mJy\,ms \\
  Event rate in FOV & 1200/day & 980/day &1300/dat \\
\end{tabular}

\caption{Table with FRB detection quantities for the \stagetwo\ experiment. All numbers are uncertain to a factor of a few, but the ones in the lowest band are particularly uncertain given that they are based on extrapolations from higher frequencies.
  \label{tab:FRB}
}
\end{table}
It is clear that such large sample would be transformative for the
field, as it would start to approach the number of galaxies in a
typical galaxy survey (DESI will, for example, measure redshifts of
some 30 million galaxies).

\subsubsection{Pulsars: alternative probe of modified gravity}

Pulsars are highly magnetized neutron stars that, due to their anisotropic
emission and rapid spinning, are observed as lighthouse-like periodic sources
that can be used as astrophysical clocks. The extraordinary precision of these
clocks permits their use in pulsar timing arrays to search for gravitational
waves with light-year wavelengths, as would be emitted by the mergers of
super-massive black holes
\cite{2018ApJS..235...37A, 2015PhRvL.115d1101T, 2013Sci...342..334S}.
In addition, the extreme compactness of
neutron stars permits precision tests of general relativity in the strong
gravity regime by tracking the dynamics of multi-body pulsar systems using
pulsar timing \cite{1989ApJ...345..434T, 2006Sci...314...97K}. These
opportunities
to test fundamental physics depend on the discovery of new highly stable
millisecond pulsars or pulsars in exotic dynamical systems.

Like FRB searches, pulsar searches can benefit from the high mapping speed of transit
interferometers. The proposed experiment covers the 200 to
1100\,MHz band, which includes part of the spectrum that has been
identified as promising for finding the millisecond pulsars
\citep{2014ApJ...791...67S} that
permit searches for gravitational waves and the most precise tests of general
relativity.  Current state-of-the-art surveys have searched large fractions of
the sky, with a few minutes of integration time, using telescopes with order
$(100\,{\rm m})^2$ of collecting area. Current algorithms for searching for pulsars
in collected data require that data to be contiguous in time. As such, a
transit interferometer can only integrate down in sensitivity for the
duration of a transit which, for the proposed  6\,m dishes and 70\,cm
wavelength, is
roughly 27 minutes for a source at the equator and longer at higher
declinations. It would take roughly 15 days to survey most of the sky to this
depth, at which point the square kilometer of collecting area and 27 minutes of
dwell time would permit the discovery of pulsars 1000 times fainter than
current surveys.

In addition, recently proposed algorithms permit the coherent co-adding of
observations taken on consecutive days \cite{2016arXiv161006831S}, meaning the integration
time on a given patch of the sky could be dramatically increased. Depending on
the efficacy of these algorithms, which has not yet been demonstrated, this
would permit the detection of sources fainter by a few orders of
magnitude.

Compared to future surveys, the proposed experiment will be 300 times more
sensitive than CHIME, 64 times more sensitive than HIRAX, and 6 times deeper
than the maximum depth of FAST (even in a 10-year survey, FAST could only reach
its maximum depth over a small fraction of the sky, whereas we are proposing to
reach this depth over the full sky). The SKA, having a similar timeline and
comparable collecting area, will have a comparable maximum depth. However, due
to the non-compact configuration of the SKA antennas, it will only be able to
survey a small fraction of the sky to this depth.

However, beyond discovery, \stagetwo will be able to continuously
monitor the SKA discovered pulsars. The SKA1-LOW and SKA1-MID arrays
will detect of the order of \num{3000}
pulsars~\cite{2015aska.confE..40K}. It is clear that none of the
current telescope facilities, including SKA itself, would have enough
sky time to follow up the majority of these discoveries. Due to the
daily monitoring of a significant subset of these pulsars (depending
on the pointing), \stagetwo\ will be complementary to SKA.  With its
unprecedentedly high timing cadence, \stagetwo\ will be able to
characterize each of these new pulsar discoveries, and carry out a
systematic study of pulsar temporal variabilities, including nulling,
glitches, sub-pulse drifting, giant pulse emission, and potential
signatures of new fundamental physics.

\begin{figure}
  \centering
  \includegraphics[width=0.7\linewidth]{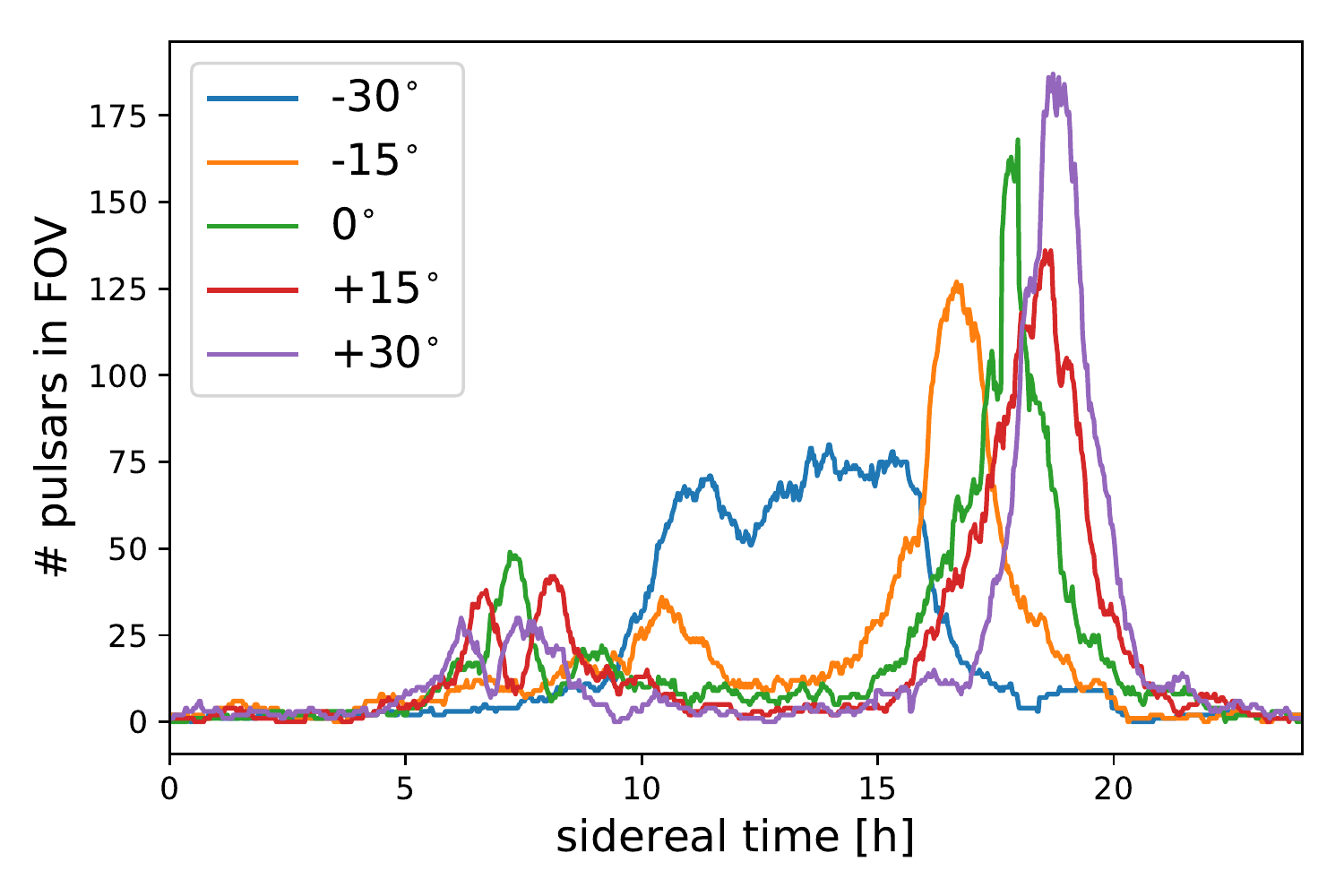}
  \caption{Known~\cite{2005AJ....129.1993M} and forecasted SKA1
    pulsars \cite{2016arXiv161006831S} in the \stagetwo\ field of view
    during a sidereal cycle for various N-S pointing offsets assuming
    observations at latitude~\SI{-30}{\degree}.}
\end{figure}

\clearpage

\section{Challenges and opportunities}
\label{sec:challenges}

While $21\,$cm intensity mapping provides an efficient
means of measuring large scale structure to high redshift, the instrument and analysis must be designed to overcome
systematic sources of contamination: terrestrial radio
signals from human-generated radio frequency interference (RFI) and
the Earth's ionosphere, and extremely bright astrophysical
synchrotron foregrounds from our own galaxy. The former can be addressed with suitable site
locations and benefits from RFI mitigation and ionospheric characterization work from current low frequency
instruments. We can address the latter by using the inherent spectral
smoothness of the foregrounds to separate them from the cosmological
signal. However, this places stringent requirements on frequency-dependent instrument calibration, and foreground removal becomes a key
design driver for instrument characterization, stability, and
uniformity. A baseline instrument configuration that can
achieve foreground removal and sensitivity limits sufficient for the
science goals outlined in the previous sections will require that we build a highly
redundant array of order 65,000 uniformly-spaced feeds, allowing
fast-Fourier transform (FFT) beamforming for data correlation and
compression, operating across a redshift range of $z\sim 2-6$
(200-500\,MHz), and utilizing real-time gain calibration. As noted below, storing the full correlation matrix is not practical, but beamforming this number of detectors as a method of data compression is possible with present-day computation resources, although it requires real-time calibration that has not yet been demonstrated with current instruments. This input from current experiments is critical to assess the trade-offs between raw sensitivity and ease of calibration. Achieving
foreground removal requirements with a sensible analysis strategy can only occur with a concerted R\&D effort along
three directed paths, described in more detail throughout this section: 

\begin{itemize}
\item \textit{Technological:} The primary technological development paths to
  build and calibrate the instrument baseline design described above (with the capability to remove foregrounds and enable data compression) include \textbf{improved
    signal processing and digital conversion electronics};
  \textbf{optimized RF analog chain design with an emphasis on uniformity}; and \textbf{gain
    stabilization and beam characterization instrumentation}.

\item \textit{Analysis:} The primary analysis path is to build on the foreground removal and RFI mitigation techniques from current generation experiments and develop FFT beamforming compression and associated instrument design specifications to enable analysis at an achievable computation scale.

\item \textit{Simulations:} The primary simulation path is to build
  synthetic data for Development and Deployment, Validation and
  Verification, and Uncertainty Quantification. This must include full
  instrument characteristics to optimize instrument design and fully
  explore cosmological parameter constraints, particularly for analysis
  involving cross-correlations and other survey data. The minimum
  required inputs to form a sky map for this process are mock catalogs
  with galactic foregrounds and point sources. By the time this
  project becomes reality, our understanding of the low-frequency sky will
  be considerably improved from \stageone\ and Epoch of Reionization experiments.
\end{itemize}

In Section~\ref{subsec:technical_challenges} we review the outstanding design requirements for $21\,$cm cosmological mapping, heavily informed by the experience of the current generation of experiments. In Section~\ref{subsec:enabling_technologies} we summarize the main technological R\&D areas to address these, and then describe specific technology advances in more detail. In Sections ~\ref{subsec:data_analysis} and~\ref{subsec:simulation_challenge} describe the analysis and simulations challenges, respectively. Finally, in section~\ref{subsec:DOE_capabilities} we relate the technical needs of a $21\,$cm experiment to historical DOE strengths and capabilities, as well as pointing out opportunities for growth.

\subsection{Design Drivers and Requirements}
\label{subsec:technical_challenges}

\begin{figure}
  \centering
  \includegraphics[width=1.0\linewidth]{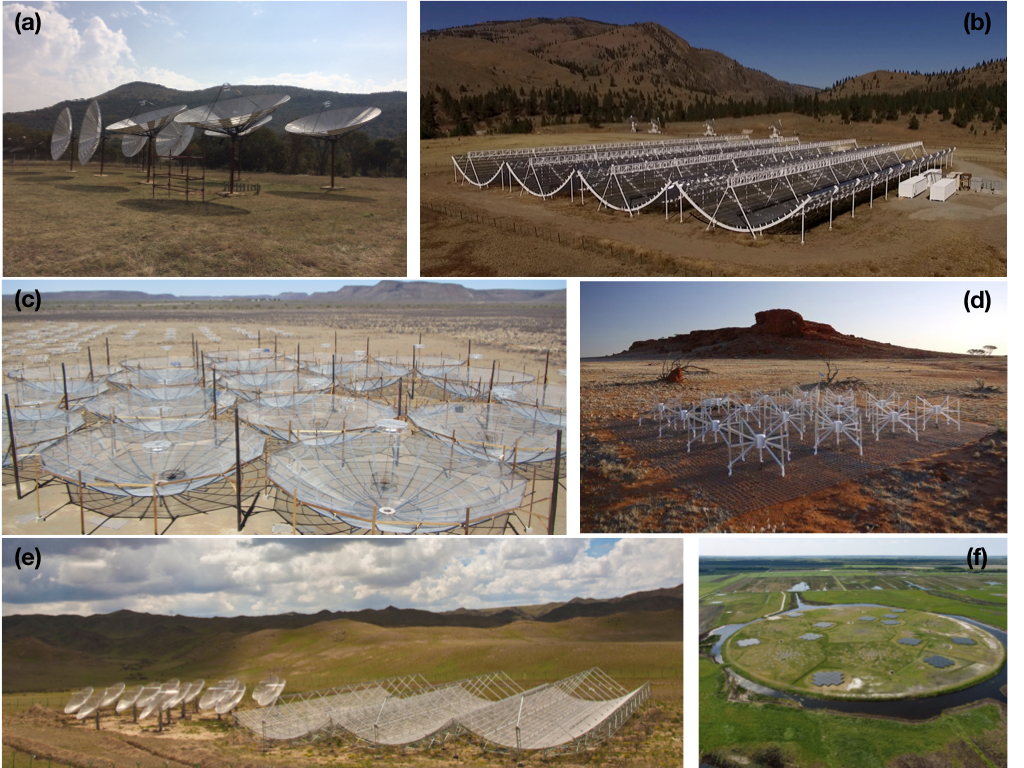}
  \caption{A sample of relevant $21\,$cm interferometric experiments currently fielded: \textbf{(a)} 8-element HIRAX prototype array operating at 400--800\,MHZ \cite{2016SPIE.9906E..5XN}; \textbf{(b)} CHIME experiment operating at 400--800\,MHz \cite{2014SPIE.9145E..22B}; \textbf{(c)} HERA \cite{2017PASP..129d5001D} operating at 50--250 \,MHz with PAPER \cite{pober_et_al2013b} in the background; \textbf{(d)} MWA operating at 80--300\,MHz \cite{2012rsri.confE..36T}; \textbf{(e)} Tianlai \cite{2012IJMPS..12..256C} operating at 700--800\,MHz; \textbf{(f)} LOFAR \cite{2013A&A...556A...2V} operating at 10--230\,MHz. EDGES \cite{2017ApJ...835...49M} is not included because it is targeting a global signal. LEDA is relevant but not pictured ~\cite{6328797}.}
  \label{fig:21cmexpts}
\end{figure}

Using radio surveys of galaxies to probe the BAO scale and constrain dark energy has a long history in the literature (see \cite{2004NewAR..48.1013R} and references therein). It was realized more than a decade ago~\cite{2006astro.ph..6104P,2004NewAR..48.1013R} that low resolution 
radio telescopes in an `intensity mapping mode' could be sensitive to redshifted neutral hydrogen with enough resolution to be resolve the BAO signature and could be used to transform
constraints on Dark Energy and other cosmological parameters. The
first measurements were made on large, steerable dishes, choosing a
survey region which overlaps with high-redshift galaxy surveys,
allowing a detection of highly redshifted  neutral hydrogen via cross-correlation
\cite{2010Natur.466..463C,2013ApJ...763L..20M,2013MNRAS.434L..46S}. Following on this success, new radio interferometers have been built that are dedicated to measuring neutral
hydrogen at high redshift, in principle overcoming the limitations of a single dish measurements at higher redshifts. In this section, we outline the
primary design drivers for a $21\,$cm cosmology survey
instrument. Experience from current generations of experiments already taking data (HIRAX,
CHIME, LOFAR, PAPER, HERA, MWA, and Tianlai among others -- see
Figure~\ref{fig:21cmexpts}) has shown that the most challenging
requirements come from a tackling bright astrophysical
foregrounds. The experiments populate a wide space of instrument configurations, and the largest instrument on the sky in the phase-2 redshift range is CHIME~\cite{2014SPIE.9145E..22B,2016JAI.....541004B,Berger:2016ci,2014SPIE.9145E..4VN},
which has chosen a cylindrical dish design to give the instrument a
wide field of view in one direction, but which requires an intricate
calibration scheme~\cite{Berger:2016ci}. \textbf{ Below we outline the design drivers for a \stagetwo~experiment, but it should be emphasized that foreground contamination is almost always setting the requirement, and adequately removing it must include dedicated efforts across all of instrument design, data analysis, and simulation.}

\newcommand{\maybebullet}{}

\medskip
\noindent{\bf \maybebullet Astrophysical Foregrounds.}
Astrophysical foregrounds, primarily synchrotron emission from the galaxy and unresolved point sources, have much higher
intensity than the cosmological signal of interest.  These foregrounds have a smooth spectral shape and hence can in principle be distinguished from the $21\,$cm emission from large scale structure~\cite{2017PASA...34...33P,pober_et_al2013b,seo_and_hirata2016,2012MNRAS.419.3491L}. However, any frequency dependence in the instrument response, for example from the instrument beam or gain fluctuations, can complicate our ability to differentiate between the smooth foreground and the essentially Gaussian cosmological signal~\cite{Shaw:2014vy,Shaw:2013tb}. Removing these foregrounds drives design choices including element uniformity, array redundancy, assessment of instrument stability and stabilization methods; provides opportunities for new calibration techniques in both beam and gain measurements; and requires analysis and simulations to fold in calibration measurements and assess their impact on cosmological parameter estimation.

\medskip
\noindent{\bf \maybebullet Instrument Calibration.}
Work in $21\,$cm calibration focuses on instrument gain and beam
measurement for the goal of removing astrophysical foreground
power. Simulations for CHIME have provided a scale to the problem: the
instrument response on the sky (`beam') must be understood to 0.1\%,
and the time-dependent response of the instrument (`gain') must be
calibrated to 1\% ~\cite{Shaw:2013tb,Shaw:2014vy}.  Current
instruments rely primarily on sky signals for both types of calibration, however
this has not yet been demonstrated to adequately remove foregrounds with these instruments. Throughout this chapter we outline design choices to meet uniformity and stability specifications that must be carefully integrated into the instrument design, verified during testing and deployment, as well as develop or advance new methods of calibration for this removal.

\medskip
\noindent{\bf \maybebullet FFT beamforming requiring real time calibration and array redundancy.}
The correlation cost and data rate from the \stagetwo\  array will
require implementing an FFT beamforming correlator. Such correlators
use FFT-based sampling of the interferometric
geometry~\cite{tegmark2009} to reduce the computational
correlation cost from order $N^{2}$ to $N\log N$ and output data
volume from $N^{2}$ to order $N$. Taking advantage of this technique
requires that all elements of the array be redundant (that their beams
are similar), placing stringent requirements on element uniformity. In
addition, this correlation will be performed in real time, and so this
requires that we employ real-time calibration to account for
instrumental changes (or that the instrument remains extremely
stable). This technique will be attempted on current generation
telescopes, and we expect work on those experiments will inform
requirements and algorithms for \stagetwo\  instrument.

\medskip
\noindent{\bf \maybebullet Environmental considerations.} In addition to astrophysical foregrounds, two sources of terrestrial signal contaminants must be eliminated or otherwise mitigated: human generated radio-frequency interference (RFI) and Faraday rotation in the ionosphere.

Radio bands within the $21\,$cm redshift range $0.1<z<6$ are popular
as communications frequencies. This forms a bright RFI signal at
discrete frequencies within our measurement band. RFI can be reduced
or eliminated by a suitable choice of radio-quiet
observation site such as the middle of South Africa or western
Australia~\cite{2015PASA...32....8O}, which are remote areas with
limited communications in countries with suitable infrastructure. Even
if RFI must be removed, various experiments operating in locations
with high degrees of interference, notably LOFAR (located in the
Netherlands), have built impressive RFI removal techniques
\cite{2012A&A...539A..95O} that the \stagetwo\ experiment can draw from.

The ionosphere is a plasma and acts in concert with the Earth's magnetic field to rotate the polarization vector of incoming light. The rotation is proportional to $\lambda^{2}$ as well as the number of free electrons present in the ionosphere, which vary across all time scales. While the cosmological signal is unpolarized, most foreground emission from the galaxy is polarized, and so this adds a time variable component to the foreground characterization and removal. The $\lambda^{2}$ dependence means it is not expected to impact the shorter wavelengths (frequencies above 500\,MHz, $\sim z<2$), but it will impact longer wavelengths relevant to a \stagetwo~experiment. Work towards measuring and removing this rotation using accurate maps of the magnetic field and GPS data to infer free electron content is ongoing for experiments at long wavelengths. Because signal propagation through the ionosphere is critical for satellite telecommunications, it is well modelled and current low frequency radio telescopes are working to remove signal variability from the ionosphere~\cite{2015PASA...32...29A}.

\medskip

\noindent{\bf \maybebullet Required Sensitivity.} Instrument noise stems from a combination of intrinsic amplifier noise (noise temperatures for state-of-the-art radio telescopes range from 25\,K cryogenic to 100\,K uncooled) and sky brightness temperature (which span between
10K - 1000K depending on pointing and frequency). Because synchrotron emission increases at lower frequencies, at high redshifts
(above $z\sim$ 3) the system noise is dominated by the sky and no
longer by the amplifier, thus improved noise must be achieved by
fielding more antennas rather than better performing front-end amplifiers. In the absence of systematic effects, detecting the $21\,$cm signal requires fielding instruments including thousands of receivers to achieve mapping down to the mean brightness
temperature of the cosmological $21\,$cm signal of $\sim$0.1 - 1\,mK in the redshift range $0.1 < z < 6$ within a few years.

\medskip
\noindent{\bf \maybebullet Computing Scale.}
Radio astronomy has always been at the forefront of `big data' in
astronomy. Current generation $21\,$cm instruments produce $\gtrsim \SI{100}{\tera\byte}$ of
data per day without any compression, natively generating an amount of data $\propto N^{2}$ where $N$ is the number of elements (currently $N\sim 10^{3}$),
representing challenges in data reduction, transfer, storage, analysis, distribution, and simulation. Compression by a factor of $\sim N$ is achievable by exploiting redundancy within the interferometer, but requires the use of real-time, in-situ calibration and places strong constraints on the uniformity of the optics between elements. For the straw-man experiment with $N = 256^2$ ($\sim$65k) elements, this compression would reduce the data rate from \SI{1350}{\peta\byte} per day to \SI{100}{\tera\byte} per day, but still produce a data set that is \SI{200}{\peta\byte} over a multi-year observation campaign. To aid in data transport, analysis, and data quality assessment, we plan to compress our data further, co-adding maps into a weekly cadence. This reduces to data size but increases pressure on real-time instrument calibration. In addition, to enable transient science we will need fast triggers, already deployed at current generation instruments.

\subsection{Technologies Enabling Science}
\label{subsec:enabling_technologies}

Understanding the instrument requirements illustrated above allows us
to identify dedicated, targeted, and coordinated research and  development areas that
will enable  a $21\,$cm \stagetwo\ experiment sufficient for the science case presented throughout this document. We propose a multi-pronged development effort: early digitization for improved stability and uniformity, optimizing the analog radio receiver elements, and new methods in beam and gain calibration.

\subsubsection{Early digitization and signal processing}
\label{subsubsec:Electronics}

\medskip

Most generally {\it gain} refers to the scaling between the incoming signal and the digitized signal, typically from the analog system (feeds, amplifiers, cables) and digitizer. Analog components are subject to gain
variation, typically due to temperature changes, as the signal travels
from the focus of the dish to the later digitization and correlation
stages. As noted above, gain variation is one of the limiting factors
in removal of astrophysical foreground power. One avenue of
development is to digitize directly at the focus of the dish because
signal information is ``vulnerable'' at all points along the analog
stages, so the imperative is to digitize as early as possible, after
which the signal is (nearly) ``invulnerable''. The resulting digital
signal has more resiliency against time-variable changes in the signal
chain (while some of these are simply moved from analog signal
transfer into the clock distribution, the latter is inherently
narrow-band), offers the possibility of more flexibility in
calibration injection signal algorithms to make gain solutions more
robust, and allows us to use commodity or other well-established
protocols developed for timing and data transfer. However, this comes
at the expense of overcoming the RFI from the digitization in the
field, potentially increased cost, and will require all amplification
to occur at the focus and thus we may find we need carefully designed amplifiers and thermal regulation
at the focus as well.

Several technology developments make receiver electronics with integrated digitizers (early digitization) a promising technology for $21\,$cm projects. Critical components that are now available commercially include:
\begin{itemize}
\item
Room temperature amplifiers with noise temperatures below sky brightness requirements from 100MHz to 1.2GHz.

\item
Low cost digitizers operating in the gigahertz regime with up to 14-bit resolution are readily available. This allows a trade-off: high bandwidth
direct digitization provides the ability to oversample and design high performance digital band selection filters and high order frequency equalizers, but analog conditioning is simpler to implement and model. The final
design will be decided by cost trade-offs while still meeting
stability requirements for foreground removal.
\item
Low cost programmable logic devices capable of interfacing with a high-speed ADC, providing digital filtering to the frequency range of interest, and interfacing to high speed networks.
\item
Similarly, the availability of integrated RF / ADC / FPGA devices in the near future may provide a path to very compact high-performance receivers.
\end{itemize}

By digitizing at the focus we broaden the possibilities for instrument calibration, bandwidth, and signal processing, however there are a few additional considerations:

\begin{itemize}
\item As noted, one of the technical challenges for $21\,$cm telescopes is the need for $<$1\% gain stability over at least 24 hours.  The primary culprits of gain variation with temperature come from the amplifiers and any analog transmission (either coaxial cable loss or radio-frequency-over-fiber). By digitizing at the focus, the analog transmission is unnecessary and then any variation will be dominated purely by the amplifiers. The resulting temperature variation can be either mitigated by use of thermal regulation of the circuitry at each dish focus or removed by injecting a calibration signal, or both. Because noise diodes have a gain stability of $2\times 10^{-3}/^\circ$C, achieving the required gain stability still requires thermal regulation of $\sim ^\circ$C. Amplifiers have roughly similar thermal regulation requirements, however they are more difficult to decouple from the environment because they are either connected or embedded in the antenna. Thus, development should be placed towards building calibration sources, digital or otherwise, to enable gain stabilization.

\item We must isolate the sensitive RF input with signals in the -100dBm range from the high power digital outputs from the ADC which typically operate near 0\,dBm. In addition, RF radiation from the digital processing system must be shielded from the input and from any other antennas.

\item The raw data rate from the digitizer is large, a few
  $\times$10Gbit/second. This can be substantially reduced depending
  on the oversampling level, with digital filtering in the FPGA that
  receives the digitizer data, followed by transmitting only the
  bandwidth containing useful physics data. For some correlator
  architectures it may also be useful to transmit data separated by
  frequency band to an array of correlation processors. The system can
  trade off oversampling at a few gigahertz and digitally filtering
  down to the band of interest for a more complex analog system.  In
  theory a digital filter can do significantly better than an analog
  filter in terms of stability and out of band rejection, and may
  become more cost effective on the time scale of the \stagetwo\ instrument.

\item There must be a very precise clock distribution system sent out to each of
the digitizers.  This moves the instrument phase calibration problem from the
analog system into the clock distribution system.  Numerous techniques exist for
synchronizing a distributed clocking system, and these must be adapted to enable
a low cost calibration system. This has been found to be challenging even with the digitizers in only two locations, and so carefully designing and testing this timing system, including mitigations and estimates for cable reflections, will be a critical R\&D task for any distributed digitization across the instrument. 

\end{itemize}

\begin{figure}
  \centering
  \includegraphics[width=0.7\linewidth]{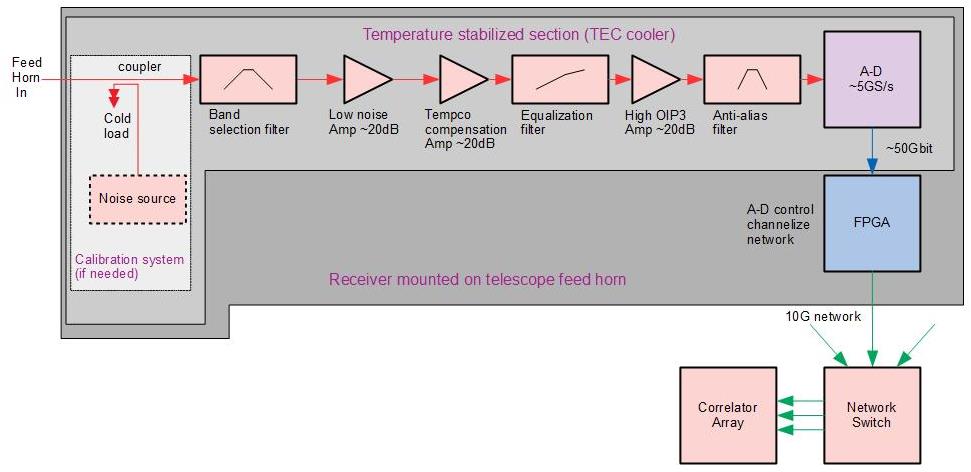}
\caption{Block diagram for a proposed early digitization
front-end.}
\label{fig:front_end_block_diagram}
\end{figure}

\subsubsection{Analog design}
\label{subsubsec:optics}

\begin{figure}
  \centering
  \includegraphics[width=0.7\linewidth]{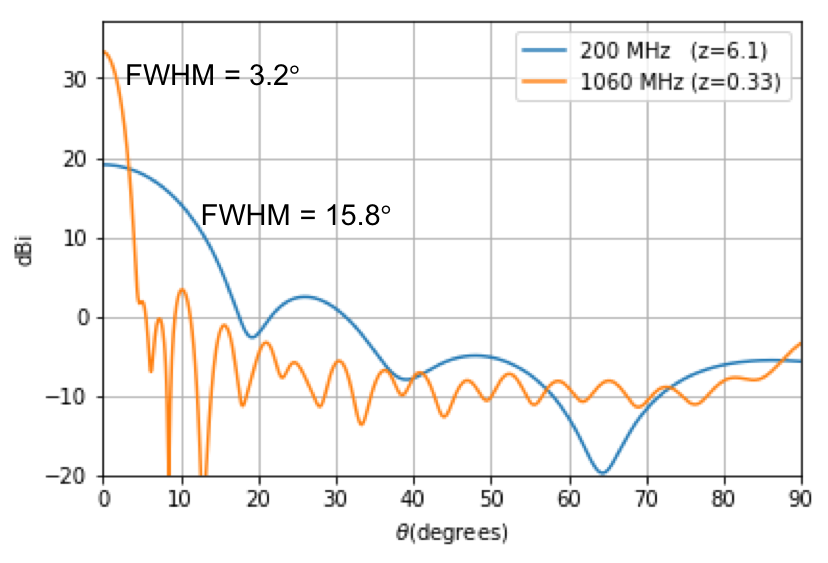}
\caption{Simulated antenna gain patterns (co-polar, H-plane). An on-axis, 6~m diameter, parabolic dish reflector is illuminated by a dipole feed antenna designed to provide an edge taper of $\sim -10~$dB over a wide (5:1) bandwidth. Wide band feeds of this type are under development. Even at the lowest frequency (200~MHz), where the reflector is only 4 wavelengths in diameter, the beam spillover to the ground is less than the expected level of foreground emission in the main beam.}
\label{fig:gain_patterns}
\end{figure}

\begin{figure}
  \centering
  \includegraphics[width=0.5\linewidth]{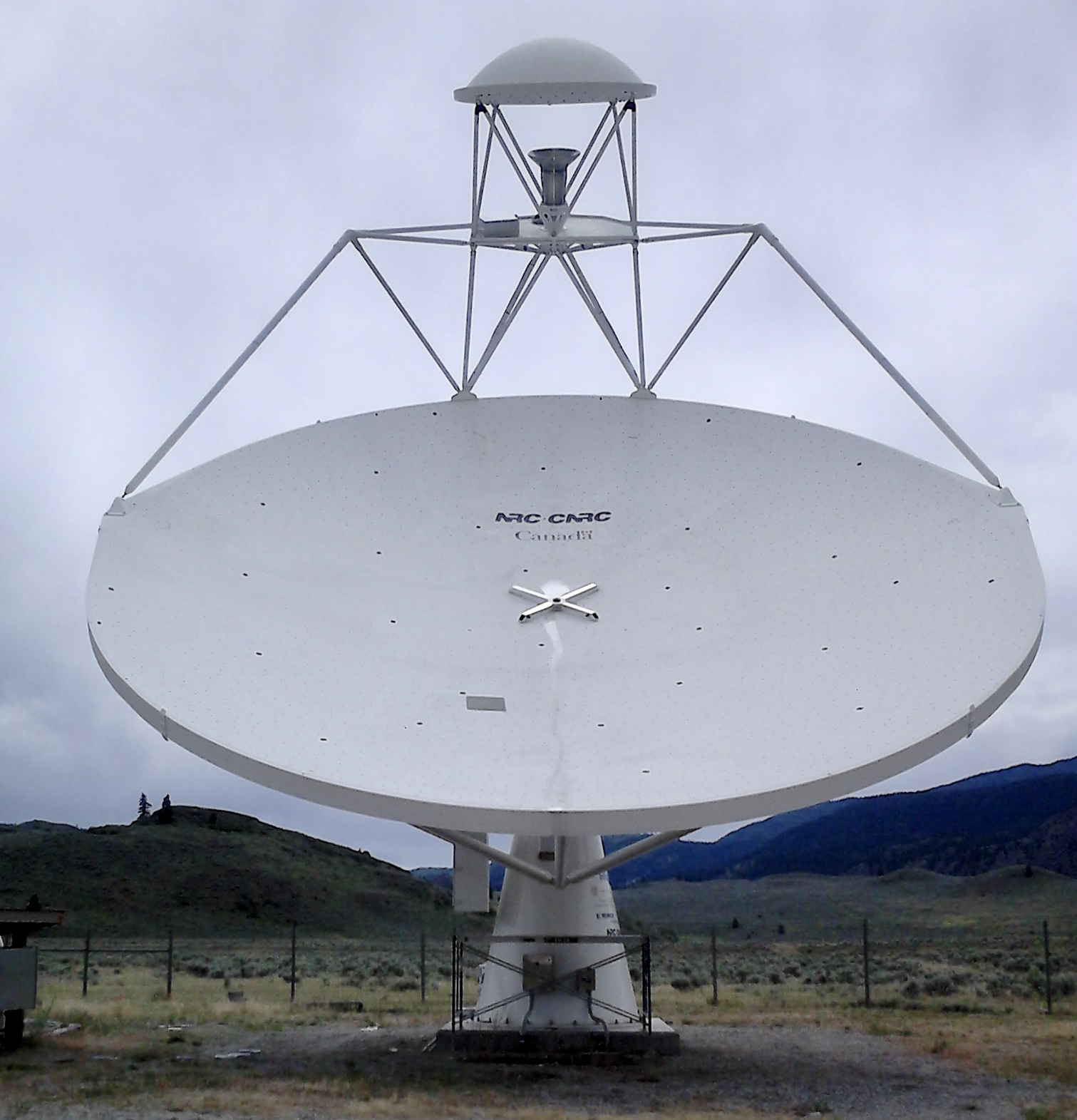}
  \caption{Prototype SKA fiberglass dish, located in Canada at the Dominion Radio Astrophysical Observatory}
  \label{fig:fiberglass}
\end{figure}

\medskip
\noindent
{\bf \maybebullet Optimization of optical design.}
Most existing and near-future $21\,$cm experiments, e.g. CHIME~\cite{2014SPIE.9145E..22B}, Tianlai~\cite{2012IJMPS..12..256C}, HIRAX~\cite{2016SPIE.9906E..5XN}, and HERA~\cite{2017PASP..129d5001D}, all have chosen parabolic reflectors with the receivers supported at the focus with metal legs, leading to some diffraction and reflections. To illuminate the dish, they have also designed variants of dipole receivers with wide beams that have non-negligible cross-talk and frequency-dependence. These choices are typically made as a cost- and complexity savings, but make calibration more difficult. Further study for optimization, including options such as off-axis geometries (like SKA-mid and ngvLA) and possibly horn/Gregorian receivers, will be important particularly since many of those experiments will have greater experience with the parabolic reflector geometries in the near-term.  These experiments also span a range of wavelength-to-size ratios and we would use these experiments along with simulations to form a specification on the dish diameter. Wide bandwidth optical systems are under development. See Figure~\ref{fig:gain_patterns}. The optimization would include keeping marginal costs low while also meeting uniformity and  bandwidth flatness specifications, and exploring new dish fabrication techniques (such as using a fiberglass-based design~\cite{7303193}, see Figure~\ref{fig:fiberglass}). \newline

\medskip
\noindent
{\bf \maybebullet Front-end sensitivity and bandpass}.
When properly designed, the receiver noise temperature is dominated by loss in the analog feed as well as the noise in the first stage amplification. HIRAX has chosen to reduce the system noise by up to 30\% by fabricating the first stage amplifier directly in the antenna itself, reducing the transmission loss and taking full advantage of low-noise transistors available in these bands. In addition, current generations of $21\,$cm experiments~\cite{2017PASP..129d5001D} have found that their bandpass shape is a limitation of their foreground removal, and are actively working on new feed designs that have a more carefully shaped bandpass. One development path for the active circuitry in the HIRAX feed would be to add additional RF circuitry to flatten the bandpass to remove ripple and other features, allowing an easier path for foreground removal. This introduces more stringent oscillation conditions on all amplification stages to reduce the possibility of amplifier oscillation and we will learn more about the feasibility of this technique for mass production as additional prototypes are fabricated for HIRAX. \newline

\medskip
\noindent
{\bf \maybebullet Uniform interferometric elements for calibration and FFT correlation.}
Similar interferometric baselines should see the same sky signal and so differences between them can be used to assess relative instrument gains over time. This technique is known as `redundant baseline' calibration and has been developed as a method of meeting the gain stability requirements~\cite{2010MNRAS.408.1029L,2018arXiv180500953C,2018MNRAS.477.5670D,2014ASInC..13..393R,2016ApJ...826..181D,2017arXiv170101860S}. This requires both a decision to space the interferometer dishes the same distance apart, and also have highly uniform interferometric elements. Most $21\,$cm instruments have chosen their baseline spacing to use this technique, however have been limited by the fact that their interferometric elements are not identical enough to achieve precision calibration. To overcome this, we would investigate dish fabrication tolerances required for this calibration as well as how we might use new dish fabrication techniques (for example, fiberglass dishes with embedded mesh conductors, currently being prototyped for SKA and HIRAX, see Figure~\ref{fig:fiberglass}) to meet these needs.

In addition, the requirements that we use FFT or similar beamforming~\cite{2004NewA....9..417P,tegmark2009,2010PhRvD..82j3501T,2017arXiv171008591M} to compress that data forces stringent requirements on the uniformity of response, beam shape, mechanical construction and alignment, gain control, etc.~across what will ultimately be on the order of $\sim$65k detector copies. The requirements for this uniformity and how to achieve it will be part of the instrument design process.

\subsubsection{Instrument Calibration}
\label{subsubsec:beam_characterization}

\medskip
\noindent
{\bf \maybebullet
Gain Stability.} Each antenna has a characteristic response to an input sky signal, which varies with both time and frequency, known as the instrument gain. The frequency-dependent gain for each input must be known to $\sim$ 1\% on time scales between the integration period ($<5$s scales) and a few hours (depending on the frequency of on-sky radio calibrator sources)~\cite{Shaw:2014vy}. The two primary techniques for achieving this are to design an instrument which is inherently stable enough to meet this specification or to design a calibration plan which can ensure we meet this specification, or (ideally) both. CHIME~\cite{2014SPIE.9145E..22B,2014SPIE.9145E..4VN} is updating a classic radio noise-injection scheme which can be used to calibrate many signal chains at once. To implement such an active calibration technique for dishes will require development of stablized transmission algorithms and may be made easier with early digitization and development of calibration sources which may be independently fielded at the focus or flown on a quadcopter drone. We will also require passive models of gain and beam variation with temperature and dish pointing. This modeling is essentially standard for radio telescopes and precision modelling has been demonstrated with at least one instrument (CHIME).

\medskip
\noindent
{\bf Beam Characterization.} Each antenna also has a characteristic response on the sky, known as the instrument beam. Because this response (main beam and sidelobes, as well as polarization) is capable of mixing frequency dependence and sky location, it is expected to be the primary source of contamination from foreground emission into the signal band, and so must be known even more accurately than the gain ($\sim$0.1\%)~\cite{Shaw:2014vy}. This level of calibration is difficult for $21\,$cm telescopes because they are stationary and designed to have large beams for improved survey speed \cite{2017PASA...34...62S}. In addition, some instruments (such as CHIME) have large dish sizes which can be difficult to model and simulate, requiring exceedingly detailed knowledge of support structures and surface mesh.  Many $21\,$cm instruments are beginning to use quadcopter drones to map the beam shape (HERA\cite{Jacobs_drones}, SKA\cite{Virone_drones, Pupillo_drones}, LOFAR\cite{Chang_drones}) and while this technique seems promising to meet the needs for $21\,$cm cosmology it is unlikely we will be able to measure all of the beams from all of the dishes in an instrument with 65k dishes, and so this beam calibration requirement also forces a specification on uniformity in dish fabrication.

\subsubsection{Data flow and processing}
\label{subsubsec:data_flow}


Computing requirements for a 65k-element interferometer come from both the correlation burden and the data reduction, transfer, storage, analysis, and synthetic data production. For the correlator computation, we will need to pursue development in computing approaches which
can improve the cost scaling both for equipment and power. Examples could
include using commodity-hosted FPGA's, using/developing dedicated
ASIC's \cite{7436238}, or GPUs to smoothly take advantage of the fast-paced hardware updates for correlator computation.

\begin{figure}
  \centering
  \includegraphics[width=0.9\linewidth]{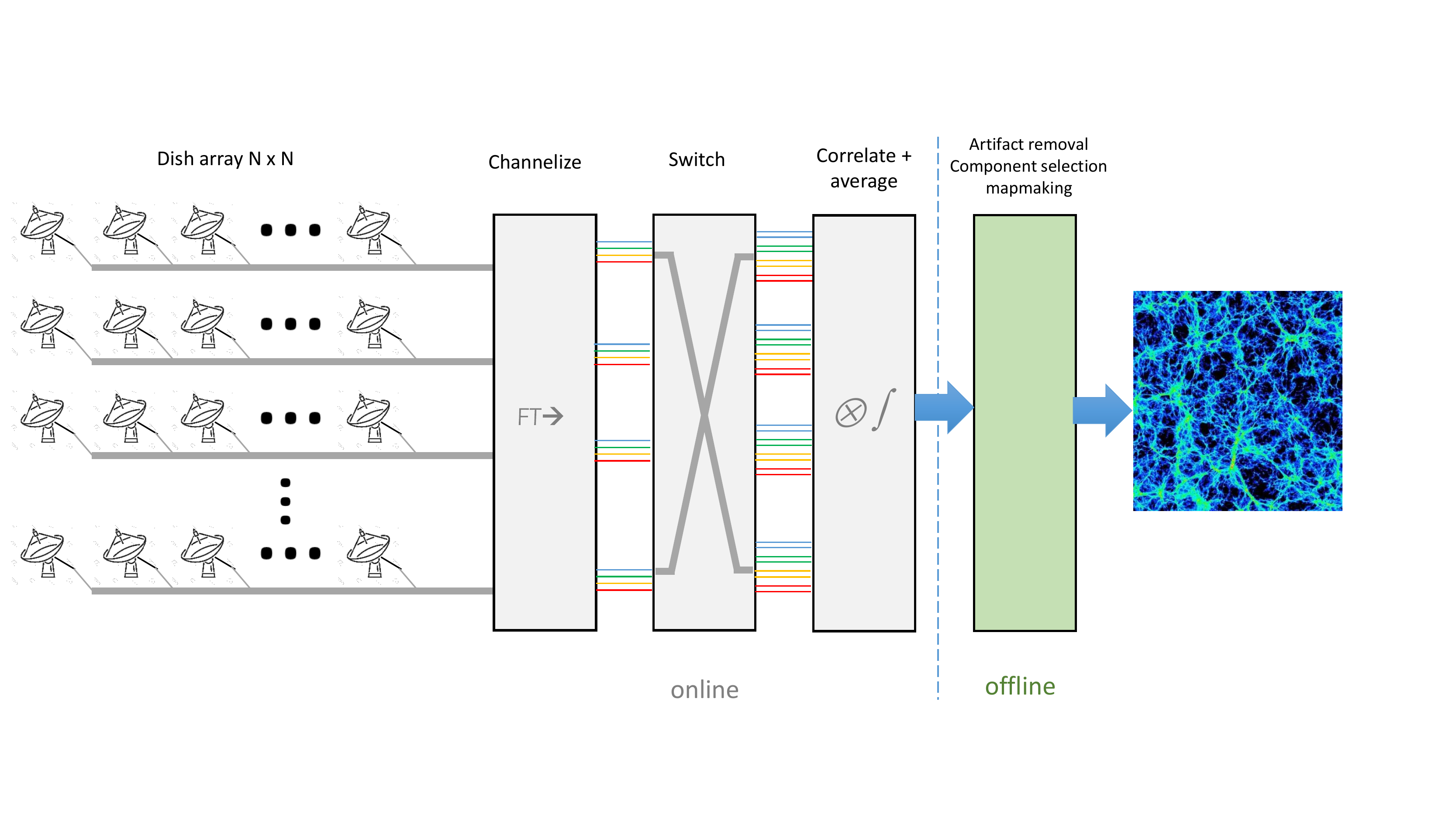}
\caption{Illustration of anticipated data flow in a large interferometric array.  Conversion of waveform data to frequency space, e.g. channelization, is accomplished close to each receiver; coincident data for each frequency bin are collected from all stations through a cross-bar switch (also called a ``corner-turn'' operation); correlations are constructed for each frequency bin, which can then be time-averaged and stored, followed by physics analysis.}
\label{fig:data_flow_block_diagram}
\end{figure}

\subsection{Data Analysis}
\label{subsec:data_analysis}

\newcommand{\tcite}[1]{[{\color{green}#1}]}  
\newcommand{\rscomment}[1]{{\color{red} \textbf{[RS:  #1]}}}


\medskip

Releasing science deliverables for the community from a $21\,$cm experiment depends crucially on developing and deploying, including validation and verification, an analysis pipeline that can ingest vast quantities of data and transform it into well characterized frequency maps and power spectra. This is a computationally costly and varied exercise, but does not require continuous real time processing, and thus can be performed at an external high performance computing site.
We can divide the analysis up into three broad areas discussed below.

\medskip
\noindent {\bf Flagging, Calibration and Pre-processing at scale.} In this area
the data is processed to reduce the remaining systematics which may
effect our ability to access the cosmological signal.
Of particular importance is cleaning of any RFI by flagging times and frequency channels that have been contaminated. This is a well understood problem within radio astronomy \citep{2012A&A...539A..95O}, though the effects of residual RFI at the small level of the $21\,$cm signal is only starting to be addressed \citep{2018MNRAS.479.2024H}.
Though much of our calibration must be done in real-time (see \ref{subsec:technical_challenges}) to enable FFT correlation, there are still degrees of freedom that must be corrected, particularly degeneracies that may not have been fully fixed (including but not limited to an overall gain scale for the entire observation, \citep{2010MNRAS.408.1029L}), and calibration of the bandpass (the array-wide frequency dependent gain). Again these are problems that are well understood within radio astronomy.\\[-0.8em]

\noindent {\bf Astrophysical Foreground Removal.} Along
with the sensitivity requirements for measuring a faint signal, the
key analysis problem for $21\,$cm intensity mapping is the need to
remove contaminants that are many orders of magnitude larger than the
cosmological signal. Though foreground cleaning is a common problem
across cosmology, the required dynamic range is unique to $21\,$cm intensity
mapping.

In principle the foregrounds can be separated from the signal using their smooth frequency dependence \citep{2010ApJ...721..164S}. However, even an ideal instrument couples anisotropy in the astrophysical foregrounds into spectral structure with an amplitude generally significantly larger than the cosmological signal. This extremely challenging problem is called mode-mixing and is exacerbated by instrumental systematics such as gain variations and optical imperfections which must be minimised (see the discussion in Section~\ref{subsec:technical_challenges}). There exist in the literature many proposed techniques to separate the cosmological signal from the foregrounds, but these have only demonstrated success in simulations.

Foreground mitigation falls broadly into two classes: foreground
avoidance and foreground cleaning. Foreground avoidance is the
simplest of these two approaches, relying on the fact that
contamination produced by a typical interferometer configuration is
strongest in certain regions of $k$-space (see Appendix \ref{app:wedge}). Producing cosmological results only using the cleanest modes is a simple and effective technique. This technique, however, becomes deeply unsatisfactory at low frequencies, particularly in the dark ages. Here galactic synchrotron and extragalactic point source radiation quickly becomes very bright, typically hundreds of Kelvin at \SI{100}{\mega\hertz}, even at high galactic latitudes. At the same time the window of clean modes dramatically narrows due to the relative scaling of the angular diameter distance and Hubble parameter with redshift \cite{2015MNRAS.447.1705P}. Combined, this means that at a given threshold for contamination we exclude increasingly large regions of $k$-space at high redshifts, significantly degrading any cosmological result.

Foreground cleaning instead of (or in conjunction with) foreground avoidance then becomes an attractive option. A general feature of foreground cleaning methods is that they rely on detailed knowledge of the instrument response to predict and subtract the actual foreground signal. For instance, given perfect knowledge of the complex beam of each individual antenna, a tomographic map of the sky can be effectively deconvolved to remove the spectral structure induced by the instrument's beam. The residual contamination is set by both the amplitude of the raw contamination and the accuracy with which the beam has been measured. This is similar in spirit to the residual temperature-to-polarization leakage produced by mismatched beams of orthogonal polarizations in CMB $B$-mode searches, which can be accurately predicted and removed given beam measurements despite the fact that the CMB temperature anisotropy ``foreground'' is orders of magnitude larger than the $B$-mode signal.\\[-0.8em]

\medskip
\noindent {\bf Cosmological Processing.} Having cleaned the foregrounds out of the data we then need to process it to quantities useful for cosmology such as power spectra and sky maps. Though this has been done within the CMB and LSS communities for many years, the fact that we are dealing with radio interferometric data after foreground cleaning brings unique challenges. The source of these is that the measured data is abstract: it is a complex, spatially and spectrally non-local measurement of the sky. This adds significant complexity in generating maps and power spectra from the data, but also tracking which modes have been measured (and which are missing) to allow us to accurately measure uncertainties. Regardless, we expect to be able to significantly draw on the conceptual frameworks used for cosmological data analysis to be able to tackle these problems \cite{Shaw:2014vy,2014PhRvD..90b3018L,2004ApJ...609....1J}.\\[-0.8em]

Although we can create a broad outline of how the analysis pipeline,
and we are able to draw on many mature and well understood techniques,
there are several areas that will require research investment to
ensure the success of a large scale $21\,$cm intensity mapping survey.

\medskip
\noindent {\bf Scaling.} While we can draw on existing techniques for all stages of the analysis, a significant challenge is scaling these to be able to work with the vast increase in data that we will generate in an energy-constrained/post-Moore's computing landscape. This will require optimizations in algorithms and implementations to reduce the computational cost of the processing, and ensuring that the techniques can scale in parallel to run on leading edge supercomputers.

\medskip
\noindent {\bf Systematic Robustness.} Both astrophysical uncertainties (such as the exact nature of foregrounds) and instrumental uncertainties (such as calibration and beam optics) cause foreground contamination. Developing more robust cleaning techniques will reduce systematic biases, but potentially allow us to reduce the instrumental tolerances leading to cost savings.

\medskip
\noindent {\bf Improving signal recovery.} Significant numbers of modes are lost to foregrounds, which reduces our constraining ability generally, but particularly affects science that needs access to the largest scales. Improved foreground removal that reduces the effect of the wedge could improve this, as would methods like tidal reconstruction \cite{Zhu:2015zlh,Zhu:2016esh,Foreman:2018gnv}, but these techniques need substantial development. Similarly, traditional reconstruction techniques \cite{2007ApJ...664..675E,2009PhRvD..79f3523P} that recover non-linear modes need work adapting them for the peculiarities of $21\,$cm intensity mapping.

\subsection{Simulation Needs and Challenges}
\label{subsec:simulation_challenge}

The challenges facing $21\,$cm surveys are significant but, at least to $z=6$, well understood. However, our ability to tackle them requires a sophisticated approach to overcome them both through instrumental design and offline analysis. It is therefore essential to use simulations to close a feedback loop that allows us to predict, and thus refine, the effectiveness of a design and analysis strategy.

Producing realistic simulations of data from any instrument configuration and propagating these to final cosmological results is a conceptually straightforward prospect:
\begin{enumerate}
  \item Produce a suite of full-sky maps of the ``true'' sky, with one map per frequency and at each frequency bin observed by the instrument. There are a variety of approaches to form full-sky maps of the signal and foreground, and full exploration of the data should include common sky models to include other observables (e.g. galaxy surveys) for form estimates of cross-correlations. 
  \item ``Observe'' these maps with a simulation pipeline that contains sufficient realism to capture any and all non-idealities that might produce contamination in the data.
  \item Feed these mock observations into the data analysis pipeline discussed in the previous section, and the same pipeline that would be used on real data, and produce reduced data and cosmological analyses.
\end{enumerate}

\begin{figure}
  \centering
  \includegraphics[width=0.49\linewidth]{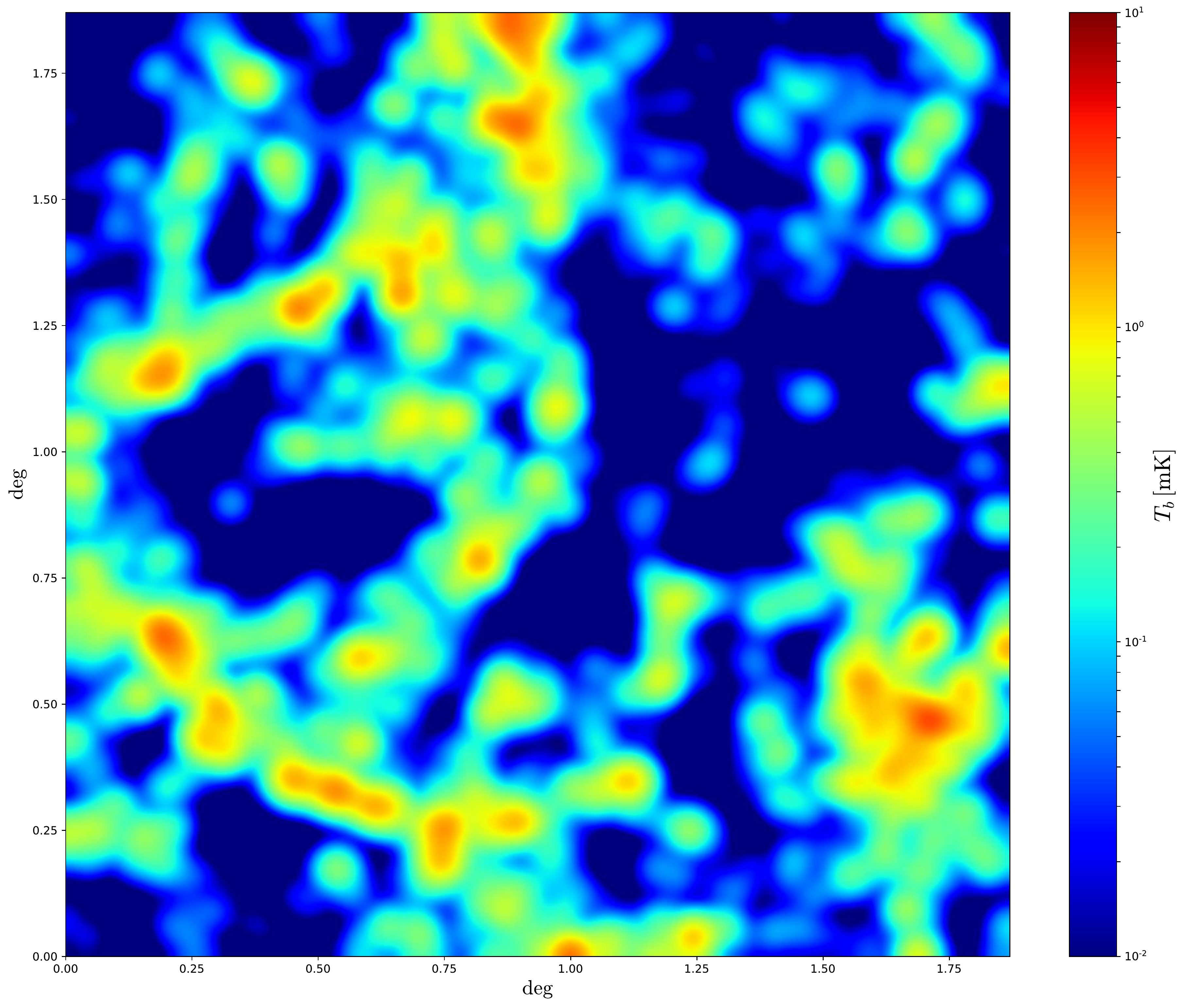}
  \includegraphics[width=0.49\linewidth]{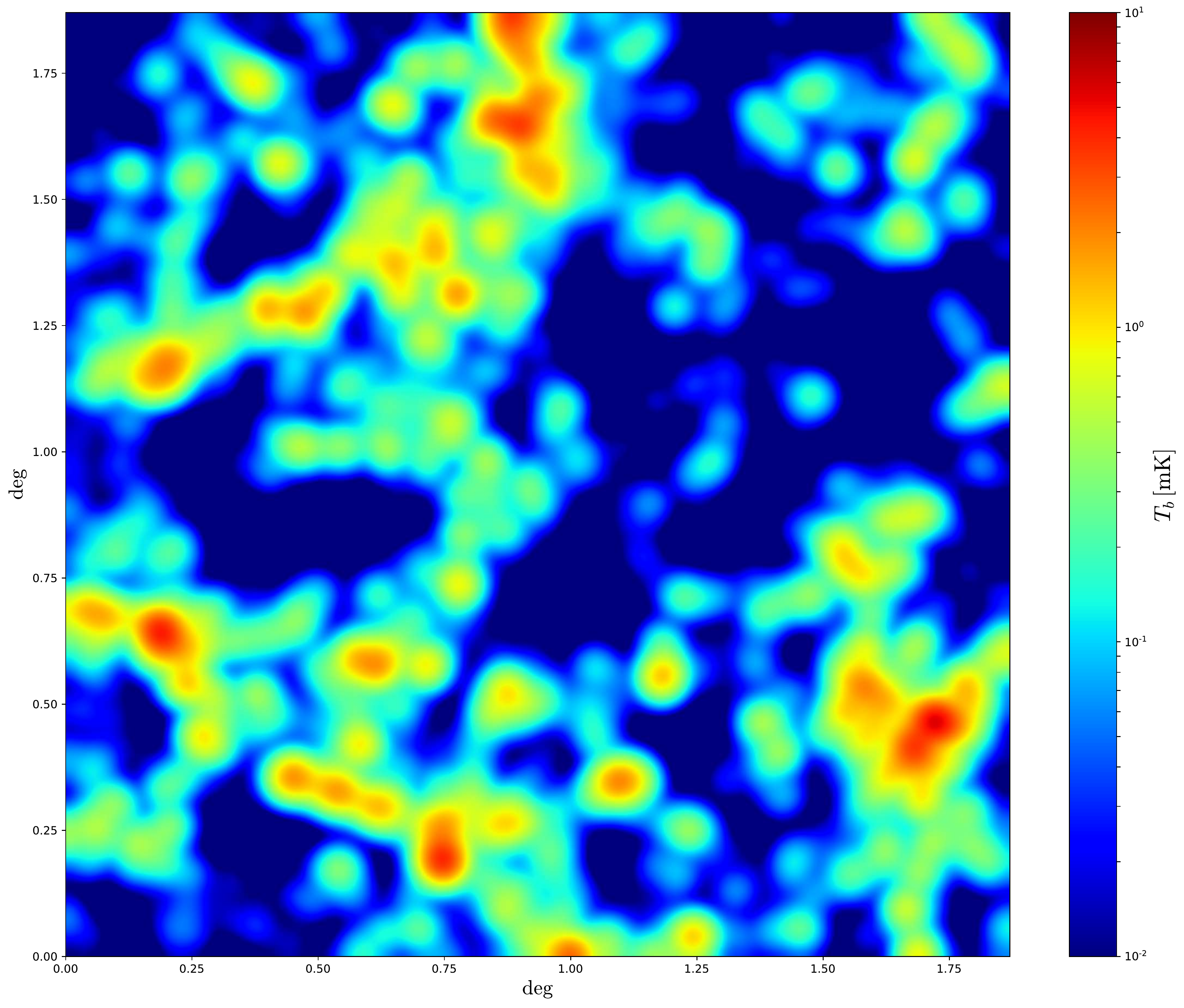}\\
  \caption{21cm maps at a frequency of 710 MHz over a channel width of 1 MHz with an angular resolution of 1.5' over an area of $\simeq4~{\rm deg}^2$. The map on the left has been created from the state-of-the-art magneto-hydrodynamic simulation IllustrisTNG with a computational cost of $\simeq$18 million cpu hours. The map on the right panel has been generated by assigning HI to dark matter halos of an N-body simulation using the a simplification of the ingredients outlined in \citep{2018ApJ...866..135V}. The computational cost of the N-body simulation is much lower than that of the full hydrodynamical simulation, and allow us to model the HI field in a very precise and robust manner.}
  \label{fig:21cm_sim}
\end{figure}

For verification of foreground removal effectiveness gaussian or pseudo-Gaussian $21\,$cm simulations are largely sufficient \cite{2014MNRAS.444.3183A,Shaw:2014vy}. However, for targeting sensitivity to specific effects (e.g. non-Gaussian initial conditions), or in cross-correlation with other probes, more accurate simulations constructed from mock-catalogues will be required. This allows us to produce correctly correlated maps for additional tracers (e.g. LSST photometric galaxies), and also for radio point source contribution to the foregrounds.

Though the relation between HI density and total matter density involves complex environment dependent processes, simulating it can be done efficiently. Recent work has shown that one can take advantage of the fact that neutral hydrogen in the post-reionization era resides almost entirely inside dark matter halos~\citep{2018ApJ...866..135V}. It is possible to combine the presently available HI data to constrain an analytical halo model framework for the abundance and small-scale clustering of HI systems \citep{Padmanabhan2017a, Padmanabhan2017b}. Thus, one can calibrate the relation between dark matter halos and HI using hydrodynamic simulations and create 21cm maps via less expensive methods such as N-body or fast numerical simulations like Pinocchio \cite{Pinocchio}, ALPT \cite{ALPT}, HaloGen \cite{HaloGen}, EZMock \cite{EZmock}, PATCHY \cite{Patchy}, COLA \cite{COLA}, QuickPM \cite{QuickPM}, FastPM \cite{FastPM}, or Peak Patch~\cite{stein-peak-patch}.  It may even be possible to adopt a simple perturbation-theory-inspired approach \cite{Modi:2019hnu}, which would allow very large volumes to be simulated very efficiently.

As the dominant foreground contribution, simulating the galactic synchrotron must be done with care to ensure that the simulations are not artificially easy to clean. A simple approximation can be produced by proceeding from a full sky map at a radio frequency (typically the Haslam radio survey map at 408~MHz) and scaling this map to different frequencies based on the known spectral index of galactic synchrotron radiation. However this is not sufficient at the dynamic range between the foregrounds and the $21\,$cm signal and we must be careful to include: spectral variations about a pure power law; small scale angular fluctuations not captured in existing surveys; and polarization, including the effects of emission at a wide range of Faraday depths which generates significant spectral structure in the polarized emission \cite{Shaw:2014vy}. More sophisticated galactic models, for example from MHD simulations, could also be developed and used here.

Regarding (2), a realistic instrument simulation pipeline would take
the maps discussed and convolve them with the complex beam for each
antenna in the interferometer. This can be done by direct convolution
utilising the fact that for a transit telescope it is sufficient to
generate a single day of data. However for wide-field transit
interferometers this can be more efficiently performed in harmonic
space using the $m$-mode formalism ($O(N\log{N})$ instead of
$O(N^2)$). Some of the required code would be similar and could in
principle built upon similar codes used in the CMB science, such as
the TOAST
package\footnote{\url{http://hpc4cmb.github.io/toast/intro.html}}
using fast numerical techniques for beam convolution
\cite{2010ApJS..190..267P}.

For these simulations we need to generate realistic simulations of the telescope beams. Electromagnetic simulation codes such as CST, GRASP and HFSS can be used for this, but achieving the accuracy required is challenging computationally \cite{2017JAI.....650003C,2017arXiv170808521D,2016ApJ...831..196E}. An alternate approach is to generate synthetic beams with sufficient complexity to capture the challenges posed by real beam, these are computationally easier to produce, but must be informed by real measurements and electromagnetic simulations to ensure their realism, and may be aided by machine learning algorithms.

Capturing non-idealities in the analog system, particularly gain variations, is mostly straightforward as these can be applied directly to the ideal timestreams. Additionally we need to include time-dependent beam convolution (including position and brightness) for temporally varying sources such as solar, jovian and lunar emission as well as the effects of RFI at low levels~\cite{2018MNRAS.479.2024H}.

Including calibration uncertainties poses a particular challenge, because of the realtime calibration and compression of the instrument, simulating these effects requires either: generating data at the full uncompressed rate, applying gain variations, and then performing the calibration and compression processes; or the computationally easier alternative of generating models of the effective calibration uncertainties.

After the first two stages, mock observations are then fed to the proposed data analysis pipeline, and propagated through to final cosmological products, to assess analysis systematics, instrument design, real-time calibration, and data processing to determine whether the pipeline is sufficient to meet our science goals.

Though the simulation program is well defined, there are already many
open challenges discussed below.

\medskip \noindent {\bf Understanding the HI distribution.} To map the HI distribution to the cosmologically useful matter distribution requires cutting edge hydrodynamic simulations to capture the small halos that HI favours over a cosmologically interesting volume. This additionally improves the efficiency and accuracy with which we can produce mock skies.

\medskip \noindent {\bf Scale.} To assess and understand a proposed design we need to be able to produce large numbers of emulators that Monte-Carlo over the experimental uncertainties. The number, size and complexity of these simulations requires a large scale effort to plan, generate and manage them. 

\medskip \noindent {\bf Improving the Feedback Loop.} While a straightforward version of the simulation loop above can tell us whether a proposed design does or doesn't meet our needs, it does not tell us how to improve the design to ensure that it does. For a complex instrument with many design parameters is is essential to be able to guide this process by using simulations to infer the most relevant combinations of changes.

\subsection{Relation to DOE capabilities}
\label{subsec:DOE_capabilities}

 This chapter has enumerated the technology and analysis challenges for studying cosmic acceleration by mapping the large-scale structure of the universe using $21\,$cm radiation. As with other large DOE-HEP experiments, it requires data from imperfect detectors to be turned into useful scientific output by application of multi-level calibration schemes that incorporate the as-built instrument characteristics and thorough end-to-end numerical simulation of the physics of the measurement process. DOE has a unique heritage in successfully constructing large experiments of this type, making it a particularly appropriate home for the development of a $21\,$cm intensity mapping experiment. Capabilities found in the DOE Laboratory complex in technical, computing, and management categories are discussed below.

\medskip \noindent \textbf{Technical capabilities.} DOE has long experience with RF systems for its hadron and electron accelerators. Hardware for manipulating RF modes to efficiently couple sources to accelerator waveguides and cavities has much in common with the matching optics used for radio telescope receivers. High channel-count, fast RF digitization and processing is also used extensively in control and beam diagnostics.
 Large accelerators such as LCLS-II can include over
a thousand channels of RF front ends and high-performance digitizers
connected to a distributed data network.  Although optimization of dynamic range, bandwidth, and noise characteristics differ from those needed for the $21\,$cm experiment, many commonalities between the designs remain.

Data acquisition systems at large HEP and photon science experiments generate
enormous volumes of digital data that must be transported over networks that
may comprise tens of thousands of high-speed links. Data transport, real-time
processing, and interface to commodity server farms requires DAQ developers to
have specialized expertise in the most modern microelectronics families
(ASICs, FPGAs, optical transceivers, etc.) and to be aware of rapidly
advancing trends that open opportunities for greater performance in future
projects. The Front End Link Exchange (FELIX) and global Feature Extractor
(gFEX) platforms being developed for the ATLAS experiment are examples of
state-of-the-art hardware coming out of the DOE labs; evolved versions of such
platforms can find very direct applications in real-time $21\,$cm signal
processing.

Table \ref{tab:datarates} shows a comparison of data rates in some current and future
experiments drawn from HEP, photon science, and radio astronomy. Data rates in
the $21\,$cm \stagetwo\  experiment proposed here, although challenging, are not out
of the range of some of the more ambitious projects shown.

\begin{table}
  \centering
  \renewcommand{\arraystretch}{1.4}
  \begin{tabular}{|l|l|l|l|l|}\hline
\multirow{2}{*}{Experiment}  & Data rate &  \multirow{2}{*}{~Year~~~} & \multirow{2}{*}{Note} & \multirow{2}{*}{Ref.} \\
& [GB/s] & & & \\ \hline
VLA & 0.3 & 2013 & Resident Shared Risk Observing mode & \cite{VLAdr}\\
ALMA & 1.8 & 2021 & Overall & \cite{ALMAdr}\\
LHC & 25 & 2018 & Average rate, all 4 experiments after triggering~~ & \cite{LHCdr}\\
LSST & 6.4 & 2022 & Peak rate & \cite{LSSTdr}\\
LCLS & 10 & 2009 & CXI instrument & \cite{Thayer2017}\\
LCLS-II & 320 & 2027 & High frame-rate scattering detector & \cite{LCLSIIdr}\\
XFEL & 13 & 2017 & 2D area detector & \cite{XFELdr}\\
SKA1 & 8,500 & 2022 & Overall & \cite{SKAIdr}\\
CHIME & 13,000 & 2017 & Input to F-engine & \cite{CHIMEdr}\\
\textbf{21cm \stagetwo}~ & \textbf{655,000} & 2030 & \textbf{Input to F-engine} & \\
\hline
  \end{tabular}
\caption{Rates for current and proposed data-intensive experiments, drawn from HEP, photon science, and radio astronomy. The data rates for the \stagetwo experiment going into the F-engine are expected to be manageable within the time-frame of the experiment. \label{tab:datarates}
}
\end{table}

\medskip \noindent  \textbf{Computing capabilities.}
All stages of developing this experiment will require the
involvement of large computing facilities. The full system simulation as
well as actual data processing will require high-performance computing
and efficient storage, handling and processing of data volumes in the petabyte
range. This can be efficiently addressed through existing and planned
infrastructure facilities within the DOE laboratory complex that will
also drive new developments in network connectivity between DOE
sites. DOE runs NERSC, one of the world's largest high-performance
computing systems, ALCF and ORCF (limited access) and has put significant investment into exascale
computing across all centers. It also
hosts two CERN Tier-1 data centers.

In addition to challenges presented by the data volumes alone, there
are massive algorithmic challenges that can be efficiently addressed
using existing DOE structures present within Advanced Science Computer
Research (ASCR) and SciDAC. On the simulation side these includes
running large simulations of the universe. On the data analysis sides,
the calibration problems and foreground removal problems can be recast
in terms of large-scale linear solvers, error analysis, kernel estimation,
machine learning, etc.  These problems will benefit from developments
in the current exascale initiative and work that has been done on
hybrid compute architectures that can be particularly efficient ith
large data rates.

\medskip \noindent \textbf{Management capabilities.}
A $21\,$cm \stagetwo\  experiment will need to follow organizational models similar to
those that have evolved in DOE's other recent HEP programs. These may include
coordination with other agencies and/or international partners, setting up
scientific collaborations and a formal structure responsible for executing the
project plan, and arranging for appropriate levels of oversight. During the
construction phase, test systems for quality assurance and metrology will be
essential for mass-produced components to meet performance requirements.
Predecessor projects such as US-ATLAS/CMS silicon detector modules, LSST focal
plane raft towers, and CMB-Stage 4 detectors and readout will provide
useful models and lessons. Finally, DOE has experience in organizing
collaboration-wide scientific activities to generate high-fidelity simulations
of system performance. The LSST Dark Energy Science Collaboration's Data
Challenges are a recent example. As stated earlier, it will be absolutely
essential to perform end-to-end simulations for a $21\,$cm \stagetwo\  experiment.

\medskip \noindent \textbf{Current DOE laboratory efforts.} There are currently
several small path-finder efforts at various labs not directly funded by DOE HEP.

At BNL, a small test-bed experiment, BMX, has been set-up operating at
1.1-1.5GHz. It has been taking data since Fall 2017 in single dish
mode and was upgraded to a 4 dish interferometer in the Summer
2019.  The results are promising despite the experiment being situated
at the lab site, which is an extremely poor location in terms of
RFI. Early science results include characterization of out-of-band
emission from global navigation satellite services that will act as a
potential systematic for low-redshift $21\,$cm experiments. As a
test-bed, the system will be used to test various approaches towards
beam and gain calibration and to gather on-sky data from early
digitization prototypes. It will thus continue to provide a convenient
bridge between laboratory testing and a test deployment on a real
radio telescope which often involves significant travel costs and
limited time allocation.

The Fermilab Theoretical Astrophysics Group has been closely involved
with $21\,$cm intensity mapping for the past decade.  Early work
included forecasting and technical design studies for $21\,$cm arrays
\cite{2010ApJ...721..164S} and development of analysis techniques
\cite{Shaw:2014vy,Shaw:2013tb}.  Currently, with NSF support, the
group hosts the Tianlai Analysis Center (TAC), which analyzes data for
the Tianlai instrument in China.  The current, ``Pathfinder" version of
Tianlai includes an array of 16, 6-meter diameter dishes and 3, 15 m x
40 m cylinder telescopes operating in the 650-1420 MHz range and acts
as a useful test-bed instrument for future efforts \cite{Das2018}.
Near-term goals include determining the optimal design of future
arrays (cylinders, dishes or both), and detecting HI at low and high
redshift ($z \sim$ 0.15 and 1.0).  The effort includes data storage,
calibration, RFI removal, data quality assessment, mapmaking, power
spectrum analysis, and development and testing of the Tianlai analysis
pipeline.  These tasks are partly enabled by the substantial computing
resources at Fermilab's Scientific Computing Division.

\clearpage

\section{\texorpdfstring{$21\,$cm}{21cm} measurements beyond redshift \texorpdfstring{$z\sim6$}{z of 6}}
\label{sec:intro-highz}

In this document we have so far talked about the $21\,$cm intensity
mapping as mapping of the aggregate emission from many unresolved
galaxies. However, this is a correct picture only in the universe at
redshifts lower than $z\lesssim 6$, where the universe is mostly
ionized with a few pockets of neutral hydrogen residing in
galaxies. 

Going to earlier times and higher redshifts, we encounter two distinct
regimes. The epoch between $z\sim30$ and $z\sim6$ is also known as the
Epoch of Reionization.  During these periods, first-generation stars
and galaxies were formed and begun the process of reionizing the
universe. This process is highly non-linear and driven by astrophysics
rather than cosmology.  This epoch experimentally interesting, because
the signal is boosted by large region of completely ionized
``bubbles'' residing in sea of otherwise largely neutral hydrogen.
Therefore, significant effort is dedicated to measuring this regime
and we  describe it in the Section \ref{sec:cosmic-dawn-epoch}.

Going even further, to redshift higher than $z\gtrsim 30$, we see the
universe as it was before the formation of the first stars. The pristine
hydrogen, untainted by the messy start and galaxy formation promises
the ultimate frontier, but it is experimentally daunting as we discuss
in Section \ref{sec:dark-ages}.

\subsection{Cosmic Dawn and Epoch of Reionization}
\label{sec:cosmic-dawn-epoch}

$21\,$cm techniques have been used for studying the Cosmic Dawn and
the Epoch of Reionization (EoR). A number of experiments such as
HERA~\cite{2017PASP..129d5001D}, PAPER~\cite{2010AJ....139.1468P},
LOFAR~\cite{2013A&A...556A...2V}, MWA~\cite{2013PASA...30....7T}, and
GMRT~\cite{2013MNRAS.433..639P} are seeking to make the first
measurements of how the first luminous objects affected the
large-scale distribution and ionization state of hydrogen. While these
efforts target a currently unexplored phase of galaxy formation, they
do not have P5 goals as primary science and thus we are not proposing
these for consideration by the DOE. However, they do have indirect
relevance to the goals outlined in this roadmap, for two
reasons. First, these experiments may detect signatures of exotic
physics that are relevant to P5 goals, provided these signatures
cannot be easily be explained by $\Lambda$CDM, even when allowing for
extreme astrophysical scenarios.  Second, these experiments face many
of the same technical challenges as the experiments proposed in this
roadmap, and thus any breakthroughs on either side in instrumentation,
observation, or data analysis will be mutually beneficial.

A prime example of possible exotic physics would be the recent results from the EDGES
experiment \cite{2018Natur.555...67B}. EDGES has claimed a first detection of a large dip in spectral energy distribution of the cosmic radio monopole at
around $78$MHz, corresponding to $z \sim 17$ if this is due to the $21\,\textrm{cm}$ line.
While such an absorption feature is predicted by most theories of Cosmic Dawn, the dip
measured by EDGES is anomalously large, implying hydrogen gas that is considerably cooler
than is allowed by $\Lambda$CDM or an additional source of background besides the CMB \cite{2018arXiv180207432F}. This discovery has yet to be confirmed, and there are some  concerns related to the foreground modeling \cite{Hills:2018vyr}. However, if true, it would present a remarkable measurement which has already generated considerable interest within the high-energy physics community.
The EDGES result, if validated, could potentially point to the first hints of interactions
between baryons and the dark sector \cite{2018arXiv180303245F,2018arXiv180303091B,2018arXiv180302804B,2018Natur.555...71B,2018arXiv180309734S,2018arXiv180310671H,2018arXiv180508794N}, or place constraints on the primordial power spectrum \cite{2018arXiv180511806Y}, relic neutrino decays \cite{2018arXiv180511717C}, dark energy \cite{2018arXiv180306944C,2018arXiv180307555H}, axions \cite{2018arXiv180410378M, 2018arXiv180504426H,2018arXiv180505577S}, interactions between dark matter and dark energy \cite{2018arXiv180511210W}, dark matter annihilations \cite{2018arXiv180303629D,2018arXiv180305803Y,2018arXiv180309398C},
decaying dark matter \cite{2018arXiv180309390C}, primordial black holes \cite{2018arXiv180309390C,2018arXiv180309697H}, fuzzy dark matter \cite{2018arXiv180501253L}, and warm dark matter \cite{2018arXiv180308039S,2018Natur.555...71B,2018arXiv180500021S}.

Fundamentally, a $21\,\textrm{cm}$ experiment aims to make large, three-dimensional maps
of the distribution of hydrogen, regardless of the epoch it is probing. Thus, breakthroughs
with Cosmic Dawn and EoR experiments also represent breakthroughs for any experiment
described within this roadmap. In this respect, discoveries like the EDGES result  could potentially be  significant steps forward. A confirmed EDGES detection would be analogous
to the first measurements of the CMB blackbody spectrum, while follow-up measurements of
the spatial fluctuations of the $21\,\textrm{cm}$ line would be analogous to the first measurements
of CMB anisotropies. Just as with the CMB, such measurements would herald the beginning
of a new standard probe of cosmology.

\subsection{Dark Ages}
\label{sec:dark-ages}

After recombination\footnote{Recombination is really a misnomer for
  this epoch since protons and electrons were never combined
  before. Primordial combination might be a more appropriate
  phrase.} of hydrogen, when the Cosmic Microwave Background (CMB) was
created at redshifts around $z\sim 1150$, the universe was completely
neutral, with neutral hydrogen the dominant component. As matter continued
to cluster in the post-recombination universe, peaks in the matter density
were enhanced and eventually led to the formation of the first generation of
stars and galaxies, which emitted radiation capable of reionizing the
ambient neutral hydrogen. Between
recombination and the formation of the first stars, there is a high-redshift epoch that
is ideal for the cosmological mapping of density fluctuations through
$21\,\textrm{cm}$ intensity mapping, during which hydrogen is neutral and
and traces the overall matter distribution. This epoch is generally referred to as the Dark Ages. 

\begin{figure}[t]
  \centering
  \includegraphics[width=\linewidth]{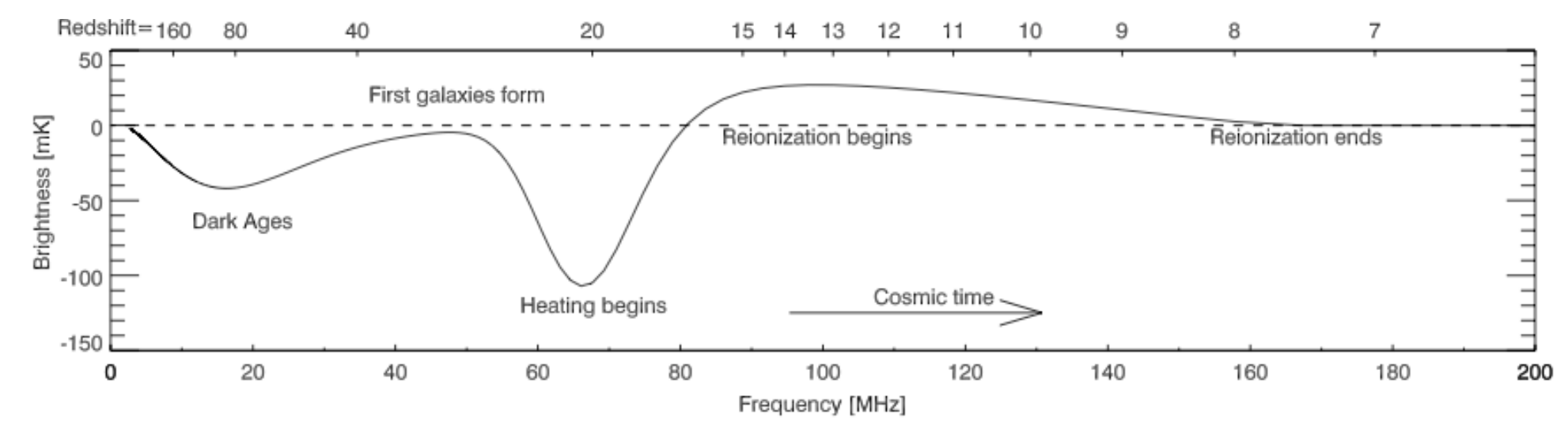}
  \caption{The $21\,$cm monopole intensity through cosmic times (plot
    adapted from \cite{2013PhRvD..87d3002L}).}
  \label{fig:mono}
\end{figure}

Several physical details prevent the mapping of density fluctuations
over the entire redshift range from recombination and
reionization. For instance, when $ z \gtrsim 150$, residual free
electrons from recombination provide a coupling between the CMB and
the temperature of the hydrogen gas through Compton scattering. In
turn, collisions drive the spin temperature (which quantifies the
relative number of hydrogen atoms in the ground versus the excited
hyperfine state) to the gas temperature.  With the CMB temperature in
equilibrium with the spin temperature, there is no net absorption or
emission from the $21\,\textrm{cm}$ line, and therefore no signal to
observe. At $z \sim 150$, Compton scattering becomes inefficient. The
spin temperature and the gas temperature remain coupled to one
another, but together decouple from the CMB temperature. The gas then
cools adiabatically as $(1+z)^{-2}$, in contrast with the CMB's
cooling as a $(1+z)^{-1}$, which results in a net absorption
signal. This continues until $z \sim 30$, at which point the neutral
hydrogen is sufficiently dilute that the collisional coupling between
the gas temperature and the spin temperature become
ineffective. Direct absorption of emission of $21\,\textrm{cm}$
photons then couples the CMB temperature to the spin temperature once
again, and the signal disappears.  The observed brightness temperature
of the 21\,cm signal as a function of redshift is shown in
Figure~\ref{fig:mono}.

A redshift window in the range
$30 \lesssim z \lesssim 150$ could potentially be used for $21\,\textrm{cm}$ intensity
mapping and would provide large-scale maps of pristine density
fluctuations. There are several advantages to doing so. First,
the regime is too high in redshift for the first luminous objects
to have formed yet, and therefore the signal is driven by cosmology
rather than astrophysics. Second, the signal is not Silk damped,
and thus density perturbations can in principle be mapped to
extremely small scales (with perhaps the Jeans scale being the only limitation). Third, these small-scale structures are
still in the linear regime at such redshifts, making theoretical
modeling efforts considerably simpler than analogous efforts
for $z\sim 0$ galaxy surveys. Finally, the volume of our observable
Universe that falls in the range $30 \lesssim z \lesssim 150$ is
 substantial, leading to exquisite statistical errors on parameters.

\subsubsection{A new window into the Universe}

In the CMB, well-understood linear processes are sufficient to relate
observed anisotropies in temperature and polarization to energy
density perturbations generated during the early universe. This is
what makes the CMB such a powerful probe of fundamental physics,
limited mainly by diffusion damping~\cite{SilkDamping} that erases
anisotropies (and therefore primordial information) on small
scales. On the other hand, lower-redshift large scale structure in
principle offers many more accessible modes, but a large portion of
these modes is affected by nonlinear processes that are difficult to
model. These nonlinearities are less severe at higher redshift: in
particular, before the first collapsed objects formed at $z\sim 30$,
the limiting scale is the Jeans scale, $k_{\rm J} \sim 300$
Mpc$^{-1}$~\cite{AliHaimoud:2013hpa}. Since the number of linear modes
scales as the cube of the maximum linear wavenumber, observations at
this epoch hold great promise for increasing our knowledge of
fundamental physics.

The only observable available to us during this epoch is the $21\,$cm
hyperfine transition of neutral hydrogen.\footnote{There is also a
  hyperfine transition in deuterium nuclei, corresponding to photons
  with wavelength $92\,$cm. In principle, this is observable with the same
  interferometers designed for $21\,$cm, and would yield a pristine
  measurement of the primordial deuterium abundance, but will be a
  much more challenging observation than
  $21\,$cm~\cite{Sigurdson:2005mp}.} The theory of the high-redshift $21\,$cm
signal is very well understood~\cite{Loeb:2003ya,Lewis:2007kz}, and
for most purposes is well described by linear perturbation
theory~\cite{AliHaimoud:2013hpa}. From a practical standpoint, the
signal, which is in absorption against the CMB back-light, will be
very hard to observe for many reasons that are similar to those that
hinder the detection of $21\,$cm emission at lower redshifts. In addition,
a $21\,$cm photon originating at these very high redshifts will redshift
into the low MHz wavebands, which will be hard to observe from the
ground due to reflection by the ionosphere. It is estimated that this
limitation becomes significant for $z \gtrsim 45$
($\nu\lesssim 30$MHz~\cite{2009NewAR..53....1J}), and any signal beyond that would
require an experiment outside of the ionosphere, such as in space, or,
as has been proposed in Refs.~\cite{BurnsMoon,2009NewAR..53....1J, DEX}, on the far side of the moon. 

This certainly implies that any measurement will be very far in the future. For this reason, we will not suggest a specific experiment (which would come with a unique set of limitations), but instead remark upon the general potential of an experiment targeting these observations, that would inevitably build on the progress made with lower-redshift detections. Simply put, the high-redshift $21\,$cm signal will provide a three dimensional window into the linear Universe, providing access to of order $10^{10}$ more modes than the CMB\footnote{Assuming $10^4$ independent redshift slices in this redshift range, each with for $\ell_{\rm max} = 10^6\simeq \ell_{\rm Jeans}$ \cite{Loeb:2003ya}.}. This tremendous amount of statistical power makes $21\,$cm measurements from the Dark Ages the ultimate probe of the conditions in the early Universe. Exquisite constraints could be expected on many quantities of interest~\cite{Loeb:2003ya}, such as the scalar spectral index~\cite{Adshead:2010mc} and primordial non-Gaussianities \cite{Pillepich:2006fj,2015PhRvD..92h3508M,2017JCAP...03..050M}. 

Before we present a unique science target, let us briefly highlight two observables discussed earlier, namely primordial features (Section~\ref{sec:feat-prim-power}) and non-Gaussianities (Section~\ref{sec:prim-non-gauss}), that a probe of the Dark Ages could significantly improve.

The detectability of features at high reshifts depends critically on the amplitude,
frequency and scale-location of the features, as well as the angular
and redshift resolution of the experiment. Forecasts show~\cite{Chen:2016zuu} that a cosmic
variance limited $21\,$cm experiment measuring fluctuations in the
redshift range $30\leq z \leq 100$ with a 0.01-MHz bandwidth and
sub-arcminute angular resolution could potentially improve bounds by
several orders of magnitude for most features compared to current
Planck bounds. At the same time, $21\,$cm tomography also opens up a
unique window into features that are located on very small scales
($k\gg 1$ Mpc$^{-1}$).

Besides features in the power spectrum, the same physics generally
produces features in all primordial correlation functions. The $21\,$cm
field as a probe of non-Gaussianities, and the bispectrum in particular,
 has been explored in Ref.~\cite{Xu:2016kwz}. Of
particular interest is the possible detection of massive particles in
the early Universe. Heavy particles with higher spin can leave
distinct features on higher-order correlation function~\cite{Baumann:2017jvh,Lee:2016vti}. The
signal is predicted to be very small, but a detection would present the
first evidence for a mass hierarchy as predicted by string theory~\cite{Arkani-Hamed:2015bza}. Because of the smallness of the signal, $21\,$cm has been
suggested~\cite{2017JCAP...03..050M} to provide the only realistic
observable to constrain the presence of these particles. We refer to
Ref.~\cite{2017JCAP...03..050M} for details of the models that could
potentially be observed with $21\,$cm.

Now, we will present a single example that is rather unique, concerning the potential signatures of primordial gravitational waves in fluctuations of the observed $21\,$cm intensity. We describe these signatures below, and provide estimates for their constraining power on the amplitude of gravitational wave power left over from the early Universe.

\subsubsection{Gravitational tensor modes}

One of the holiest grails in our attempt to understand the physics of the early Universe is the possible detection of primordial gravitational waves. These can be generated by the same early-universe process that generates the seeds for the (scalar) density fluctuations that we observe in the CMB and large scale structure. Within the paradigm of inflation, the expected level of primordial gravitational waves generated during inflation is measured with respect to the production of scalar fluctuations by a relation of the two primordial power spectra:
\bea
P_{\zeta} &=& A_s k^{-3}\left(\frac{k}{k_*}\right)^{n_s-1}\ , \\
P_h &=& r A_s k^{-3}\left(\frac{k}{k_*}\right)^{n_t} \ .
\eea
In single-field slow-roll inflation, some of the parameters above are related by $n_s = 1-2\eta -6\epsilon$, $r = 16 \epsilon$, and $n_t = -r/8$. Here~$\eta$ and~$\epsilon$ are two slow-roll parameters, which are proportional to the second and first derivative of the scalar potential, respectively, and are required to be much less than unity for inflation to last a sufficient time to solve the horizon and flatness problems~\cite{PhysRevD.23.347}. In more complicated models, including those with multiple fields, deviations from slow-roll, and non-canonical kinetics, these relations will be altered, pick up additional degrees of freedom, or break altogether. The relation between the scale dependence and the amplitude of primordial waves is particularly interesting. A deviation from a red spectrum would indicate a violation of the null energy condition, and suggest the spectrum was not generated from the vacuum (see e.g. \cite{Cook:2011hg,Namba:2015gja}), or could rule out inflation as the source of gravitational waves~\cite{Khoury:2001wf}.

Current attempts using the B-mode polarization signal in the CMB aim to detect $r$ as low as $10^{-3}$ \cite{Abazajian:2016yjj}, providing an interesting science target in terms of the field excursion during inflation~\cite{Lyth:1996im}. Unfortunately, it is quite possible given the nature of $r$, which effectively describes the energy scale of inflation, that the actual level of primordial gravitational waves is orders of magnitude below $10^{-3}$. Measurements beyond this level will be difficult using CMB B-modes, mostly due to B-modes generated through lensing of E-modes, which obscure primordial B-modes at the level of $10^{-2}$ for ground-based observations. Delensing methods can mitigate a large fraction, but this becomes increasingly hard for smaller values of $r$. Furthermore, for very low values of $r$, patchy screening and scattering of CMB photons around reionization can generate B-modes which will be hard to disentangle (although the maps could in principle be de-screened~\cite{Meerburg:2017lfh}) from primary B-modes. Many other probes of primordial gravitational waves face significant challenges. For example, direct detection using interferometers (e.g.~LIGO and (E)LISA) is unlikely given the relatively small scales probed by such experiments~\cite{Meerburg:2015zua}, and methods utilizing the polarized Sunyaev-Zel'dovich effect require very low noise levels in the CMB and an exquisite measurement of  free electrons in the Universe~\cite{2012PhRvD..85l3540A}.

Measurements of large-scale structure during the Dark Ages will be affected by a gravitational wave background in several ways, and observations over a large enough volume have the potential to see these effects at high significance. We will highlight two such effects here: 
\begin{enumerate}
\item {\bf Tidal fossils}: After a large-scale tensor mode enters the horizon, it will induce a specific kind of inhomogeneity into the statistics of the density field, similar to what happens with the tidal field generated by scalar perturbations at second order. While the original tensor mode will decay with time, its imprint on large-scale structure will not, leaving behind a ``fossil" that can be detected at later times using an appropriate estimator~\cite{2010PhRvL.105p1302M,Schmidt:2013gwa,Masui:2017fzw}. The power spectrum of this estimator is directly connected to the primordial tensor power spectrum, and therefore to the tensor-to-scalar ratio, with constraining power scaling with the inverse of the number of observed modes. Ref.~\cite{2010PhRvL.105p1302M} has argued that a Dark Ages survey could use this effect to constrain~$r$ to the $10^{-6}$ level.
\item {\bf Curl lensing}: Like density fluctuations, gravitational waves can affect the paths of photons as they travel through the universe. Unlike density fluctuations, however, gravitational waves generate a curl component of a reconstructed deflection field. The potential of these curl modes as a probe of gravitational waves has been studied e.g.\ in Refs.~\cite{Dodelson:2003bv,Li:2006si,Book:2011dz,Chisari:2014xia,Sheere:2016yqu}. The constraining power of this method also scales with the inverse of the number of modes, and in Ref.~\cite{Book:2011dz} it was argued that in principle a measurement of curl lensing from the Dark Ages could provide a constraint as low as $r=10^{-9}$.
\end{enumerate}
A full treatment of all effects induced due to the presence of large-scale tensor perturbations, including the two effects above, was performed in Refs.~\cite{Schmidt:2012nw,Schmidt:2013gwa}. Observationally, it is not evident that all of these effects can be easily separated. In our forecast below, we will assume that tidal fossils and curl lensing can be distinguished. We hope to report in the near future to what extent these effects can indeed be separated (for example, through bias-hardened estimators, as recently explored in Ref.~\cite{Foreman:2018gnv} for the case of scalar lensing).

We consider a Dark Ages $21\,$cm survey over $30<z<150$, corresponding to a comoving volume of roughly $900\, (h^{-1}{\rm Gpc})^3$. The number of modes is set by the maximum observable wavenumbers along and perpendicular to the line of sight,~$k_{\parallel{\rm max}}$ and~$k_{\perp{\rm max}}$, and we assume that the statistics of these modes are amenable to theoretical predictions at the necessary precision. We assume sufficient frequency resolution to access the Jeans scale in the line-of-sight direction, $k_{\parallel{\rm max}} \sim 300 {\rm Mpc}^{-1}$. In the transverse direction, we map $k_{\perp{\rm max}}$ into the corresponding baseline $b$ that can observe that wavenumber. This mapping is redshift-dependent; for the tidal fossil forecast, we evaluate it at $z=30$ since this is where the signal to noise peaks. For the curl lensing forecast, we split the survey into four equal redshift bins, evaluate the mapping (and any other relevant redshift-dependent quantities) at the midpoint of each bin, and combine the separate forecasts from the different bins. Note that $b$ is not necessarily the longest baseline present in the instrument, but rather the maximum baseline at which all shorter modes are signal-dominated.

For tidal fossils, we adopt the quadratic estimator from Ref.~\cite{Masui:2017fzw}, using their expression for the estimator noise with the survey properties given above. For curl lensing, we use a modification of the formalism from Ref.~\cite{Foreman:2018gnv}, which simply amounts to a change in filters applied to the observed $21\,$cm fluctuations. We ignore nonlinearities in the $21\,$cm field, which will slightly degrade the signal to noise at the longest baselines we consider. The ability to detect lensing is affected by shearing of coordinates at the source redshift by gravitational waves present at that redshift; we incorporate this ``metric shear" in our forecasts, following Ref.~\cite{Dodelson:2003bv}.\footnote{Important differences between our forecast and that of Ref.~\cite{Book:2011dz} include the incorporation of metric shear, which degrades the signal to noise, and the use of a fully 3-dimensional formalism that accounts for correlations caused by long modes along the line of sight.} The curl lensing power spectrum is computed using a modified version of CAMB~\cite{Lewis:1999bs}, and we compute the $21\,$cm brightness temperature power spectrum following Ref.~\cite{2015PhRvD..92h3508M}.

\begin{figure}[t]
  \centering
  \includegraphics[width=0.7\linewidth]{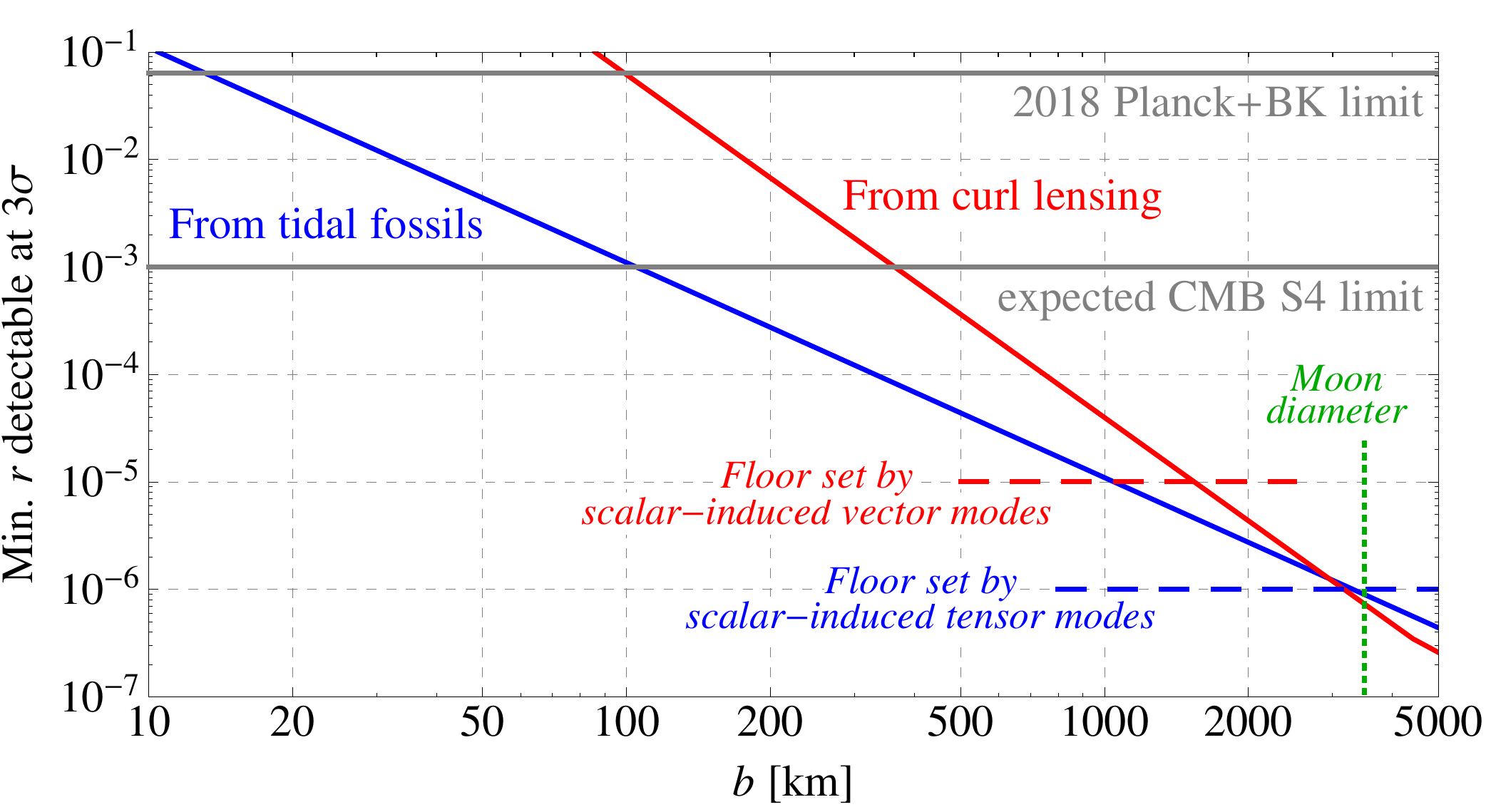}

  \caption{\label{fig:DarkAgesR} The minimum value of the
    tensor-to-scalar ratio $r$ detectable with a Dark Ages $21\,$cm
    survey, as a function of the maximum baseline $b$ for which $21\,$cm
    observations are signal-dominated. Blue and red curves correspond
    to the tidal fossil and curl lensing methods discussed in the main
    text. The corresponding dashed lines indicate floors at which the
    primordial GW signal becomes dominated by the next-strongest
    signal in each method. The horizontal grey lines show the current upper
    limit, $r\leq 0.064$ (95\%~CL), from a combination of Planck 2018 and
    BICEP2/Keck 2014 data~\cite{Akrami:2018odb}, and the expected limit from CMB-S4 
    ($r\lesssim 10^{-3}$~\cite{Abazajian:2016yjj}).
     We find that for
    $b\gtrsim \mathcal{O}(100{\rm km})$, $r$ can be detected at a
    lower level than with CMB-S4, while an
    interferometer covering a large portion of the moon can detect $r$
    as low as~$10^{-6}$. Achieving even a fraction of this precision
    would be challenging for any other known probe of primordial
    GWs. 
  }

\end{figure}

In Fig.~\ref{fig:DarkAgesR}, we plot the minimum value of $r$ detectable at 3$\sigma$ by either method, assuming that primordial gravitational waves are the dominant signal in each case. We have also indicated the levels at which other effects begin to dominate the primordial signal. For curl lensing, vector perturbations generated at second order by primordial scalar perturbations produce the dominant signal if $r\lesssim 10^{-5}$~\cite{Saga:2015apa,Saga:2016cvt}. Contaminants in the tidal fossil estimator have not been extensively investigated, but tensor perturbations generated by second-order scalar couplings have been found to enter other observables at the level of $r\sim 10^{-6}$ (e.g.~\cite{Baumann:2007zm}), so we take this to be the relevant floor.\footnote{These second-order contributions can be exactly computed once the amplitude of scalar perturbations is known, and could then be subtracted from a measurement of tidal fossils or curl lensing to access values of $r$ smaller than the floors we have quoted. However, cosmic variance will prevent us from obtaining sufficiently precise measurements  for this procedure to work. Our forecasts do not include cosmic variance; in other words, for each maximum baseline, we have computed maximum values of $r$ for which the null hypotheses of ``no tidal fossils'' or ``no curl lensing'' could be rejected at $3\sigma$. If $r$ is below either of the quoted floors, a rejection of these null hypothesis will not inform us about the value of $r$.}

We find that an interferometer with baselines of at least a few hundred kilometers would be able to constrain $r$ to the level of $10^{-3}$, equivalent to the target for CMB-S4, with even larger arrays being able to beat this target. Such arrays are clearly a highly ambitious notion, but currently represent the only feasible way to detect primordial gravitational waves at a lower level than CMB-S4. At the extreme limit of feasibility, an array covering a large fraction of the Moon's surface (corresponding to a maximal baseline of $3500\,$km, the Moon's diameter), could in principle detect $r$ as low as $10^{-6}$. Achieving even a fraction of this goal would result in a large scientific payoff, which motivates further research and development in this direction.

\clearpage

\section{Conclusions}
\label{sec:conclusions}

In this white paper, we have provided an overview of $21\,$cm cosmology, and
argued that there is a unique opportunity for the US cosmology community
to take a leading role in this field by beginning to plan for a second-generation experiment.
We reiterate three main reasons for doing so:

\begin{itemize}
\item \textbf{The experiment will address pressing science questions.}
  There have been no major discoveries revealing new physics in the
  past two decades. Collider experiments, while achieving
  important milestones such as direct detection of the Higgs boson,
  have not detected supersymmetry or other signatures that would
  directly indicate new physics beyond the
  standard model. In cosmology, the minimal $w=-1$ $\Lambda$CDM
  model has avoided any definitive observational challenge, while minimal progress has been made to uncover the physics of the early Universe. 
   We are proposing a \stagetwo\  $21\,$cm experiment
  that could advance three possible avenues for finding new physics:
   deviations from
  the standard expansion history and growth of cosmic structure at high redshift, features in the
  primordial power spectrum, and measurements of primordial
  non-Gaussianity. The first item has the potential to directly address
  some pressing dark energy questions, such as the timing of dark energy domination, while
  the second and third items are theoretically well-motivated searches
  that a large $21\,$cm array is particularly suited to address and would present groundbreaking discoveries if detected. In addition
  to these cornerstone measurements, the experiment will open up a
  trove of new scientific capabilities, such as providing a unique source screen for
  gravitational lensing and tidal reconstruction, real-time measurements of the cosmic expansion,
  and identifying or characterizing exotic transient phenomena in the radio.
  Finally,
  a \stagetwo~experiment would constitute a pivotal test ground towards
  the ultimate goal of opening up the cosmic Dark Ages for direct observations.

\item \textbf{Now is the time to do it.} After the current-generation
  flagship dark energy experiments LSST and DESI, there is not an
  obvious path to continue following optical dark energy
  studies. Pivoting to $21\,$cm would allow the US to become a leader in a
  fundamentally new and different cosmological observable. Moore's law
  improvements in the corresponding technology will continue to make
  this possibility attractive and cost-effective in the foreseeable
  future.

\item \textbf{The DOE HEP program is the natural home for this
    experiment.}  As argued in the text, the success of such a
  \stagetwo\ experiment lies in a tightly integrated instrument
  design, calibration and data analysis. The traditional radio
  astronomy projects are designed to be multi-purpose observatories on
  which time is allocated through a PI-driven process and are therefore
  not appropriate for achieving the science goals presented here. On
  the other hand, DOE has a long pedigree in building and managing
  large production programs and scientific communities in large
  HEP-style collaborations. This makes the DOE a natural home for an
  experiment like this. As argued in Section \ref{sec:sciencecase},
  the science case naturally extends beyond dark energy and here other
  agencies will probably join the effort in a mode similar to how LSST
  is being built and operated.

\item \textbf{The US national lab complex has the right expertise.} A
  \stagetwo\ $21\,$cm experiment will be a large experiment requiring
  significant R\&D and a large analysis collaboration, and will have
  significant infrastructural and production
  components. Traditionally, such experiments were done under auspices
  of the DOE as the main mission-driven high-energy physics
  agency. In particular, the DOE brings know-how in RF technology from
  accelerator and light-source facilities, as well as considerable
  expertise in high-performance computing (which is crucial, given the
  potentially enormous data volumes of a \stagetwo\ experiment).

\end{itemize}

In the core of this white paper, Sections~\ref{sec:sciencecase} and~\ref{sec:challenges},
 we have made a case
for a concrete experimental design that is an order of magnitude
larger than the current generation of $21\,$cm experiments. We have provided forecasts
and listed the numerous technical challenges. These first steps
elucidate the work which lies ahead and should progress on three
main fronts:

\begin{itemize}
\item \textbf{Strengthen the science case.} More work needs to be done
  to strengthen the science case. All science forecasts should be performed
  with the same forecasting code that will use a concrete observing
  strategy and baseline distributions rather than idealized
  approximations. Special emphasis must be paid to the modelling of
  instrumental \emph{systematics} to push beyond forecasts that
  assume all measurements are thermal noise limited beyond some
  simple (though conservative) data cuts to deal with foregrounds.
  These detailed forecasts should be used to optimize the design and
  understand the pros and cons of
  different choices for array parameters. The
  full scientific implications of specific measurements,
  such as lensing and tidal reconstruction, as well as synergies with
  other probes and planned surveys, should be better
  understood. Moreover, alternative avenues for recovering information lost to
  foreground should be explored.

\item \textbf{Develop a robust science traceability matrix.} A well
  documented flow-down from science requirements to instrument
  properties and key performance parameters is a necessary ingredient
  of a successful project. First steps for the \stagetwo\ project have
  already been performed in a basic traceability matrix presented
  in~\cite{PUMAAPC}. This work will be resolved in more detail as the
  project approaches reality.
  
\item \textbf{Research and develop hardware and calibration systems.}
  In Section~\ref{sec:challenges}, we have outlined a number of
  developments that must occur before a \stagetwo\ experiment. Some of
  them will improve the systematics, and some of them simply control
  the cost and reliability of such a large experiment. Some of these
  developments can be designed and tested in laboratory environments,
  but some will have to employ either $21\,$cm test-beds, such as the
  BMX experiment at BNL or actual \stageone\ experiments. These
  developments need to start as soon as possible in order to to be able
  to converge on an actual design in time. Some of the systematic
  budgets will have to be distributed between hardware, calibration
  and data analysis -- what is the most efficient and robust way to
  achieve this?

\item \textbf{Fully understand implications of \stageone\ experiments.}
  \stageone\ experiments will provide invaluable experience that
  should be absorbed. Have they achieved not just the primary
  scientific goals, but also the expected noise performance and
  control of systematics? What were the dominant issues? On this
  front, one should take advantage of the considerable US presence
  in $21\,$cm experiments targeting the Cosmic Dawn and reionization. While
  the scientific output of these experiments lies beyond the DOE purview,
  the resulting lessons in hardware and data analysis are directly
  transferable to our proposed \stagetwo\ experiment.

\item \textbf{Ensure that programmatic aspects are solid.} The writing
  of this white paper helped to generate a kernel collaboration and
  identify core issues. The next steps are submission to the Astronomy
  and Astrophysics Decadal Survey (cf.~\cite{PUMAAPC}) and later to the 
  Snowmass and P5 processes.
  
\end{itemize}

This whitepaper is the first step on a path towards harnessing the
considerable power of $21\,$cm cosmology. We hope you have enjoyed
reading it as much as we have enjoyed writing it.

\acknowledgments

We thank Chris Carilli, Kyle Dawson and Matt Dobbs for reviewing a
draft document and providing many useful comments.  We thank Joel
Meyers for providing the CMB-S4 noise computation used in
Section~\ref{sec:lensing}. BNL scientists acknowledge generous support
of BNL LDRD program which enabled work presented in this
whitepaper. DM acknowledges support from the Senior Kavli Institute
Fellowships at the University of Cambridge and from the Netherlands
organization for scientific research (NWO) VIDI grant (dossier
639.042.730). AO acknowledges support from the INFN grant PD 51 INDARK. 
ASt acknowledges support from the Fermi Research Alliance, LLC, under 
DOE Contract No.~DE-AC02-07CH11359. BW~gratefully acknowledges support 
by the Marvin L.\ Goldberger Membership at the Institute for Advanced 
Study, from NSF~Grant PHY-1820775 and the Simons Foundation.

We acknowledge the use of the ATNF pulsar catalogue located at
\url{http://www.atnf.csiro.au/research/pulsar/psrcat/}.

\clearpage

\appendix
\suppressfloats


\section*{Appendices}

\section{Counting linear modes}
\label{app:forecasts-modes}

\begin{figure}
   \centering
   \includegraphics[width=0.45\linewidth]{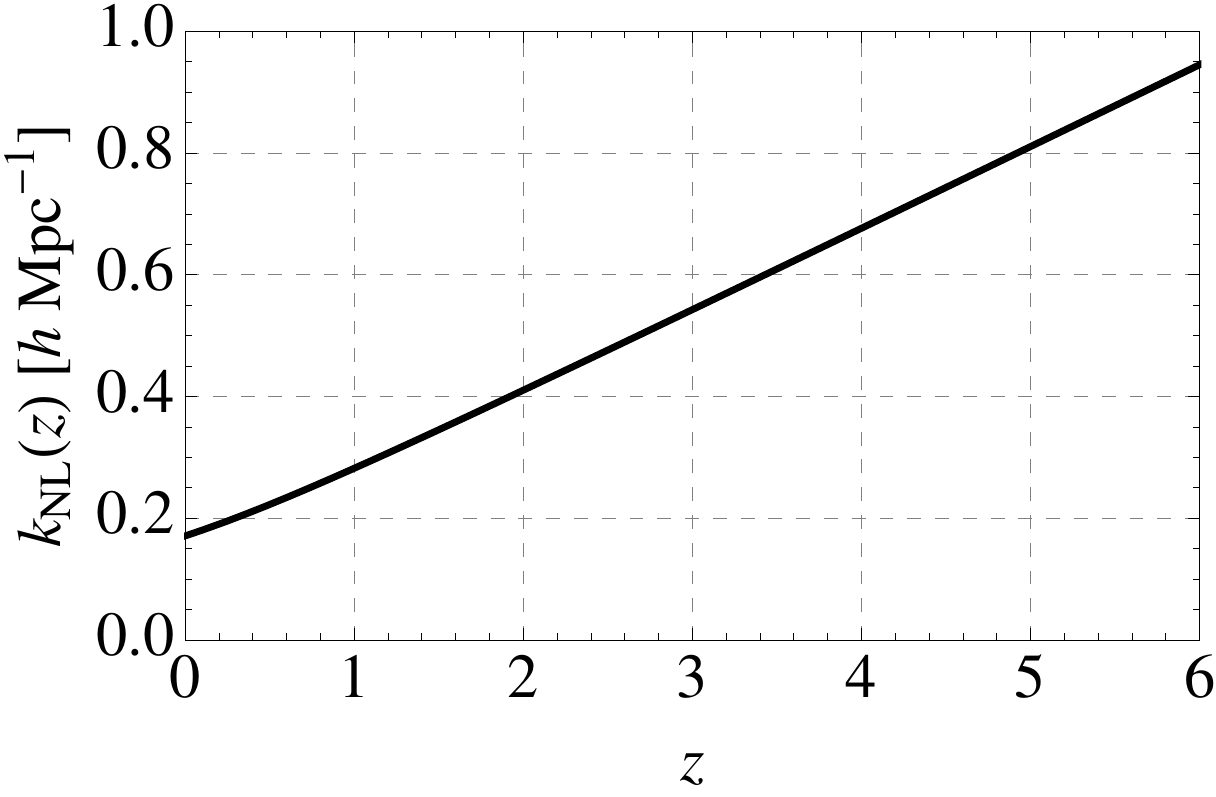}
   \caption{Numerical results based on our definition of the nonlinear scale $k_{\rm NL}(z)$ (see main text), which is an estimate for where the statistics of modes with $k<k_{\rm NL}(z)$ can be predicted with few-percent precision.}
   \label{fig:knlz}
 \end{figure}

A mode of the density field is classified as ``linear" if its wavenumber $k$ falls below some (redshift-dependent) ``nonlinear scale" $k_{\rm NL}(z)$, typically defined as the scale at which the variance of the density field becomes order unity. In this document, we use a rather stricter definition, taking $k_{\rm NL}(z)$ to be the scale below which we expect to be able to predict the measured clustering statistics {\it at the few-percent level}. A conservative estimate of this scale can be obtained from the rms displacement $\Sigma$ in the Zel'dovich approximation:
\begin{equation}
k_{\rm NL}(z) \approx \Sigma(z)^{-1} = \left[ \frac{1}{6\pi^2} \int_0^\infty dk\, P_{\rm lin}(k,z)  \right]^{-1/2} \ .
\end{equation}
We show the associated $k_{\rm NL}(z)$ curve in Figure~\ref{fig:knlz}. Note that this is the scale we estimate for the validity of one-loop perturbation theory; calculations carried out to higher order (e.g.~\cite{Foreman:2015lca}) indicate that higher values of $k_{\rm NL}(z)$ may be achievable, which would imply a substantial increase in the number of linear modes, but further work will be required before these calculations are ready to apply to data.

The cumulative number of linear modes below redshift $z_{\rm max}$ is given by (e.g.~\cite{Ma:2015jua})
\begin{equation}
N_{\rm modes} = \frac{1}{(2\pi)^3} \int_0^{z_{\rm max}} dz \frac{dV}{dz}
	\int_{k_{\rm min}}^{k_{\rm NL}(z)} d^3\boldsymbol{k}
	\approx \frac{2}{3\pi} \int_0^{z_{\rm max}} dz\,
	 \chi(z)^2 \frac{d\chi}{dz} k_{\rm NL}(z)^3\ .
\end{equation}
In the second equality, we have taken $k_{\rm min} =0$ (which has a
negligible effect on the results) and used $dV/dz = 4\pi \chi(z)^2
d\chi/dz$, where $\chi(z)$ is the comoving distance to redshift
$z$. In the presence of a foreground wedge (Appendix \ref{app:wedge}), we multiply the integrand above by the factor
\begin{equation}
\Theta\!\left( k^2\mu^2 
- k^2(1-\mu)^2\left[ \frac{\chi(z)H(z)}{c(1+z)} \sin(\theta_w) \right]^2 \right)\ ,
\end{equation}
where $\Theta(\cdot)$ is the Heaviside function and $\mu=\hat{k}\cdot\hat{z}$.

\section{Assumptions about the \texorpdfstring{$21\,$cm}{21cm} signal}
\label{app:forecasts-signal}

The $21\,$cm brightness temperature is assumed to be 
\begin{equation}
T_{\rm b} = 188 \ {\rm mK} \ h  \ (1+z)^2 \left(H(z)/H_0\right)^{-1}\times
  \left(4\times10^{-4} (1+z)^{0.6} \right),
  \label{eq:21cm_brightness}
\end{equation}
where the expression in the last bracket approximates the cosmic evolution
of $\Omega_{\rm HI}$. This is consistent with
\cite{2017arXiv170907893O} and other recent literature \cite{2019MNRAS.486.5124O,2019arXiv190710375H}. For derivation
of the brightness temperature, see e.g. \cite{2013MNRAS.434.1239B}. We
have in addition assumed evolution of cosmic $\Omega_{\rm HI}$ from
\cite{2015MNRAS.452..217C}.

The total power-spectrum signal observed by the radio interferometer
is approximately given by 
\begin{equation}
  P (\vk) = T_{\rm b}^2[(b + f\mu^2)^2 P(k) + P_{\rm SN}] + P_{\rm N},
\end{equation}
where the first term is the large-scale power spectrum modeled using
linear biasing and redshift-space distortions, $P_{\rm SN}$ is the
shot-noise contribution from halos making up the neutral hydrogen
signal (and usually irrelevant) and $P_{\rm N}$ is the noise coming from the finite system
temperature of the instrument (see App.~\ref{app:forecasts-noise}).

\begin{figure}
  \centering

  \begin{tabular}{cc}
  \includegraphics[width=0.49\linewidth]{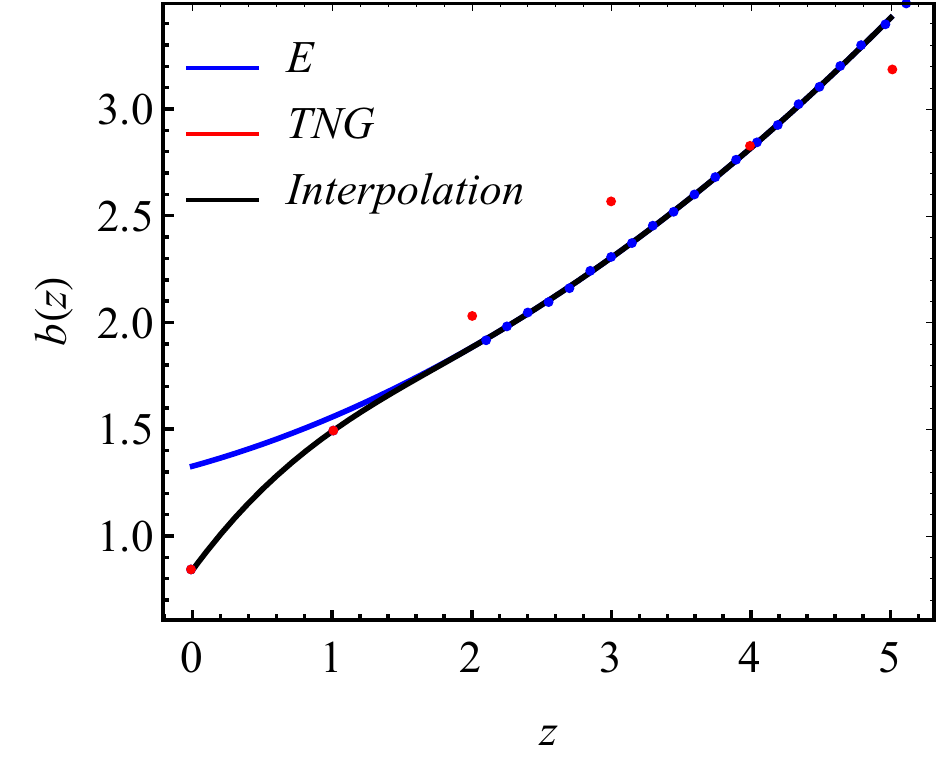}&
  \includegraphics[width=0.49\linewidth]{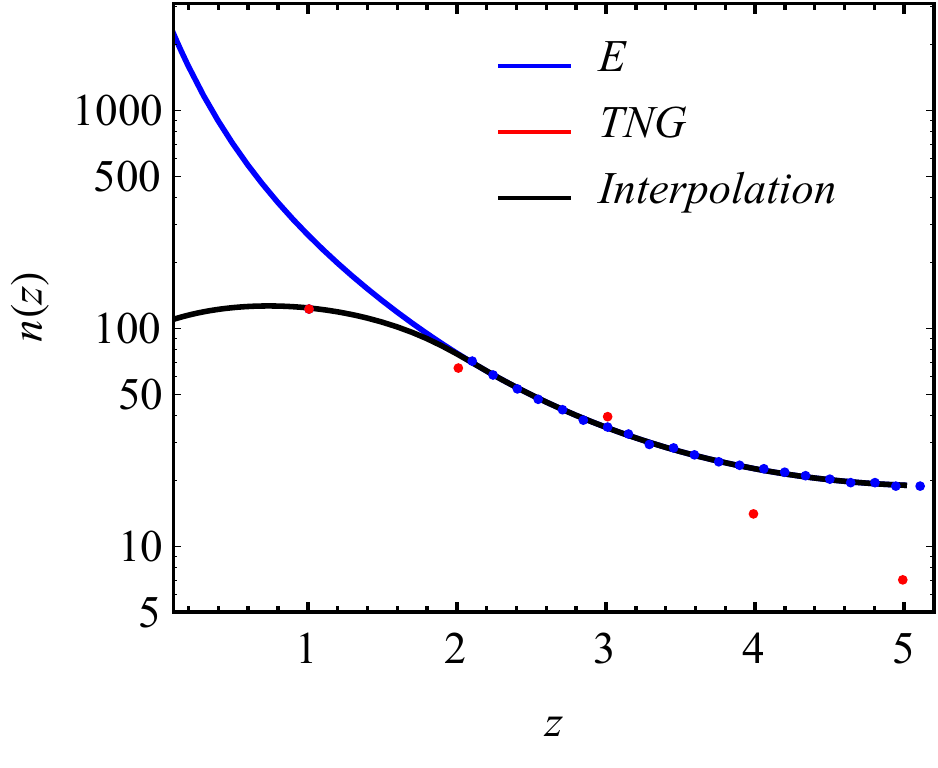}\\
  \end{tabular}
  \label{fig:tnginterp}
  \caption{Interpolation of bias values (left) and shot noise (right)
    between the halo model at high redshift and simulation results at
    low redshift. See text for discussion.}
\end{figure}

For the neutral hydrogen large-scale bias and shot-noise, we used results
from \cite{2017MNRAS.471.1788C} at redshifts beyond $z\sim 2$,
interpolating to results from the Illustris simulation
\cite{2018ApJ...866..135V}. Using the Illustris simulation helps in understanding (and, in future, constraining) the details of the physics ingredients that lead to the observed HI abundances and clustering, and also enables comparison with halo model tools. While the Illustris TNG
is also likely to be an imperfect description, it is the best simulation we currently have.
Shot-noise is also highly uncertain, but is also very
sub-dominant and does not significantly affect results. This
interpolation is illustrated in the Figure \ref{fig:tnginterp}.

At the high redshifts considered here, the linear bias assumption
should be a decent approximation down to considerably smaller scales
than for galaxies at lower redshift. Following
\cite{2014JCAP...05..023F}, we assume an \emph{effective} maximum
wave-number $k_{\rm max,eff}=0.4\,h\,{\rm
  Mpc}^{-1}$. The idea is that in practice one will fit the data to
somewhat larger $k$, which would allow one to constrain and
marginalise beyond-linear order bias parameters.  We assume a Planck 2015 best-fit cosmology, an
assumption that should not affect the results in any significant way.

\section{Foreground filtering and foreground wedge considerations}
\label{app:wedge}

\begin{figure}
  \centering
  \includegraphics[width=0.8\linewidth]{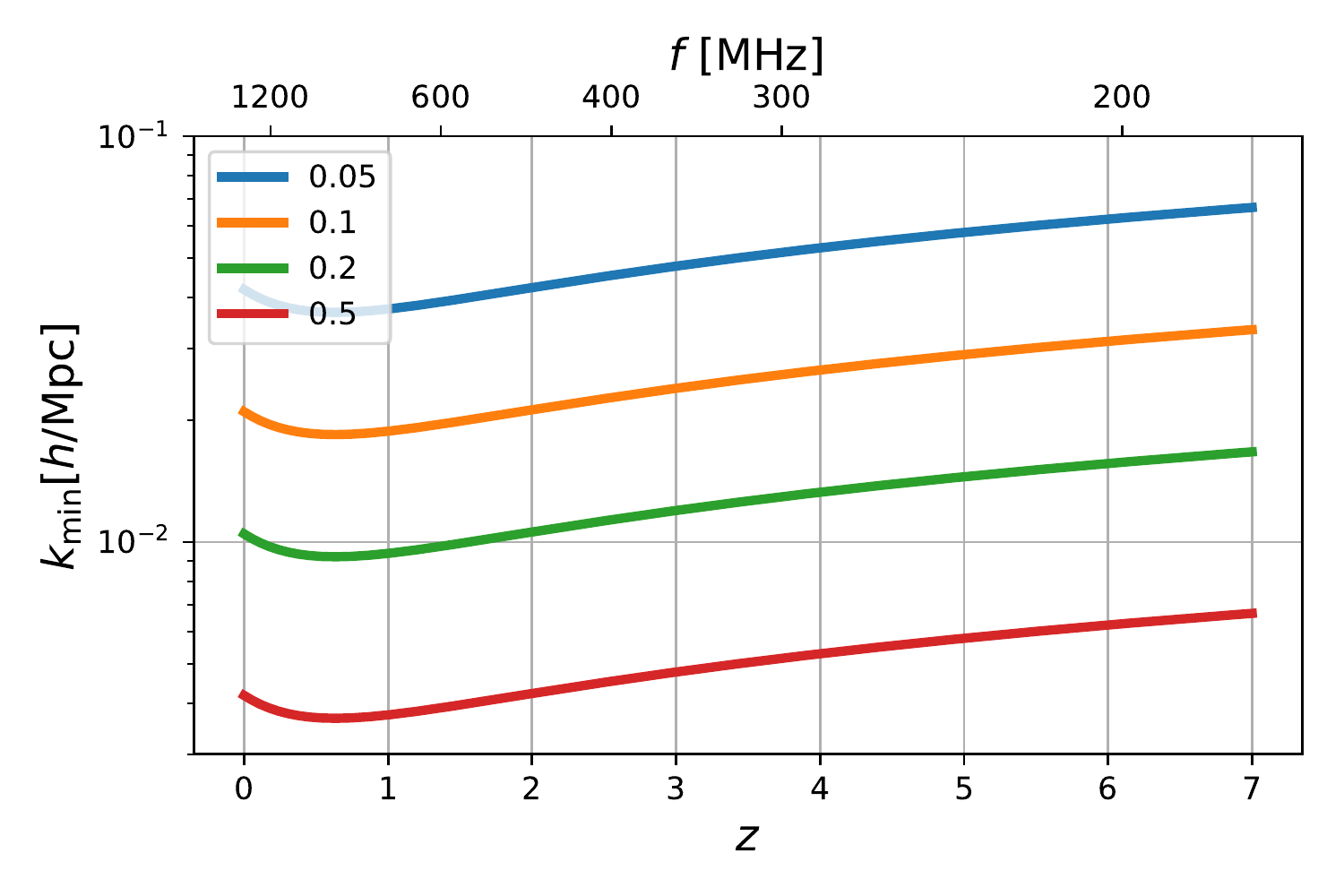}
  \caption{The minimum $\kpar$ accessible as a function of redshift
    for different choices of fractional bandwidth.  Note that the
    curves are quite flat as a function of redshift.  For
    $\Delta f/f=0.2$, $k_{\rm min}\simeq 10^{-2}\,h\,{\rm Mpc}^{-1}$
    which we shall assume for our forecasts.  }
  \label{fig:kmin}
\end{figure}

Foregrounds present a major calibration issue for $21\,$cm cosmology. At
a minimum, one loses low $\kpar$ modes due to filtering of smooth
foregrounds. Many foregrounds on the sky are (within a crude approximation) 
slowly varying functions of frequency \cite{2008MNRAS.388..247D,2016A&A...594A..10P,2008MNRAS.389.1319J}, so a perfectly calibrated instrument will have a minimum accessible (i.e., not foreground contaminated) value of $\kpar$ corresponding to the fundamental mode that fits in the
radial range under consideration. In practice, however, amplifier
gain stability and beam response changes due to changing environmental factors
(e.g.\ temperature affecting the shape of the reflector),
mean that the lowest accessible $\kpar$ will be somewhat
higher. It is useful to parameterize this in terms of the fractional
bandwidth over which we consider the instrument can be perfectly calibrated,
since both mechanical and analog electronic drivers
scale with $\Delta f/f$.
In Figure~\ref{fig:kmin} we plot the minimum value of $\kpar$ (and thus
total $k=\sqrt{\kpar^2+\kperp^2}$) accessible as a function of
fractional bandwidth. We find that it is only a weak function of
redshift. For 20\% fractional bandwidth we find that
$k_{\rm min}\simeq 10^{-2}\,h\,{\rm Mpc}^{-1}$ is appropriate over a wide
range of redshifts.  We shall assume this $k_{\rm min}$ in our forecasts.

A different issue, first discovered in the context of the epoch of
reionization experiments is the the foreground wedge
\cite{Datta2010,Vedantham2012,Morales2012,2012ApJ...756..165P,Trott2012,Thyagarajan2013,pober_et_al2013b,dillon_et_al2014,Hazelton2013,Thyagarajan_et_al2015a,Thyagarajan_et_al2015b,2014PhRvD..90b3018L,2014PhRvD..90b3019L,2015MNRAS.447.1705P,chapman_et_al2016,pober_et_al2016,seo_and_hirata2016,jensen_et_al2016,kohn_et_al2016}
(see also Section~\ref{sec:intro-pract-limit}). It has mainly been studied
for interferometric $21\,$cm experiments, although a related issue
also exists for single-dish experiments.  The foreground wedge results
from the fact that a given interferometric baseline has a fixed
physical length, which implies that it probes different angular scales
at different frequencies ($\theta \propto\lambda^{-1}$).
Interferometers are therefore inherently chromatic, and intrinsically
smooth-spectrum foregrounds can appear to have significantly more
complicated spectra.  This effect can be reduced by careful
inter-baseline calibration, which could in principle be achieved by a
carefully designed array with a sufficient density of baselines.
Achieving such calibration requirements in existing experiments,
however, has proven elusive.

We model the `wedge' as a cut on $\mu$, the cosine of the angle along the line of sight, assuming
all signal modes with $\mu<\mu_w$ are lost. The wedge is particularly acute at higher redshifts,
since the value of $\mu_w$ increases with redshift (Eq.~\ref{eq:wedge}).
In general, the wedge effects can be thought of as being caused
by sources from different parts of the sky, with sources away from phase center
being particularly affected. The most pessimistic case (known as the ``horizon wedge'')
assumes all sources above the horizon can contaminate the signal. We take a less pessimistic assumption, and only consider contamination from sources that are no further than $N_w$ times the size of the
primary beam away from the beam center.  In Figure~\ref{fig:volModes}, we show the effective loss of observed volume and number of linear modes for these cases for an experiment with 6-m dishes.  We see that the
effect is dramatic for the horizon wedge, but even in this case our fiducial experiment achieves a fifty-fold increase in the number of measured linear modes compared to an optical survey at $z<2$.

We take the position that this systematic will have to be overcome to
fully exploit the possibilities offered by the $21\,$cm technique.  We
reiterate that it is a technical rather than fundamental
problem. Instrumental design choices are vital to support this -- for
example, dishes result in a characteristic `pitchfork'-shaped region
of foreground contamination within the wedge, which leaves modes
between the pure radial ($k_\parallel \sim 0$) and horizon boundary of
the wedge relatively uncontaminated, while dipoles have strong
contamination throughout the entire wedge region in Fourier space
(i.e., it results in the loss of all modes with $\mu<\mu_w$).
Other design choices, such as reducing sidelobes and generally improving the
stability of the primary beam response with frequency will also be
valuable for allowing modes inside the wedge to be recovered. There
have also been promising methodological advances that render full wedge
calibration realistic in the future \cite{Koopmans2017}.
Therefore, when forecasting, we use two possibilities: we either
assume that the wedge has been completely calibrated out (optimistic)
or that calibration allows us to cut at $N_w=3$ times the position of the
primary beam (pessimistic).  This is motivated by the notion that for a
typical antenna design, the beam response is suppressed at the
signal/foreground level at those distances.

The $N_w=3\times$ primary-beam wedge assumption was realised by only
considering modes that satisfy
\begin{equation}
  \kpar > k_{\perp} \frac{\chi(z)H(z)}{c(1+z)} \sin(\theta_w),
\label{eq:wedge}
\end{equation}
where $\theta_w$ is the maximum angle at which fringes from a
monochromatic point source can enter the measurement and be confused
with a non-monochromatic source at phase center. Given that the beam shape
is idealised in our experiment, we take
$\theta_w = N_w 1.22\lambda/2D_{\rm eff}$\footnote{We note that the factor of 2
  in the denominator here is ad-hoc, for an airy disk, the first null
  as measured from the center is at $1.22\lambda/D$ and we then take
  this distance to represent an effective full width.}, although other
choices can be found in the literature, e.g.
$\theta_w= N_w 1.06 \lambda/2D_{\rm eff}$
\cite{2014ApJ...782...66P}. See next section for the discussion of the
effective dish size.

\section{Instrumental noise of \stagetwo\ experiment}
\label{app:forecasts-noise}

We take the \stagetwo\ to be a compact square array of $256^2 \approx 65000$ fully illuminated dishes with diameter $D_{\rm phys}=6\,{\rm m}$. We assume an integration time of 5 years (at 100\% efficiency) over half the sky ($f_{\rm sky}=0.5$).

We take the total system temperature to be
\begin{equation}
T_{\rm sys} = \frac{1}{\eta_{\rm c} \eta_{\rm s}} T_{\rm ampl} + \frac{1-\eta_{\rm s}}{\eta_{\rm s}} T_{\rm ground} + T_{\rm sky} \ ,
\end{equation}
with the following contributions:
\begin{itemize}
\item We assume the amplifier noise temperature to be $T_{\rm ampl}=50{\rm
  K}$, which is conservative compared to the best available amplifiers (which
already reach better noise figures). However, the ultra-wide-band feeds considered in this experiment will pose their own
set of challenges. In particular, due to compromises necessary to
achieve sufficient coupling to vacuum, the optical efficiency is reduced from unity and is expected to be $\eta_{\rm c} = 0.9$. A somewhat subtle point is that this optical efficiency factor decreases both signal and the sky noise by the same
factor, and as such, we model it by increasing just the effective
amplifier noise.
\item We expect that a non-negligible $1-\eta_s\sim 0.1$ fraction of our our primary beam
hits the ground at $T_{\rm ground}=300$K instead of being coupled to the sky.
\item We take the sky temperature to be
\begin{equation}
T_{\rm sky} (f) =  \left( \frac{f}{400 {\rm MHz}}\right)^{-2.75} 25 {\rm K} + 2.7{\rm
  K}. 
  \label{eq:sky_temperature}
\end{equation}
This approximation is consistent with assumptions made in the SKA
forecasting exercise \cite{skaf1, skaf2} and also with effective temperature
derived by averaging $T^{-2}$ over the Haslam 408~MHz galaxy map
\cite{2015MNRAS.451.4311R} (i.e. approximately taking into account the inverse variance
weighting one might do in practice).
\end{itemize}

These forecasts can be compared to achieved system temperatures. For
example, BMX prototype at BNL has an achieved system temperature at
1300MHz of around 70K which compares well to prediction of 80K
assuming $T_{\rm ampl}\approx 40K$, $\eta_s\approx 0.9$ and
$\eta_c\approx 1$ which are more aggressive numbers than what we
assume for \stagetwo. Similarly, CHIME has achieved a system
temperature without sky contribution of around 60K, which is also
somewhat better than these equations predict.

The non-uniform illumination of the primary reduces the effective size
of the dish.  One might naively expect that the dish illumination will
further decrease with frequency at fixed physical OMT size. However,
practical modelling has shown that this is not the case, as the portion
of the total device that is active decreases with frequency, making
the beam hitting the primary reflector nearly frequency-independent. We thus assume that the effective dish area is just a
scaled version of the physical dish area with aperture efficiency
factor of $\eta_a=0.7$ (see Chapter 9.6 of \cite{stutzman2013antenna}
for discussion of practically achievable aperture efficiencies),
namely
\begin{equation}
  D^2_{\rm eff}  = \eta_{\rm a} D^2_{\rm phys}\ .
  \end{equation}

The power spectrum of system noise is then given by (e.g.~\cite{2017arXiv170907893O})
\begin{equation}
  P_{\rm N}(\vk,z) = T_{\rm sys}(z)^2 \chi(z)^2 \lambda(z) \frac{ 1+z}{H(z)}
  \left(\frac{\lambda(z)^2}{A_{\rm e}}\right)^2 \left(\frac{1}{N_{\rm pol} t_{\rm survey}
  	\times n_{\rm b}(u=k_\perp \chi(z)/2\pi)
      } \right) \left(\frac{S_{\rm area}}{{\rm FOV}(z)}\right),
\label{eq:pnoise}
\end{equation}
where $\chi(z)$ is the comoving distance to the observed slice,
$\lambda_0\approx 21$cm is the transition rest-frame frequency,
$\lambda(z)=\lambda_0(1+z)$ is the observing wavelength, $S_{\rm area}=4\pi f_{\rm sky}$ is the total survey area, $N_{\rm pol}=2$ is the number of polarizations per feed, and $n_{\rm b}(u)$ is the
number density of baselines in the $uv$ plane. The effective collecting area per feed and effective field of view are given by

\begin{equation}
A_{\rm e}  =\pi \left( \frac{D_{\rm eff}}{2}\right)^2\ , \quad 
	{\rm FOV} =  \left( \frac{\lambda}{D_{\rm eff}} \right)^2\ .
\end{equation}

\begin{figure}
   \centering
   \includegraphics[width=0.6\linewidth]{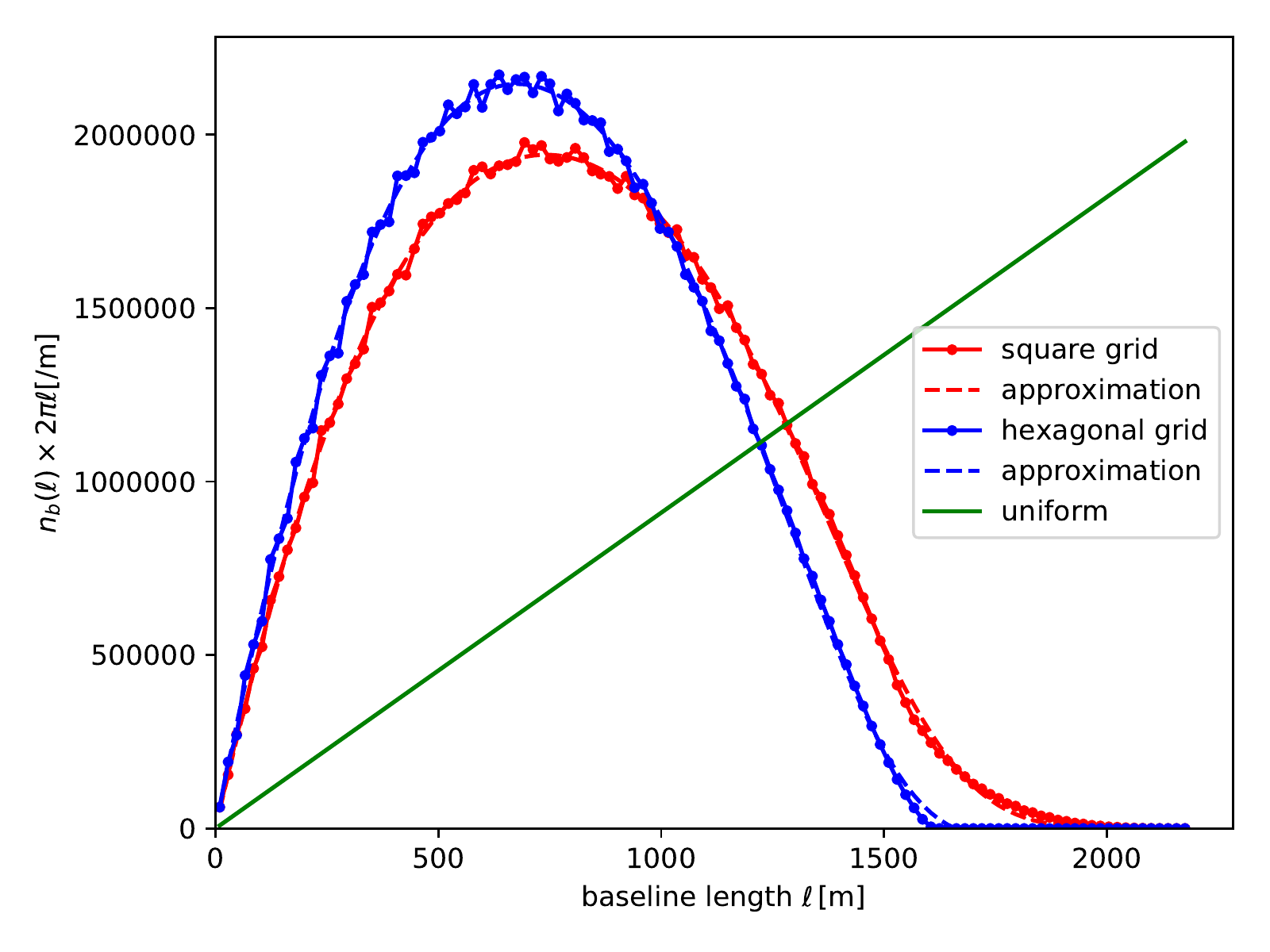}

   \caption{The number of baselines per unit radial distance (i.e. the
     integral under the above curve equals to the total number of
     baselines) for the \stagetwo\ experiment for square close packing (red) arranged in a $256^2$ square hexagonal close-packing arranged in a circle (blue) for the same number of antennas.  We plot the exact numerical results as points, our
     fitting formula as dashed lines and the approximation of constant
     $n_b(u)$ in green.  }
   \label{fig:nufit}
 \end{figure}

Many results in the literature rely on the approximation that the baseline density $n_{\rm b}(u)$
is independent of $u$ up to some maximum baseline length $u_{\rm max}$: that is, for a square array with $N_{\rm s}^2$ receivers, 
\begin{equation}
n_{\rm b}(u) = \frac{N_{\rm s}^2/2}{\pi u_{\rm max}^2}\ .
\end{equation}
 We have found that this is a surprisingly poor approximation (see
 Figure \ref{fig:nufit}, also discussed below). Instead,
 we use the following fitting formula for the number of baselines as a
 function of physical distance of antennas
 \begin{equation}
   n_{\rm b}^{\rm phys}(l) = n_0\frac{a+b (l/L)}{1+c (l/L)^d} e^{-(l/L)^e},
 \end{equation}
 where $n_0=(N_{\rm s}/D_{\rm phys})^2$, $L=N_{\rm s} D_{\rm phys}$, and the $uv$-plane density  is
 \begin{equation}
 n_{\rm b}(u) = \lambda^2\, n_{\rm b} ^{\rm phys}(l=u\lambda) \ .
 \end{equation}
 This formula has been fitted
 to our fiducial case and calibrated so that
 $\int n_{\rm b}(u) d^2u = N_{\rm baselines}^2/2 \approx N_{\rm s}^4/2$. The fitting parameters are
 \begin{align*}
\text{square close-packing:}&\quad a=0.4847,\, b=-0.3300, \, c=1.3157, \, d=1.5974, \, e=6.8390, \\
\text{hexagonal close-packing in a compact circle:}&\quad a=0.5698, \, b=-0.5274, \, c=0.8358, \, d=1.6635,  \,e=7.3177.
 \end{align*}
This formula works well
 down to a very small value of $N_{\rm s}$; even with $N_{\rm s}=8$, the total
 number of baselines matches the exact calculation to within a few percent. Figure \ref{fig:nufit} shows
 the numerical result together with the fitting formula. The shown
 formula is only true for an observing field at zenith, but we ignore
 this effect in our forecasting.

\section{Figures \ref{fig:panella} and \ref{fig:panella2}}
\label{app:panella}

Figures \ref{fig:panella} and \ref{fig:panella2} were made as
follows. A numerical simulation with $3072^3$ particles in a box of size 300 Mpc/h has been run using the \textsc{l-picola} code \citep{l-picola}. Halos were identified using the Friends-of-Friends algorithm \citep{FoF}, with a value of the linking length parameter $b=0.2$. Neutral hydrogen was then assigned to halos according to \cite{2018ApJ...866..135V}.

For LSST we assumed a photometric error of $\sigma_z=0.032(1+z)$ and
number density according to the fitting formula from the Appendix of \cite{2018arXiv180901669T}.

For dropout survey we assumed number densities of
$1.6\times 10^{-4}/({\rm Mpc}/h)^3$ ($m_{\rm UV}<24$) and
$6.0\times 10^{-4}/({\rm Mpc}/h)^3$ ($m_{\rm UV}<24.5$) at $z=3$ and
$5\times 10^{-6}/({\rm Mpc}/h)^3$ ($m_{\rm UV}<24$) and
$4\times 10^{-5}/({\rm Mpc}/h)^3$ ($m_{\rm UV}<24.5$) at $z=5$
respectively, following \cite{Stephen18}.

For Stage 2 we have assumed foreground filtering of modes with
$\kpar<0.01h$/Mpc, which for the simulation size of this box filters
just modes with $\kpar=0$ and $\kperp\geq 0$. For beam filtering we have
applied a simple Gaussian filtering with variance given by the linear
size of the array.

\clearpage

\bibliography{whitepaper}

\end{document}